\documentclass[a4paper,11pt]{article}
\pdfoutput=1 
\usepackage{jcappub}
\usepackage{amsmath}
\usepackage{graphicx}
\usepackage{latexsym}
\usepackage{xspace}
\usepackage{color}
\usepackage{hyperref} 
\usepackage{bm}
\usepackage{relsize}
\usepackage{tabularx}
\usepackage{multirow}
\usepackage{amssymb}
\usepackage[table]{xcolor}
\usepackage{slashbox}
\usepackage{braket}
\usepackage{comment}
\usepackage[utf8]{inputenc}
\usepackage{slashed}

\allowdisplaybreaks

\makeatletter
\gdef\@fpheader{}
\g@addto@macro\bfseries{\boldmath}
\makeatother

\newcommand{\ds}{\displaystyle}
\newcommand{\ovl}[1]{\overline{#1}}

\newcommand{\CPT}{CPT}

\newcommand{\ie}{\textsl{i.e.~}}

\newcommand{\eg}{\textsl{e.g.~}}

\newcommand{\etc}{\textsl{etc.~}}

\newcommand{\dd}{\mathrm{d}}

\newcommand{\sss}[1]{{\scriptscriptstyle{#1}}}

\newcommand{\uPl}{\mathrm{Pl}}

\newcommand{\usssPl}{\sss{\uPl}}

\newcommand{\Mp}{M_\usssPl}

\newcommand{\efolds}{$e$-folds~}

\newcommand{\beq}{\begin{equation}}
\newcommand{\eeq}{\end{equation}}
\newcommand{\bea}{\begin{eqnarray}}
\newcommand{\eea}{\end{eqnarray}}

\newlength{\wsingfig}
\newlength{\wdblefig}
\newlength{\wquadfig}
\newlength{\wtriplefig}
\newcommand{\Eq}[1]{Eq.~(\ref{#1})}
\newcommand{\Eqs}[1]{Eqs.~(\ref{#1})}

\newcommand{\Refc}[1]{Ref.~{\cite{#1}}}
\newcommand{\Refs}[1]{Refs.~{\cite{#1}}}
\newcommand{\Sec}[1]{Sec.~\ref{#1}}
\newcommand{\Secs}[1]{Secs.~\ref{#1}}
\newcommand{\App}[1]{Appendix~\ref{#1}}

\subheader{}

\title{Hamiltonian formalism for cosmological perturbations: the separate-universe approach}

\author[a,b]{Danilo Artigas,}
\emailAdd{danilo.artigas-guimarey@universite-paris-saclay.fr}

\author[a,c]{Julien Grain,}
\emailAdd{julien.grain@universite-paris-saclay.fr}

\author[c]{Vincent Vennin}
\emailAdd{vincent.vennin@apc.in2p3.fr}

\affiliation[a]{Universit\'e Paris-Saclay, CNRS, Institut d'Astrophysique Spatiale, 91405, Orsay, France}
\affiliation[b]{Institute of Theoretical Physics, Jagiellonian University, \L ojasiewicza 11, 
30-348 Cracow, Poland, EU}
\affiliation[c]{Laboratoire Astroparticule et Cosmologie, CNRS Universit\'e de Paris, 75013 Paris, France}

\date{today}

\begin{document}

\sloppy

\abstract{The separate-universe approach provides an effective description of cosmological perturbations at large scales, where the universe can be described by an ensemble of independent, locally homogeneous and isotropic patches. 
By reducing the phase space to homogeneous and isotropic degrees of freedom, it greatly simplifies the analysis of large-scale fluctuations. It is also a prerequisite for the stochastic-inflation formalism.
In this work, we formulate the separate-universe approach in the Hamiltonian formalism, which allows us to analyse the full phase-space structure of the perturbations. Such a phase-space description is indeed required in dynamical regimes which do not benefit from a background attractor, as well as to investigate quantum properties of cosmological perturbations. 
We find that the separate-universe approach always succeeds in reproducing the same phase-space dynamics for homogeneous and isotropic degrees of freedom as the full cosmological perturbation theory, provided that the wavelength of the modes under consideration are larger than some lower bound that we derive. We also compare the separate-universe approach and cosmological perturbation theory at the level of the gauge-matching procedure, where the agreement is not always guaranteed and requires specific matching prescriptions that we present. }


\arxivnumber{2110.11720}

\maketitle

\flushbottom

\section{Introduction}
\label{sec:intro}
Cosmological-perturbation theory (CPT) is a pillar of our modern understanding of cosmology. It consists in describing small deviations from highly-symmetric background space-times by means of perturbative techniques, while accounting for the fundamental invariance of general relativity under changes of coordinates. When CPT is developed around a homogeneous and isotropic background, an important simplification may occur at large scales (\ie on distances larger than the length scale associated with the universe expansion -- or contraction -- rate) if the universe can be described by an ensemble of independent, locally homogeneous and isotropic patches. This picture is called the separate-universe approach~\cite{Salopek:1990jq,Sasaki:1995aw,Wands:2000dp,PhysRevD.68.103515,PhysRevD.68.123518,Lyth:2005fi, Tanaka:2021dww} and is also known as the quasi-isotropic picture \cite{Lifshitz:1960,Starobinsky:1982mr,PhysRevD.49.2759,Khalatnikov_2002}. When applicable, it implies that studying the large-scale cosmological perturbations boils down to solving the homogeneous and isotropic problem with different initial conditions, which allows one to track only a subset of the relevant degrees of freedom. This represents a very substantial technical simplification.

Another simplification that occurs on large scales is when the background evolution features a dynamical phase-space attractor. This is for instance the case in inflating backgrounds, if inflation proceeds in the so-called slow-roll regime. In that case, if the separate-universe approach can be used, perturbations are subject to the same attractor, which makes them collapse on a phase-space subset (the dimension of which equals the number of matter fields), removing the dependence on initial field velocities. This further reduction of the effective phase-space makes the use of the Lagrangian framework convenient, which explains why most analyses of the separate-universe approach and of the conditions for its validity have been carried out in the Lagrangian framework.

However, there are situations in which the background dynamics is not endowed with such an attractor, hence the full phase-space structure must be considered. For instance, this is the case if inflation proceeds in the so-called ultra-slow-roll regime (which may or may not be stable, see \Refc{Pattison:2018bct}, but which always retains dependence on the initial field velocities), or for some contracting cosmologies (see \eg \Refs{Miranda:2019ara, Grain:2020wro}). In particular, bouncing cosmologies, where the classical expansion is preceded by a contracting phase and a regular bounce, are typical examples of alternatives to inflation~\cite{Wands:1998yp,Finelli:2001sr} arising in various contexts such as string theory or loop quantum gravity, see \eg \Refs{Khoury:2001wf,Barrau:2013ula,Brandenberger:2016vhg,Agullo:2016tjh}. 
 
In such situations it is important to be able to describe and compare \CPT~and the separate-universe approach in the phase space, \ie using the Hamiltonian formalism. While \CPT~has already been investigated in this framework, see \eg \Refs{Langlois:1994ec, Domenech:2017ems}, the aim of the present work is to discuss the Hamiltonian version of the separate-universe approach. Our goal is both to construct the separate-universe formalism in the Hamiltonian picture, and to establish the conditions under which it properly describes the full phase-space properties of cosmological perturbations. 

Note that a phase-space formulation of cosmological perturbations (either in CPT or in the separate-universe approach) is also crucial when it comes to describing them at the quantum-mechanical level. For instance, as shown in \Refc{Grain:2019vnq}, the choice of an initial vacuum state is intimately related to the choice of a phase-space parametrisation. In slow-roll inflation, a large class of parameterisations leads to the same vacuum state, namely the Bunch-Davies vacuum, but the situation is less clear in general and makes a phase-space formulation appropriate. 

Let us mention that the present work lays the ground for upcoming articles in which we will further investigate the gauge formalism (\ie transformation under changes of coordinates, the gauge-fixing procedure, and the construction of a gauge-invariant parametrisation of the phase space) in the Hamiltonian framework, both in full CPT and in the separate-universe approach. In this article, we will discuss the gauge-fixing procedure in most commonly-used gauges, since it plays an important role in comparing the separate-universe formalism with CPT, but one should bear in mind that this discussion will be complemented by a more systematic analysis in follow-up publications.

Another motivation behind this analysis is the so-called stochastic-inflation formalism~\cite{Starobinsky:1982ee,Starobinsky:1986fx}, which heavily relies on the separate-universe framework. In this approach, quantum cosmological fluctuations act as a stochastic noise on the large-scale evolution as they cross out the Hubble radius (either during inflation or during a slowly contracting era). The stochastic formalism has been extensively used in the context of slow-roll inflation where it has been shown to be in very good agreement with predictions from quantum-field-theoretic calculations (see \eg \Refs{Starobinsky:1994bd,Tsamis:2005hd,Finelli:2010sh,Garbrecht:2013coa,Onemli:2015pma,Burgess:2015ajz,Kamenshchik:2021tjh}) and to preserve the attractor nature of the slow-roll regime~\cite{Grain:2017dqa}. 
Combined with the $\delta N$ formalism, where curvature perturbations on large scales are related to the fluctuations in the local amount of expansion, it gives rise to the stochastic-$\delta N$ formalism~\cite{Vennin:2015hra, Fujita:2013cna}, which allows one to incorporate quantum backreaction in the calculation of the density field of the universe. This plays an important role notably in the analysis of primordial black hole production, which usually requires a phase of strong stochastic effects~\cite{Pattison:2017mbe, Biagetti:2018pjj, Ezquiaga:2019ftu}. 

Given that primordial black holes form in models where deviations from the slow-roll attractor are also observed (in particular along the ultra-slow-roll regime), the stochastic-$\delta N$ formalism has been recently extended beyond the slow-roll setup~\cite{Nakao:1988yi, PhysRevD.46.2408, Rigopoulos:2005xx, Tolley:2008na, Weenink:2011dd, Grain:2017dqa, Firouzjahi:2018vet, Ezquiaga:2018gbw, Pattison:2019hef, Pattison:2021oen}, where the full phase-space structure of the fields needs to be resolved. The present analysis will therefore confirm the validity of this approach by studying the separate-universe in the Hamiltonian framework. In practice, \Refc{Pattison:2019hef} pointed out that the derivation of the Langevin equations of stochastic inflation requires to perform a gauge transformation, from the spatially-flat gauge where the free scalar-field correlators are computed, to the uniform-expansion gauge where the stochastic noise needs to be expressed. Our goal is also to derive the tools required to perform such a transformation on generic grounds, in the full phase space of the separate-universe system. 
As mentioned above, another situation where a dynamical attractor is not always available is the case of slowly contracting cosmologies, so the present work can be seen as a prerequisite for the development of a ``stochastic-contraction'' formalism~\cite{Miranda:2019ara, Grain:2020wro}, which will be the topic of future works.

The paper is organised as follows. In \Sec{sec:CosmoHam}, we briefly review the basics of the phase-space (or Hamiltonian) formulation of general relativity, and apply it to the case of homogeneous and isotropic cosmologies. We then incorporate cosmological perturbations in the formalism, using \CPT~in \Sec{sec:cosmopert} and with the separate-universe setup in \Sec{sec:SepUniv}. We compare the two approaches in \Sec{sec:suvspert}, both at the level of the phase-space dynamics and of the gauge-fixing procedure. Our results are summarised and further discussed in \Sec{sec:conclusion}, and we end the paper with four appendices to which various technical aspects of the
calculations presented in the main text are deferred.
\section{Cosmology in the Hamiltonian formalism}
\label{sec:CosmoHam}
\subsection{Hamiltonian description of general relativity}
\label{sec:Hamiltonian:GR}
Let us start by reviewing the basics of the Hamiltonian formulation of general relativity (see \eg \Refc{thieman_book} for a detailed mathematical introduction and \Refc{Langlois:1994ec} for an application to the cosmological context). Since our work takes place in the context of primordial cosmology, for explicitness, we consider the case where the matter content of the universe is given by a single scalar field, $\phi$, minimally coupled to gravity in a four-dimensional curved space-time with metric $g_{\mu\nu}$. The total action then reads 
\bea
\label{eq:action}
	S=\ds\int\dd^4x\sqrt{-g}\left[\frac{\Mp^2}{2}R-\frac{1}{2}g^{\mu\nu}\partial_\mu\phi\,\partial_\nu\phi-V(\phi)\right],
\eea
where $g$ is the determinant of $g_{\mu\nu}$, $R$ is the four-dimensional Ricci scalar, $V(\phi)$ is the scalar field potential, and $\Mp$ is the reduced Planck mass. The Hamiltonian formulation is obtained by foliating the four-dimensional space-time into a set of three-dimensional space-like hypersurfaces, $\Sigma_\tau$, where the foliation is defined by the lapse function, $N(\tau,\vec{x})$, and the shift vector $N^i(\tau,\vec{x})$. Here, $\tau$ stands for the time variable and $\vec{x}$ for the spatial coordinates on the hypersurfaces. The line element is then expressed in the ADM form~\cite{Arnowitt:1962hi}
\bea
\label{eq:metric:ADM}
	\dd s^2=-N^2(t,\vec{x})\dd\tau^2+\gamma_{ij}(\tau,\vec{x})\left[\dd x^i+N^i(\tau,\vec{x})\dd\tau\right]\left[\dd x^j+N^j(\tau,\vec{x})\dd\tau\right].
\eea
In the above expression, $\gamma_{ij}$ is the induced metric on the spatial hypersurfaces $\Sigma_\tau$. Indices are lowered by $\gamma_{ij}$ and raised by its inverse $\gamma^{ij}$.

The canonical variables for the scalar field are $\phi$ and its conjugate momentum $\pi_\phi:=\delta S/\delta \dot\phi$, where we have introduced the notation $\dot{f}:=\dd f/\dd\tau$ for the time derivative, and $\delta S/\delta\dot\phi$ is the functional derivative. The associated Poisson bracket is $\left\{\phi(\tau,\vec{x}),\pi_\phi(\tau,\vec{y})\right\}=\delta^3(\vec{x}-\vec{y})$.
Similarly for the gravitational sector, the canonical variables are the induced metric, $\gamma_{ij}$, and its associated momentum $\pi^{ij}:=\delta S/\delta\dot\gamma_{ij}$, with the Poisson bracket reading $\left\{\gamma_{ij}(\tau,\vec{x}),\pi^{mn}(\tau,\vec{y})\right\}=\frac{1}{2}\left(\delta^m_i\delta^n_j+\delta^n_i\delta^m_j\right)\delta^3(\vec{x}-\vec{y})$.
Here, $\delta^3(\vec{x})$ stands for the Dirac distribution while $\delta^i_j$ is the Kronecker symbol. Since the time derivatives of the lapse function and of the shift vector do not appear in the action~\eqref{eq:action}, they are Lagrange multipliers, corresponding to the freedom in the choice of the coordinate system. The dynamics of the gravitational and scalar-field degrees of freedom is then derived from the following total Hamiltonian
\bea
\label{eq:full:Hamiltonian}
	C\left[N,N^i\right]=\ds\int\dd^3\vec{x}\left[N\left(\mathcal{S}^{(\mathrm{G})}+\mathcal{S}^{(\phi)}\right)+N^i\left(\mathcal{D}^{(\mathrm{G})}_i+\mathcal{D}^{(\phi)}_i\right)\right], 
\eea
which is obtained from \Eq{eq:action} by performing a Legendre transform, and
where $\mathcal{S}:=\mathcal{S}^{(\mathrm{G})}+\mathcal{S}^{(\phi)}$ is the scalar (or energy) constraint and $\mathcal{D}_i:=\mathcal{D}^{(\mathrm{G})}_i+\mathcal{D}^{(\phi)}_i$ is the vector (or momentum/diffeomorphism) constraint, which both receive contributions from the gravitational sector and from the scalar-field sector.\footnote{The term ``smeared constraint" often appears in the literature and refers to the spatial integral of the corresponding constraint and its associated Lagrange multiplier (for instance, $\int\dd^3 \vec{x} N \mathcal{S}$ is the smeared scalar constraint).\label{footnote:smeared:constraints}} As functions of the canonical variables, these constraints are given by
\bea
	\mathcal{S}^{(\mathrm{G})} & = &\frac{2}{\Mp^2\sqrt{\gamma}}\left(\pi^{ij}\pi_{ij}-\frac{\pi^2}{2}\right)-\frac{\Mp^2\sqrt{\gamma}}{2}\,\mathcal{R}(\gamma_{ij}), \label{eq:scgg} \\
	\mathcal{D}^{(\mathrm{G})}_i & = & -2\partial_m\left(\gamma_{ij}\pi^{jm}\right)+\pi^{mn}\partial_i\gamma_{mn}\, ,\label{eq:dcgg}
\eea
where $\gamma$ stands for the determinant of $\gamma_{ij}$, $\pi:=\gamma_{ij}\pi^{ij}$ is the trace of the gravitational momentum, and
\bea
	\mathcal{S}^{(\phi)} & = & \frac{1}{2\sqrt{\gamma}}\pi^2_\phi+\frac{\sqrt{\gamma}}{2}\gamma^{ij}\partial_i\phi\,\partial_j\phi+\sqrt{\gamma}V(\phi), \label{eq:scsg} \\
	\mathcal{D}^{(\phi)}_i&  =& \pi_\phi\partial_i\phi\, .\label{eq:dcsg}
\eea
The ``gravitational potential term'' is given by the three-dimensional Ricci scalar, $\mathcal{R}(\gamma_{ij})$, associated to the induced metric on the spatial hypersurfaces $\Sigma_\tau$. The equation of motion for any function $F$ of the phase-space variables is thus obtained using the full Poisson bracket
\bea
\label{eq:PoissonBracket}
	\left\{F,G\right\}=\ds\int\dd^3x\left[\left(\frac{\delta F}{\delta\gamma_{ij}}\frac{\delta G}{\delta \pi^{ij}}-\frac{\delta G}{\delta\gamma_{ij}}\frac{\delta F}{\delta \pi^{ij}}\right)+\left(\frac{\delta F}{\delta\phi}\frac{\delta G}{\delta \pi_\phi}-\frac{\delta G}{\delta\phi}\frac{\delta F}{\delta \pi_\phi}\right)\right]
\eea
and the above total Hamiltonian, \ie
\bea
	\dot{F}(\phi,\pi_\phi;\gamma_{ij},\pi^{mn})=\left\{F(\phi,\pi_\phi;\gamma_{ij},\pi^{mn}),C\left[N,N^i\right]\right\}. \label{eq:HamEqGen}
\eea
In addition, the dynamics is constrained to lie on the phase-space surface where both the scalar and the diffeomorphism constraints vanish, \ie $\mathcal{S}^{(\mathrm{G})}+\mathcal{S}^{(\phi)}=0$ and $\mathcal{D}^{(\mathrm{G})}_i+\mathcal{D}^{(\phi)}_i=0$. This is so because minimisation of the action has to hold for any arbitrary choice of the lapse function and the shift vector, appearing as Lagrange multipliers in the Hamiltonian. Furthermore, one can show that the Poisson bracket between constraints yields only combinations of the same constraints, \ie these are ``first-class'' constraints in Dirac's terminology. As a consequence, the constrained surface in the phase space is preserved through the dynamical evolution generated by the total Hamiltonian.\footnote{This holds by virtue of the contracted Bianchi identities. Indeed, one can rewrite the scalar and diffeomorphism constraints in terms of the Einstein-Hilbert tensor $G_{\mu\nu}$ and the energy momentum tensor $T_{\mu\nu}$ as \cite{Arnowitt:1962hi}:
\bea
	\mathcal{S}^{(G)} + \mathcal{S}^{(\phi)} & = & G^0_{0} -  \frac{T^0_{\,0}}{\Mp^2} ,  \\
	\mathcal{D}^{(G)}_i + \mathcal{D}^{(\phi)}_i & = & G^0_{i} -  \frac{T^0_{\,i}}{\Mp^2} \,.
\eea
In terms of the covariant derivative $\nabla$, the contracted Bianchi identities read $\nabla_\mu (G^\mu_\nu -  T^\mu_\nu/\Mp^2)=0$. Under the conditions that the spatial part of Einstein equations holds $(G_{ij}- T_{ij}/\Mp^2=0)$ and that the diffeomorphism constraints are initially satisfied $(G^0_{\,i} -  T^0_{\,i}/\Mp^2=0)$, the Bianchi identities reduce to $\nabla_0 (G^0_{\, \nu}- T^0_{\,\nu}/\Mp^2)=0$. Thus they impose the time invariance of the constrained surface (see part 3.1 of \Refc{Bojowald:2010qpa} for further details).}

The full dynamics in the Hamiltonian framework is thus given by four dynamical equations, obtained by applying \Eq{eq:HamEqGen} to the phase-space variables $(\phi,\pi_\phi;\gamma_{ij},\pi^{mn})$, plus four constraint equations (one from the scalar constraint and three from the diffeomorphism constraint). The dynamical equations for the gravitational sector are rather involved (in particular due to the complexity of $\mathcal{R}$ as a function of the induced metric components) and we will not report them here, see \Refc{thieman_book} for explicit expressions [and \Eq{eq:gammaijdot} for $\dot{\gamma}_{ij}$]. For the scalar-field sector, they take the simple form
\bea
	\dot\phi & =& \frac{N}{\sqrt{\gamma}}\pi_\phi+N^i\partial_i\phi, \\
	\dot\pi_\phi & =& -N\sqrt{\gamma}V_{,\phi}+\partial_i\left(N\sqrt{\gamma}\gamma^{ij}\partial_j\phi\right)+\partial_i\left(N^i \pi_\phi\right),
\eea
where $V_{,\phi}:=\partial V/\partial\phi$. 

\subsection{Homogeneous and isotropic cosmologies}
\label{ssec:bck}
We now apply the Hamiltonian formalism to homogeneous and isotropic cosmologies. More precisely, we consider Friedmann-Lema\^itre-Robertson-Walker (FLRW) space-times, where the metric reduces to
\bea
	\dd s^2=-N^2(\tau)\dd\tau^2+p(\tau)\widetilde{\gamma}_{ij}\dd x^i\dd x^j\, .
\eea
Here, the lapse function, $N$, and $a^2:=p$, depend on time only, and we have introduced the three-dimensional time-independent metric $\widetilde{\gamma}_{ij}$ and its inverse $\widetilde{\gamma}^{ij}$, such that $\widetilde{\gamma}^{ij}\widetilde{\gamma}_{jm}=\delta^i_m$. 

 In this setting, a given choice for the lapse function corresponds to a given choice for the time coordinate. For instance, setting $N=1$ is equivalent to working with the cosmic time (denoted $t$ hereafter), $N=a$ corresponds to conformal time (denoted $\eta$ hereafter), and $N=1/H$ (where $H=\dd \ln(a)/\dd t$ is the Hubble parameter) means working with the number of e-folds, $\ln(a)$ (denoted $\mathcal{N}$ in the following), as the time coordinate. When the time coordinate is left unspecified (\ie when the lapse function is left free), we will use the generic notation $\tau$. 
Note that homogeneity imposes that the shift vector $N^i$ depends on time only, and since a uniform vector field provides a preferred direction unless it vanishes, isotropy further imposes that $N^i=0$.
	
For FLRW space-times, the canonical variables for the gravitational sector can be reduced to a single scalar, $p(\tau)$, and its conjugate momentum, $\pi_p(\tau)$. Since $\gamma_{ij}(\tau)=p(\tau)\widetilde{\gamma}_{ij}$, the link between $\pi^{ij}$ and $\pi_p$ follows from noticing that the action $\tilde{S}$ for the homogeneous and isotropic problem can be obtained by replacing
\bea
\tilde{S}[\phi,\dot{\phi},p,\dot{p}] =\left. S\left[\phi, \dot{\phi},\gamma_{\ij},\dot{\gamma}_{ij}\right]\right\vert_{\gamma_{ij}=p(\tau)\widetilde{\gamma}_{ij}} \,.
\eea
The momentum conjugate to $p$ is thus given by
\bea
\label{eq:pi_p:pi_ij}
\pi_p=
\frac{\delta \tilde{S}}{\delta\dot{p}}=\left.
\frac{\delta\dot{\gamma}_{ij}}{\delta\dot{p}} \frac{\delta S}{\delta\dot{\gamma}_{ij}}\right\vert_{\gamma_{ij}=p(\tau)\widetilde{\gamma}_{ij}}
= \widetilde{\gamma}_{ij}  \left. \pi^{ij} \right\vert_{\gamma_{ij}=p(\tau)\widetilde{\gamma}_{ij}}\, ,
\eea
which can be inverted as
\bea
\label{eq:pi_ij:pi_p}
 \left. \pi^{ij} \right\vert_{\gamma_{ij}=p(\tau)\widetilde{\gamma}_{ij}} = \frac{\pi_p}{3}\widetilde{\gamma}^{ij}\,,
\eea
where we use that $\pi^{ij}$ is proportional to $\widetilde{\gamma}^{ij}$ because of isotropy. Since $(p,\pi_p)$ forms a set of canonically conjugate variables, one can introduce a new Poisson bracket with respect to these variables, which will be denoted with the same brackets for the sake of simplicity.\footnote{The canonical nature of the couple $(p,\pi_p)$ can be further checked by noticing that  $\gamma_{ij}=p\widetilde{\gamma}_{ij}$ can be inverted as $p=\gamma_{ij}\widetilde{\gamma}^{ij}/3$, so together with \Eq{eq:pi_p:pi_ij}, \Eq{eq:PoissonBracket} gives rise to $\{ p,\pi_p\} =\widetilde{\gamma}^{ij}\widetilde{\gamma}_{ij}/3=1 $.}

For the matter sector, homogeneity imposes that $\phi$ and $\pi_\phi$ depend on time only, so phase space can be parametrised by the time-dependent variables $(\phi,\pi_\phi;p,\pi_p)$, in terms of which the scalar constraints reduce to
\bea
\mathcal{S}^{(\mathrm{G})} & = &-\frac{\pi_p^2\sqrt{p}}{3\Mp^2}\, , \\
\mathcal{S}^{(\phi)} & = & \frac{\pi^2_\phi}{2p^{3/2}}+p^{3/2}V(\phi)	\,  . 
\eea
Let us note that thanks to homogeneity, the two diffeomorphism constraints $\mathcal{D}_i^{(\mathrm{G})}$ and $\mathcal{D}_i^{(\phi)}$ identically vanish on this reduced phase space. 

An alternative description of the gravitational sector is through the set of canonical variables
\bea
	v& := &p^{3/2}\, , \\
	\theta &:= &\frac{2\pi_p}{3\sqrt{p}}\, ,
\label{eq:theta:def}
\eea
where $v=a^3$ is the volume variable and $\theta$ is related to the expansion rate of the hypersurfaces $\Sigma_\tau$ (see \App{app:expansion}). It is straightforward to check that $\left\{v,\theta\right\}=1$, and the scalar constraints are now given by
\bea
	\mathcal{S}^{(\mathrm{G})} & = &-\frac{3v\theta^2}{4\Mp^2}\, , \label{eq:BckgConsGrav}\\
	\mathcal{S}^{(\phi)} & = & \frac{\pi^2_\phi}{2v}+vV(\phi)\, . \label{eq:BckgConsPhi} 
\eea
The main advantage of these variables is to remove all the $\sqrt{p}$ dependence. In terms of $v$ and $\theta$, the induced metric and its canonical momentum are $\gamma_{ij}(\tau)=v^{2/3}(\tau)\widetilde{\gamma}_{ij}$ and $\pi^{ij}(\tau)=\frac{1}{2}v^{1/3}(\tau)\theta(\tau)\widetilde{\gamma}^{ij}$.

Let us now derive the constraint and dynamical equations using the variables $(\phi,\pi_\phi;v,\theta)$. The scalar constraint equation $\mathcal{S}=0$, known as the Friedmann equation, reads
\bea
	\theta^2=\frac{4\Mp^2}{3}\left[\frac{\pi_\phi^2}{2v^2}+V(\phi)\right]. \label{eq:ConstHom}
\eea
The equations of motion for the gravitational sector are then given by
\bea
	\dot{v} &= &-\frac{3N}{2\Mp^2} v\theta, \label{eq:Dotv} \\
	\dot{\theta} & =& N\left[\frac{3\theta^2}{4\Mp^2}+\frac{\pi^2_\phi}{2v^2}-V(\phi)\right].
	\label{eq:Dot:theta}
\eea
The first of these dynamical equations exhibits the relation between $\theta$ and the expansion rate given by $\dot{v}/v$. The second equation, known as the Raychaudhuri equation, can be further simplified using the constraint equation~\eqref{eq:ConstHom}, and one obtains
\bea
\label{eq:Raychaudhuri}
	\dot\theta=N\left(\frac{\pi_\phi}{v}\right)^2. 
\label{eq:ThetaPi}
\eea
For the scalar-field sector, the dynamics reads
\bea
	\dot\phi & =& N \frac{\pi_\phi}{v} \, , \label{eq:DotPhiPi} \\
	\dot\pi_\phi & = & -NvV_{,\phi}\, . \label{eq:DotPi}
\eea
Let us note that combining \Eqs{eq:Raychaudhuri} and~\eqref{eq:DotPhiPi} leads to $(\dot\phi)^2=N\dot\theta$, which can be viewed as the second of the Hamilton-Jacobi equations as introduced \eg in \Refs{Salopek:1990jq,Liddle:1994dx}. 

Finally, let us see how the usual form of the Friedmann and Raychaudhuri equations can be recovered. Recalling that $v=a^3$, the Hubble parameter is given by $H_N=\dot{v}/(3v)=-N\theta/(2\Mp^2)$, where we have generalised its definition to an arbitrary lapse function $N$, and where the second expression comes from \Eq{eq:Dotv}. Upon introducing $\rho=\dot{\phi}^2/(2N^2)+V(\phi)$ and $P=\dot{\phi}^2/(2N^2)-V(\phi)$, the energy density and the pressure associated to the scalar field respectively, the Friedmann equation~\eqref{eq:ConstHom} takes the usual form 
\bea
\left(\frac{H_N}{N}\right)^2=\frac{\rho}{3\Mp^2}\, ,
\label{eq:Friedmann:usual}
\eea 
where we have used \Eq{eq:DotPhiPi} to relate $\dot{\phi}$ and $\pi_\phi$. For the Raychaudhuri equation, by combining the second Hamilton-Jacobi equation $(\dot{\phi})^2=N\dot{\theta}$ and the relation $H_N=-N\theta/(2\Mp^2)$ derived above, one obtains the usual form
\bea
\label{eq:Raychaudhury}
\frac{\dot{H}_N}{N^2} = -\frac{\rho+P}{2\Mp^2} \, .
\eea
The Klein-Gordon equation for the scalar field can also be obtained by differentiating \Eq{eq:DotPhiPi} with respect to time, and further using \Eqs{eq:Dotv},~\eqref{eq:DotPi} and the relation $H_N=-N\theta/(2\Mp^2)$, leading to
\bea
\ddot{\phi}+\left(3H_N - \frac{\dot{N}}{N} \right)\dot{\phi}+N^2 V_{,\phi}=0\, .
\eea
\section{Cosmological perturbations}
\label{sec:cosmopert}
Let us now study cosmological perturbations evolving on the homogeneous and isotropic background described in \Sec{sec:CosmoHam}. This can be either done in the Lagrangian formalism, as in \Refs{Mukhanov:1990me,Malik:2008im}, or in the Hamiltonian formalism, as in \Refc{Langlois:1994ec}. In both approaches, working in Fourier space is convenient since, owing to the background invariance under spatial translations, different Fourier modes decouple at leading order in perturbation theory. In practice, any tensor field $T_{i\cdots j}(\tau,\vec{x})$ can be Fourier transformed on the spatial hypersurfaces $\Sigma_\tau$ according to 
\bea
	T_{i\cdots j}(\tau,\vec{k})=\ds\int\frac{\dd^3 \vec{x}}{(2\pi)^{3/2}}\,T_{i\cdots j}(\tau,\vec{x})\,e^{-i\vec{k}\cdot\vec{x}}\, ,
	 \label{eq:Fourier}
\eea
where $\vec{k}\cdot\vec{x}$ is the scalar product $k_ix^i$, and the inverse transform is given by
\bea
	T_{i\cdots j}(\tau,\vec{x})=\ds\int\frac{\dd^3\vec{k}}{(2\pi)^{3/2}}\,T_{i\cdots j}(\tau,\vec{k})\,e^{i\vec{k}\cdot\vec{x}}\, .
\eea
Note that the wavevector $\vec{k}$ is defined with respect to the flat metric on spatial hypersurfaces $\Sigma_\tau$, \ie it is a {\it comoving} wavevector. As a consequence, its indices are raised and lowered with the metric $\widetilde{\gamma}_{ij}$, so for instance $k^2=k_i k^i =\widetilde{\gamma}^{ij}k_ik_j=\widetilde{\gamma}_{ij}k^ik^j$. In practice, we will be considering real-valued tensor fields, for which the Fourier coefficients must satisfy 
\bea
T^\star_{i\cdots j}(\tau,\vec{k})=T_{i\cdots j}(\tau,-\vec{k})\, ,
\label{eq:reality:condition}
\eea
where a star denotes the complex conjugate. Hereafter this contraint will be referred to as the reality condition. 
\subsection{Scalar degrees of freedom}
\label{ssec:dofpert}
In general, cosmological perturbations can be expanded into scalar, vector and tensor degrees of freedom (this is the so-called SVT decomposition~\cite{Lifshitz:1945du}). 
In the following we will focus on scalar perturbations, since they are the main purpose of the separate-universe approach, and given that vector and tensor perturbations can be dealt with in a similar way.

The lapse function $N$ and the variables describing the scalar field sector, $\phi$ and $\pi_\phi$, are scalar quantities, and so are their perturbations. They can be written as
\bea
	\delta N(\tau,\vec{x})&:= &N(\tau,\vec{x}) - N(\tau) \, , \label{eq:DefPerturbations1}\\
	\delta\phi(\tau,\vec{x}) &:= & \phi(\tau,\vec{x}) -\phi(\tau) \, , \label{eq:DefPerturbations2} \\
	\delta\pi_\phi(\tau,\vec{x})&:=&\pi_\phi(\tau,\vec{x}) - \pi_\phi(\tau) \, , \label{eq:DefPerturbations3}
\eea
where the functions $N(\tau)$, $\phi(\tau)$ and $\pi_\phi(\tau)$ are solutions to the homogenous and isotropic problem studied in \Sec{sec:CosmoHam} [hereafter, quantities solving the homogeneous and isotropic problem will always be denoted with the argument ``$(\tau)$''].

The perturbations of the shift vector can be written in a similar way, 
\bea
	\delta N^i(\tau,\vec{x}) := N^i(\tau,\vec{x})- N^i(\tau) \, ,
	\label{eq:DefPerturbations4}
\eea 
where $N^i(\tau)=0$ since the shift vector vanishes in the homogeneous and isotropic setup. According to the SVT decomposition, $\delta N^i$ can be expanded into the gradient of a scalar and a divergence-free vector, namely $\delta N_i = \partial_i (\delta N_1) + (\delta N_2)_i $, where $\delta N_1$ is a scalar and $\delta N_2$ is a vector such that $\partial_i (\delta N_2)^i=0$. As explained above, we focus on scalar perturbations and thus set $\delta N_2=0$. In Fourier space, one has
\bea
\label{eq:deltaNi:deltaN1}
	\delta N^i(\tau,\vec{k})= i \frac{k^i}{k} \delta N_1(\tau,\vec{k})\, ,
\eea
where $\delta N_1(\tau,\vec{k})$ has been rescaled by an overall $k$ factor for later convenience. Note that the reality condition  ${\delta N^i}^\star(\tau,\vec{k})=\delta N^i(\tau,-\vec{k})$, see \Eq{eq:reality:condition}, translates into the same condition for $\delta N_1$, namely $\delta N_1^\star(\tau,\vec{k})=\delta N_1(\tau,-\vec{k})$, since $i\vec{k}$ is invariant under complex conjugation and sign flipping of the wavevector.

The induced metric and its conjugate momentum are perturbed as 
\bea
\label{eq:DefPerturbations5}
\delta\gamma_{ij}(\tau,\vec{x}) &:=& \gamma_{ij}(\tau,\vec{x})- \gamma_{ij}(\tau)\, , \\
\delta\pi^{ij}(\tau,\vec{x})    &:=& \pi^{ij}(\tau,\vec{x})-\pi^{ij}(\tau) \, .
\label{eq:DefPerturbations6}
\eea
These are tensors on the spatial hypersurfaces $\Sigma_\tau$, and still according to the SVT decomposition they can be expanded as $h_{ij} = h_1 \gamma_{ij} + \partial_i \partial_j h_2 + \partial_i (h_3)_j + \partial_j (h_3)_i + (h_4)_{ij}$, where $h_{ij}$ denotes a generic tensor form, $h_1$ and $h_2$ are scalars, $h_3$ is a divergence-free vector, and $h_4$ is a conserved tensor in the sense that $\partial_i h_4^{ij} = 0$ and $(h_4)_i^{\ i}=0$. Keeping only scalar perturbations amounts to setting $h_3=h_4=0$. In Fourier space, $h_1$ is proportional to $\gamma_{ij}$ while $h_2$ is proportional to $k_i k_j$. For this reason, we introduce the two basis matrices
\bea
\label{eq:Mij:def}
	M^1_{ij}:=\frac{1}{\sqrt{3}}\widetilde{\gamma}_{ij} & ~~~\mathrm{and}~~~ & M^2_{ij}:=\sqrt{\frac{3}{2}}\left(\frac{k_ik_j}{k^2}-\frac{\widetilde{\gamma}_{ij}}{3}\right)\, ,
\eea 
which are indeed linear combinations of $\gamma_{ij}$ and $k_i k_j$, and whose indices are raised and lowered using the metric $\widetilde{\gamma}_{ij}$ since $\vec{k}$ is a comoving wavevector. Note that $M^1$ captures the purely isotropic part of the perturbations. Our choice of normalisation (which slightly differs from the one in \cite{Langlois:1994ec}\footnote{As a consequence, our gravitational variables slightly differ from the ones in \Refc{Langlois:1994ec}. Denoting $(\delta\gamma_A^\mathrm{L},\delta\pi_A^{\mathrm{L}})$ the variables used in \Refc{Langlois:1994ec}, the link between the two sets of variables is given by: 
\bea
	\delta\gamma_1^{\mathrm{L}}=\frac{\delta\gamma_1}{v^{2/3}\sqrt{3}}\, ,\qquad
	\delta\pi_1^{\mathrm{L}}=\sqrt{3}v^{2/3}\delta\pi_1\, ,\qquad
	\delta\gamma_2^{\mathrm{L}}=\sqrt{\frac{{3}}{2}}\frac{\delta\gamma_2}{v^{2/3}}\, ,\qquad
	\delta\pi_2^{\mathrm{L}}=\sqrt{\frac{2}{3}}v^{2/3}\delta\pi_2.
\eea
Both sets of variables are related by a diagonal canonical transformation, which thus corresponds to a pure squeezing~\cite{Colas:2021llj}.}) is such that these two matrices form an orthonormal basis, \ie $M^{ij}_AM^{A'}_{ij}=\delta_{A,A'}$, where $A$ and $A'$ run over 1 and 2. In Fourier space, the scalar perturbations in the induced metric and its momentum can thus be expanded as
\bea
\label{eq:delta:gamma:Mbasis}
	\delta\gamma_{ij}(\tau,\vec{k})&=&\delta\gamma_1(\tau,\vec{k}) M^1_{ij}+\delta\gamma_2(\tau,\vec{k}) M^2_{ij}(\vec{k})\, , \\
	\delta\pi^{ij}(\tau,\vec{k})&=&\delta\pi_1(\tau,\vec{k}) M^{ij}_1+\delta\pi_2(\tau,\vec{k}) M^{ij}_2(\vec{k})\, .
\label{eq:delta:pi:Mbasis}
\eea
The two scalar degrees of freedom for the gravitational sector are then described by $(\delta\gamma_1,\delta\pi_1)$ and $(\delta\gamma_2,\delta\pi_2)$. They are related to the original induced metric and conjugate momentum through 
\bea
\label{eq:delta:gamma:A:delta:gamma:ij}
\delta\gamma_A=M^{ij}_A\delta\gamma_{ij}
\quad\text{and}\quad
\delta\pi_A=M^A_{ij}\delta\pi^{ij}\, .
\eea

Let us finally stress that, although the indices in $M_A^{ij}$ are lowered and raised with $\widetilde{\gamma}_{ij}$, \Eqs{eq:delta:gamma:Mbasis} and~\eqref{eq:delta:pi:Mbasis} should  not be interpreted as leading to similar rules for $\delta\gamma_{ij}$ and $\delta\pi^{ij}$. Indeed, the indices of $\gamma_{ij}$ and $\pi^{ij}$ are lowered and raised with the full induced metric $\gamma_{ij}$. For instance, at linear order in perturbation theory, this leads to $\delta\gamma^{ij}(\tau,\vec{x})=-\gamma^{im}(\tau)\gamma^{jn}(\tau)\delta\gamma_{mn}(\tau,\vec{x})$, hence
\bea
\label{eq:delta:gamma:ij:UP}
	\delta\gamma^{ij}(\tau,\vec{k})&=&-\frac{\delta\gamma_1(\tau,\vec{k})}{v^{4/3}}M_1^{ij}-\frac{\delta\gamma_2(\tau,\vec{k})}{v^{4/3}}M_2^{ij}(\vec{k})\, ,
\eea
where we have used that $\gamma^{ij}(\tau)=v^{-2/3}\widetilde{\gamma}^{ij}$. For the conjugate momentum, one obtains, still at leading order in perturbation theory, $\delta\pi_{ij}(\tau,\vec{x})=\gamma_{im}(\tau)\gamma_{jn}(\tau)\delta\pi^{mn}(\tau,\vec{x})+\delta\gamma_{im}(\tau,\vec{x})\gamma_{jn}(\tau)\pi^{mn}(\tau)+\delta\gamma_{jm}(\tau,\vec{x})\gamma_{in}(\tau)\pi^{mn}(\tau)$. In Fourier space, this leads to
\bea
\label{eq:delta:pi:ij:DOWN}
	\delta\pi_{ij}(\tau,\vec{k}) =\left[v^{4/3}\delta\pi_1(\tau,\vec{k})+v\theta\delta\gamma_1(\tau,\vec{k})\right] M^1_{ij}+\left[v^{4/3}\delta\pi_2(\tau,\vec{k})+v\theta\delta\gamma_2(\tau,\vec{k})\right] M^2_{ij}(\vec{k})\, ,\quad
\eea
where $\pi^{mn}(\tau)$ has been related to $\theta$ by combining \Eqs{eq:pi_ij:pi_p} and~\eqref{eq:theta:def}. We note that the configuration variables $\delta\gamma_1$ and $\delta\gamma_2$ also contribute to $\delta\pi_{ij}$. For expressions of $\delta\gamma_{ij}(\tau,\vec{k})$ and $\delta\pi
^{ij}(\tau,\vec{k})$ valid at second order, see \Eqs{eq:delta:gamma:ij:UP:2ndOrder} and~\eqref{eq:delta:pi:ij:DOWN:2ndOrder}.

We have thus identified the relevant scalar degrees of freedom at the perturbative level  (we note that the lapse function and the shift vector have no associated momenta, and so is the case for their perturbations). For completeness, the relationship between the perturbative degrees of freedom in the Hamiltonian framework and those defined in the Lagrangian approach are given in \App{app:lag}.
\subsection{Dynamics of the perturbations}
\label{sec:Dynamics:Scalar:Perturbations}
Let us now study the dynamics of the perturbation variables introduced in the previous section. Our starting point is to view \Eqs{eq:DefPerturbations2}, \eqref{eq:DefPerturbations3}, \eqref{eq:DefPerturbations5} and \eqref{eq:DefPerturbations6} as defining a canonical transformation, which simply consists in subtracting fixed, time-dependent functions from the phase-space variables. Such a transformation, which is a mere translation in phase space, obviously preserves the Poisson brackets, hence it is indeed canonical. Our first task is to derive the Hamiltonian for this new set of canonical variables. 

In practice, let us formally arrange the configuration variables into a vector $\vec{q}(\tau,\vec{x})$, with conjugated momentum  $\vec{p}(\tau,\vec{x})$. The perturbation variables are defined according to $\delta\vec{q}(\tau,\vec{x}) = \vec{q}(\tau,\vec{x})- \vec{q}(\tau)$ and $\delta\vec{p}(\tau,\vec{x}) = \vec{p}(\tau,\vec{x})- \vec{p}(\tau)$, where $\vec{q}(\tau)$ and $ \vec{p}(\tau)$ solve the homogeneous and isotropic problem described in \Sec{ssec:bck}. 
When evaluated on the fields $\vec{q}(\tau,\vec{x})$ and $\vec{p}(\tau,\vec{x})$, the scalar constraint can be Taylor expanded in $\delta\vec{q}$ and $\delta\vec{p}$ (for the moment to infinite order, so the analysis remains exact at this stage) according to
\bea
& &\kern-3em
\mathcal{S}\left[\vec{q}(\tau,\vec{x}),\vec{p}(\tau,\vec{x})\right] =\mathcal{S}\left[\vec{q}(t),\vec{p}(t)\right] 
+\underbrace{\delta q_\mu \frac{\partial\mathcal{S}}{\partial q_\mu}\left[\vec{q}(t),\vec{p}(t)\right] +\delta p^\mu \frac{\partial\mathcal{S}}{\partial p^\mu}\left[\vec{q}(t),\vec{p}(t)\right] }_{\mathcal{S}^{(1)}\left[\delta\vec{q},\delta\vec{p}\right]}
\nonumber \\ & &\kern1em
+\underbrace{\frac{1}{2}\sum_{\{n,m ; n+m=2\} }\left(\delta q_\mu\right)^n \left(\delta p^\nu\right)^m \frac{\partial^2\mathcal{S}}{\left(\partial q_\mu\right)^i \left(\partial p^\nu\right)^j}\left[\vec{q}(t),\vec{p}(t)\right]}_{\mathcal{S}^{(2)}\left[\delta\vec{q},\delta\vec{p}\right]}
+\sum_{n\geq 3} \mathcal{S}^{(n)}\left[\delta\vec{q},\delta\vec{p}\right]\, .
\label{eq:S:Taylor}
\eea
In this expression, the first term vanishes, $\mathcal{S}[\vec{q}(\tau),\vec{p}(\tau)] =0 $, since, by definition, $\vec{q}(\tau)$ and $\vec{p}(\tau)$ satisfy the homogeneous and isotropic scalar constraint. The other terms are organised in powers of the perturbation variables: $\mathcal{S}^{(1)}$ contains linear combinations of the perturbation variables, $\mathcal{S}^{(2)}$ contains quadratic combinations, \etc A similar expression can be written down for the diffeomorphism constraint $\mathcal{D}_i$. 

The dynamics of $\vec{q}(\tau,\vec{x})$ and $\vec{p}(\tau,\vec{x})$ is given by the Hamitonian~\eqref{eq:full:Hamiltonian}, whose Hamilton equations read
\bea
\label{eq:eom:qmu}
\dot{q}_\mu(\tau,\vec{x}) &=& N(\tau,\vec{x}) \frac{\partial \mathcal{S}}{\partial p^\mu}\left[\vec{q}(\tau,\vec{x}),\vec{p}(\tau,\vec{x})\right] + N^i (\tau,\vec{x})\frac{\partial \mathcal{D}_i}{\partial p^\mu}\left[\vec{q}(\tau,\vec{x}),\vec{p}(\tau,\vec{x})\right] \, ,\\
\dot{p}^\mu(\tau,\vec{x}) &=& - N(\tau,\vec{x}) \frac{\partial \mathcal{S}}{\partial q_\mu}\left[\vec{q}(\tau,\vec{x}),\vec{p}(\tau,\vec{x})\right] - N^i(\tau,\vec{x}) \frac{\partial \mathcal{D}_i}{\partial q_\mu}\left[\vec{q}(\tau,\vec{x}),\vec{p}(\tau,\vec{x})\right]\, .
\label{eq:eom:pmu}
\eea
Upon plugging the Taylor series~\eqref{eq:S:Taylor} (and the analogue formula for the diffeomorphism constraint) into \Eq{eq:eom:qmu}, where $\dot{q}_\mu(\tau,\vec{x}) = \dot{q}_\mu(\tau)+\delta \dot{q}_\mu(\tau,\vec{x})$, one obtains
\bea
\delta\dot{q}_\mu& =&   N(\tau) \frac{\partial \mathcal{S}}{\partial p^\mu} \left[\vec{q}(\tau),\vec{p}(\tau) \right]  - \dot{q}_\mu(\tau) 
+\delta N(\tau,\vec{x})\frac{\partial}{\partial\left(\delta p^\mu\right)}  \mathcal{S}^{(1)}\left[\delta\vec{q},\delta\vec{p}\right] 
\nonumber \\ & &
+ N (\tau,\vec{x})\frac{\partial}{\partial\left(\delta p^\mu\right)} \sum_{n\geq 2} \mathcal{S}^{(n)}\left[\delta\vec{q},\delta\vec{p}\right] 
+ N^i(\tau,\vec{x}) \frac{\partial}{\partial\left(\delta p^\mu\right)} \sum_{n\geq 1} \mathcal{D}_i^{(n)}\left[\delta\vec{q},\delta\vec{p}\right] \, .
\eea
In this expression, the first two terms in the right-hand side cancel each other out since, by definition, $\vec{q}(\tau)$ obeys the first Hamilton equation of the homogeneous and isotropic problem. Only remain the last three terms, which shows that the equation of motion of $\delta\vec{q}$ has the form of a first Hamilton equation with a Hamiltonian density given by $ \delta N \mathcal{S}^{(1)}\left[\delta\vec{q},\delta\vec{p}\right] +N  \sum_{n\geq 2} \mathcal{S}^{(n)}\left[\delta\vec{q},\delta\vec{p}\right] 
+ N^i \sum_{n\geq 1} \mathcal{D}_i^{(n)}\left[\delta\vec{q},\delta\vec{p}\right]$. The same conclusion can be drawn from plugging the Taylor series of the constraints into \Eq{eq:eom:pmu} and deriving the equation of motion for $\delta\vec{p}$, which can be cast into a second Hamilton equation with the same Hamiltonian, namely
\bea
\label{eq:Expanded:Hamiltonian}
C\left[\delta\vec{q}, \delta\vec{p} \right] = \int \dd^3 \vec{x}  & &\Bigg[N(\tau) \mathcal{S}^{(2)} + \delta N \mathcal{S}^{(1)}  + \delta N^i \mathcal{D}_i^{(1)}
\nonumber  \\ & & 
+ N(\tau) \sum_{n=3}^\infty \mathcal{S}^{(n)}+ \delta N \sum_{n=2}^\infty \mathcal{S}^{(n)}+\sum_{n=2}^\infty\delta N^i \mathcal{D}_i^{(n)}
\Bigg] \, .
\eea
In this expression, the quadratic terms have been singled out for later convenience. Although we have shown that this Hamiltonian gives the correct equations of motion, one must ensure that the correct constraints are also recovered. This is the case since the perturbed lapse function multiplies $\sum_{n\geq 1}\mathcal{S}^{(n)}$ in \Eq{eq:Expanded:Hamiltonian}, namely the full scalar constraint minus $\mathcal{S}[\vec{q}(t),\vec{p}(t)]$, which itself vanishes as already mentioned. One can also check that the perturbed shift vector multiplies $\sum_{n\geq 1}\mathcal{D}_i^{(n)}$, which is nothing but the full diffeomorphism constraint. 

Even though the above Hamiltonian provides an exact description of the perturbation variables, in practice, tractable calculations can only be performed by truncating the expansion at a finite order. At leading order in perturbation theory, only the quadratic terms in \Eq{eq:Expanded:Hamiltonian} remain (\ie those in the first line). Variation with respect to the perturbed lapse and the perturbed shift give the linear scalar constraint equation $\mathcal{S}^{(1)}=0$ and the linear diffeomorphism constraint equation $\mathcal{D}_i^{(1)}=0$ respectively, while the dynamics of the other phase-space coordinates is generated by $\mathcal{S}^{(2)}$ [given that $N(\tau)$ is already determined by the background solution]. 

An important remark is that, although $\mathcal{S}^{(2)}$ vanishes in the full theory (the full scalar constraint vanishes so it must vanish order by order), this is not guaranteed at linear order in perturbation theory (where only the linear constraints are satisfied). The reason is that, at that order, linear relationships are imposed between the phase-space variables, where all quadratic and higher-order contributions are neglected. Since $\mathcal{S}^{(2)}$ is a quadratic constraint, this explains why it is not satisfied. It may still be referred to as the quadratic ``constraint'' in what follows, although one must recall that this constraint is not satisfied at linear order. Finally, let us note that if one wanted to study higher orders, one could iterate the same procedure, and perform the canonical transformation which consists in subtracting from the perturbation variables the solutions to the linear problem that we will now derive. One would then find that, at order $n$, the scalar and diffeomorphism constraints are given by $\mathcal{S}^{(n)}=0$ and $\mathcal{D}_i^{(n)}=0$, while the dynamics is generated by $\mathcal{S}^{n+1}$.\\

As mentioned at the beginning of this section, \Eqs{eq:DefPerturbations2}, \eqref{eq:DefPerturbations3}, \eqref{eq:DefPerturbations5} and \eqref{eq:DefPerturbations6} can be seen as a canonical transformation, which preserves the Poisson brackets, hence $(\delta\phi, \delta\pi_\phi, \delta\gamma_{ij},\delta\pi^{ij})$ share the same Poisson brackets as  $(\phi, \pi_\phi, \gamma_{ij},\pi^{ij})$ and given below \Eq{eq:metric:ADM}. 
In \Sec{ssec:dofpert}, the scalar degrees of freedom were identified, and for the gravitational sector they are given by $\delta\gamma_1$, $\delta\gamma_2$, $\delta\pi_1$ and $\delta\pi_2$. Their Poisson brackets can be obtained from \Eq{eq:delta:gamma:A:delta:gamma:ij}, which leads to $\{ \delta\gamma_A(\vec{x}), \delta\gamma_{A'}(\vec{y})\} =\{ \delta\pi_A(\vec{x}), \delta\pi_{A'}(\vec{y})\}  =0$ and $\{ \delta\gamma_A(\vec{x}), \delta\pi_{A'}(\vec{y})\} = M_A^{ij} M^{A'}_{\ell m} \{ \delta\gamma_{ij}(\vec{x}),  \delta\pi^{\ell m}(\vec{y}) \} = \delta(\vec{x}-\vec{y}) (M_A^{ij} M^{A'}_{ij} + M_A^{ij} M^{A'}_{ji} )/2 =  \delta(\vec{x}-\vec{y}) \delta_{A,A'}$ where we have used that the $M$ matrices form an orthonormal basis. As a consequence, arranging the scalar perturbations into a vector $\delta\vec{ \phi}:=(\delta\phi,\delta \gamma_1,\delta \gamma_2)$ for convenience, the conjugate momentum to $\delta\vec{\phi}$ is given by $\delta\vec{ \pi}_\phi:=(\delta\pi_\phi,\delta \pi_1,\delta \pi_2)$. In real space, the Poisson brackets thus read
\bea
\left\{\delta\vec{\phi}(\tau,\vec{x}),\delta\vec{\pi}_\phi(\tau,\vec{y})\right\}= \delta^3(\vec{x}-\vec{y})\boldsymbol{I}\, ,
\eea
where $\boldsymbol{I}$ is the identity matrix (here in 3 dimensions), while in Fourier space they are given by
\bea
\left\{\delta\vec{\phi}(\tau,\vec{k}),\delta\vec{\pi}_\phi^\star(\tau,\vec{k}')\right\}= \delta^3(\vec{k}-\vec{k}')\boldsymbol{I}\, .
\eea
Note that since the two matrices $M^A_{ij}$ satisfy $M^A_{ij}(-\vec{k})=M^A_{ij}(\vec{k})$, the reality condition~\eqref{eq:reality:condition} applies for $\delta\gamma_A(t,\vec{k})$ and $\delta\pi_A(t,\vec{k})$ [in addition to holding for $\delta N(t,\vec{k})$, $\delta\phi(t,\vec{k})$, and $\delta \pi_\phi(t,\vec{k})$ as already mentioned].

The expansion of the constraints at first and second order in the perturbation variables is performed in \App{app:gloss} in the case of a flat FLRW background, and below we only quote the results. The linear diffeomorphism constraint is given in \Eq{eq:D1:app:final} and reads
\bea
\label{eq:D1i:D1}
	\mathcal{D}^{(1)}_i(t,\vec{k})=i\,k_i\,\mathcal{D}^{(1)}(t,\vec{k}),
\eea
where $\mathcal{D}^{(1)}$ is a scalar given by
\bea
	\mathcal{D}^{(1)}=\pi_\phi\,\delta\phi+\frac{1}{\sqrt{3}}v^{1/3}\theta\left(\frac{1}{2}\delta\gamma_1-\sqrt{2}\delta\gamma_2\right)-\frac{2}{\sqrt{3}}v^{2/3}\left(\delta\pi_1+\sqrt{2}\delta\pi_2\right)\, . \label{eq:diff1gen}
\eea
Note that \Eq{eq:D1i:D1} implies that the reality condition~\eqref{eq:reality:condition}, ${\mathcal{D}^{(1)}_i}^\star(t,\vec{k})=\mathcal{D}^{(1)}_i(t,-\vec{k})$, also holds for $\mathcal{D}^{(1)}$, namely ${\mathcal{D}^{(1)}}^\star(t,\vec{k})=\mathcal{D}^{(1)}(t,-\vec{k})$. In general, the diffeomorphism constraint is a vector and leads to three constraint equations. In the present case, it reduces to a single constraint equation, $\mathcal{D}^{(1)}(t,\vec{k})=0$, since only scalar perturbations are considered.

The linear scalar constraint is obtained by combining the contributions in \Eqs{eq:calT1:app:final}, \eqref{eq:calW1:app:final}, \eqref{eq:T1:app:final}, \eqref{eq:W1:app:final}, and reads
\bea
	\mathcal{S}^{(1)}(t,\vec{k})&=&-\frac{\sqrt{3}}{\Mp^2}v^{2/3}\theta\,\delta\pi_1-\frac{v^{1/3}}{\sqrt{3}}\left(\frac{\pi_\phi^2}{v^2}-V+{\Mp^2}\frac{k^2}{v^{2/3}}\right)\,\delta\gamma_1+\frac{\Mp^2}{\sqrt{6}}\frac{k^2}{v^{1/3}}\delta\gamma_2
\nonumber\\ & &	
	+\frac{\pi_\phi}{v}\,\delta\pi_\phi+vV_{,\phi}\delta\phi\, ,
	\label{eq:scal1gen:simp} 
\eea
where the background scalar constraint~\eqref{eq:ConstHom} has been used. This constraint also satisfies the reality condition~\eqref{eq:reality:condition}, $\mathcal{S}^{(1)\star}(t,\vec{k})=\mathcal{S}^{(1)}(t,-\vec{k})$, since it is a linear combination of the perturbation variables (each satisfying the reality condition) with coefficients given by real-valued functions of the background and of $k^2$.

Let us note that, when expressing the smeared constraints (see footnote~\ref{footnote:smeared:constraints}) in Fourier space, one should make sure to avoid double counting of the degrees of freedom and take into account the reality condition~\eqref{eq:reality:condition}. In practice, the integration over Fourier modes can be split into two parts, $\mathbb{R}^{3+}:=\mathbb{R}^2\times\mathbb{R}^+$ and $\mathbb{R}^{3-}:=\mathbb{R}^2\times\mathbb{R}^-$. The integration over $\mathbb{R}^{3-}$ can then be written as an integral over $\mathbb{R}^{3+}$ using the fact that $\delta N(t,\vec{k})$, $\delta N_1(t,\vec{k})$, $\mathcal{D}^{(1)}_i(t,\vec{k})$ and $\mathcal{S}^{(1)}(t,\vec{k})$ all satisfy the reality condition. This gives for the smeared constraints
\bea
\label{eq:D1:calD1}
	D^{(1)}[\delta N^i]&=&\ds\int_{\mathbb{R}^{3+}}k\,\dd^3k\,\left[\delta N_1 \mathcal{D}^{(1)\star}+\delta N^\star_1 \mathcal{D}^{(1)}\right], \\
	S^{(1)}[\delta N]&=&\ds\int_{\mathbb{R}^{3+}}\dd^3k\left[\delta N\mathcal{S}^{(1)\star}+\delta N^\star\mathcal{S}^{(1)}\right]\, ,
	\label{eq:S1:calS1}
\eea
where the extra $k$ in the smeared diffeomorphism constraint comes from \Eq{eq:D1i:D1}.

As explained below \Eq{eq:Expanded:Hamiltonian}, at quadratic order, only the perturbed scalar constraint, $\mathcal{S}^{(2)}$, is needed. Expressing the smeared constraint as an integral over $\mathbb{R}^{3+}$ in order to avoid double counting again,
\bea
\label{eq:S2.calS2}
	S^{(2)}\left[N\right]=2\ds\int_{\mathbb{R}^{3+}} \dd^3k\,N(\tau)\,\mathcal{S}^{(2)}(\tau,\vec{k}),
\eea
in \App{app:gloss} we show that [see \Eqs{eq:calT2:app:final}, \eqref{eq:calW2:app:final}, \eqref{eq:T2:app:final} and~\eqref{eq:W2:app:final}]
\bea
\label{eq:scalconst2}
\mathcal{S}^{(2)}&=&\frac{v^{1/3}}{\Mp^2}\left(2\left|\delta\pi_2\right|^2-\left|\delta\pi_1\right|^2\right)+\frac{1}{2v}\left|\delta\pi_\phi\right|^2+\frac{v}{2}\left(\frac{k^2}{v^{2/3}}+V_{,\phi,\phi}\right)\left|\delta\phi\right|^2 \nonumber\\ 
	&&+\frac{1}{3v^{1/3}}\left(\frac{\pi^2_\phi}{v^2}+\frac{V}{2} - \frac{\Mp^2 k^2}{4 v^{2/3}}\right)\left|\delta\gamma_1\right|^2 + \frac{1}{3v^{1/3}}\left(\frac{\pi^2_\phi}{v^2}+\frac{V}{2} - \frac{\Mp^2 k^2}{8 v^{2/3}}\right) \left|\delta\gamma_2\right|^2 \nonumber \\
	&&-\frac{\theta}{4\Mp^2}\left(\delta\pi_1\delta\gamma^\star_1+\mathrm{c.c.}\right)+\frac{\theta}{2\Mp^2}\left(\delta\pi_2 \delta\gamma^\star_2+\mathrm{c.c.}\right) 
	+\frac{\sqrt{2}\Mp^2}{24v}k^2\left(\delta\gamma_1\delta\gamma_2^\star+\mathrm{c.c.}\right)
	\nonumber \\
	&&-\frac{\sqrt{3}}{4}v^{1/3}\left[\left(\frac{\pi_\phi}{v^2}\delta\pi_\phi-V_{,\phi}\delta\phi\right)\delta\gamma_1^\star+\mathrm{c.c.}\right], 
\label{eq:S2:full}
\eea
where ``c.c.'' means the complex conjugate of the previous term, and where the background scalar constraint~\eqref{eq:ConstHom} has been used to further simplify the expression. The two first lines correspond to diagonal terms, the third line features cross terms within the gravitational sector (we notice that there is no cross term in the scalar-field sector), while the fourth line stands for coupling between the two sectors. In particular, one can see that the scalar field perturbations couple only to the isotropic gravitational configuration. There is however no coupling between the scalar field and the isotropic gravitational momentum $\delta\pi_1$.  The absence of coupling between $\delta\phi$ or $\delta\pi_\phi$ and $\delta\gamma_2$ or $\delta\pi_2$ is due to the fact a scalar field can only generate isotropic perturbations. The two gravitational degrees of freedom are however coupled to each other, through a term of the form $k^2(\delta\gamma_1)(\delta\gamma_2)$, that is to say via gradient interactions.

We are now in a position where we can derive the equations of motion for the perturbations, 
\bea
	&&\left\{\begin{array}{l}
		\ds\dot{{\delta\gamma_1}}=- \frac{2}{\sqrt{3}} v^{2/3} k \delta N_1 -\frac{\sqrt{3}}{\Mp^2}v^{2/3}\theta\,{\delta N}-\frac{N}{\Mp^2}\left({2v^{1/3}}\,{\delta\pi_1}+\frac{\theta}{2}\,{\delta\gamma_1}\right), \\
		\ds\dot{{\delta\pi_1}}= - \frac{v^{1/3}\theta}{2\sqrt{3}} k \delta N_1 +  \frac{v^{1/3}}{\sqrt{3}}\left(\frac{\pi^2_\phi}{v^2}-V + \Mp^2 \frac{k^2}{v^{2/3}}\right)\,{\delta N} \\
		~~~~~~~~\ds+N\bigg[-\frac{2}{3v^{1/3}}\left(\frac{\pi^2_\phi}{v^2}+\frac{V}{2} - \frac{\Mp^2 k^2}{4 v^{2/3}} \right)\,{\delta\gamma_1}+\frac{\theta}{2\Mp^2}\,{\delta\pi_1} \\
		~~~~~~~~\ds+\frac{\sqrt{3}}{2}v^{1/3}\left(\frac{\pi_\phi}{v^2}\,{\delta\pi_\phi}-V_{,\phi}\,{\delta\phi}\right) - \frac{\sqrt{2}}{12 v} \Mp^2 k^2 \delta\gamma_2 \bigg],
	\end{array}\right.  \nonumber\\
	&&\nonumber \\
	&&\left\{\begin{array}{l}
		\ds\dot{{\delta\gamma_2}}= -2\sqrt{\frac{2}{3}} v^{2/3} k \delta N_1 + N \left( \frac{4 v^{1/3}}{\Mp^2} \delta\pi_2 + \frac{\theta}{\Mp^2}\delta\gamma_2 \right) , \\
		\label{eq:eom:gen}
		\ds\dot{{\delta\pi_2}}= \sqrt{\frac{2}{3}} v^{1/3} \theta k \delta N_1 -  \frac{\Mp^2 k^2}{\sqrt{6} v^{1/3}}\delta N
		\\  ~~~~~~~~\ds
		 + N \bigg[ - \frac{\theta}{\Mp^2} \delta \pi_2 
		- \frac{\sqrt{2} \Mp^2}{12 v} k^2 \delta \gamma_1 
		- \frac{2}{3 v^{1/3}} \left( \frac{\pi_\phi^2}{v^2} + \frac{V}{2} - \frac{\Mp^2 k^2}{8 v^{2/3}} \right) \delta\gamma_2 \bigg] \,,
	\end{array}\right.  \label{eq:EOM}\\
	&&\nonumber \\
	&&\left\{\begin{array}{l}
		\ds\dot{{\delta\phi}}=\frac{\pi_\phi}{v}\,{\delta N}+N\left(\frac{1}{v}\,{\delta\pi_\phi}-\frac{\sqrt{3}}{2}\frac{\pi_\phi}{v^{5/3}}\,{\delta\gamma_1}\right), \\
		\ds\dot{{\delta\pi_\phi}}=-\pi_\phi k \delta N_1 -vV_{,\phi}\,{\delta N}-N\left[v\left(\frac{k^2}{v^{2/3}} + V_{,\phi,\phi}\right)\,{\delta\phi}+\frac{\sqrt{3}}{2}v^{1/3}V_{,\phi}\,{\delta\gamma_1}\right].
	\end{array}\right.  \nonumber
\eea

\subsection{Fixing the gauge}
\label{ssec:gaugefix}
Since the theory is independent of a specific choice of space-time coordinates, some combinations of the perturbation variables can be set to zero, which amounts to working in a specific gauge. Changes of coordinates bear four degrees of freedom (one per coordinate), made of two scalars and one vector.\footnote{In general, a change of coordinates can be written as $x^\mu\to x^\mu+\xi^\mu$, where $\xi_i=\partial_i f + f_i$, with $\partial_i f^i=0$. The two scalar degrees of freedom correspond to $\xi^0$ and $f$ while $f^i$ contains the vector degrees of freedom.\label{footnote:ChangeOfCoordinates}} In practice, they correspond to the Lagrange multipliers of the theory (one scalar in the lapse, one scalar and one vector in the shift). Since we are dealing with scalar perturbations only, this implies that two combinations of scalar perturbations can be set to zero. The vanishing of their respective equations of motion leads to two additional vanishing combinations. Together with the two linear constraint equations, this allows one to freeze six out of the eight variables (namely $\delta N$, $\delta N_1$, $\delta\phi$, $\delta\pi_\phi$, $\delta\gamma_1$, $\delta\pi_1$, $\delta\gamma_2$ and $\delta\pi_2$), such that only two variables (hence a single physical degree of freedom) remain. 

This single remaining physical degree of freedom can be parametrised in a gauge-invariant way, \eg using the so-called Mukhanov-Sasaki combination~\cite{Mukhanov:1981xt,Kodama:1984ziu}\footnote{Note that, to match conventions usually adopted in the literature, we use a different normalisation than in previous versions of this article.}
\bea
\label{eq:MS:def}
Q_{{}_\mathrm{MS}} := \sqrt{\frac{v}{N}} \delta\phi + \frac{\Mp^2\pi_\phi}{\sqrt{6N}\theta  v^{7/6}}\left(\sqrt{2}\delta\gamma_1-\delta\gamma_2\right) .
\eea
A detailed discussion of gauge transformations in the Hamiltonian formalism, and of the systematic construction of gauge-invariant combination, will be presented separately in a forthcoming article. The above constraint and dynamical equations allow one to derive an autonomous equation of motion for the Mukhanov-Sasaki variable, namely
\bea
\label{eq:MS:CPT}
\ddot{Q}_{{}_\mathrm{MS}} + \left(k^2-\frac{\ddot{z}}{z}\right)Q_{{}_\mathrm{MS}} =0\, ,
\eea
which is written in conformal time $\eta$, and where $z\equiv v^{1/3}\sqrt{2\epsilon_1}\Mp $ with $\epsilon_1\equiv 2 \Mp^2 \dot{\theta}/(N\theta^2)$ the first Hubble-flow parameter.

An alternative approach is to fix the gauge in which the calculation is performed. Although a gauge-invariant approach is more elegant in general, gauge fixing may be required in some problems (for instance in numerical approaches, see \eg \Refc{Fidler:2015npa}, or in the stochastic-$\delta N$ formalism as explained below in \Sec{sec:Uniform:Expansion:Gauge:def}).
We thus end this section by considering a few different gauge choices that are commonly used in the literature, namely the spatially-flat gauge, the Newtonian gauge, the (generalised) synchronous gauges and the uniform-expansion gauges. These gauges will also be of particular interest to discuss on how to properly match the separate universe approach to \CPT, which is why they are introduced before \Sec{sec:SepUniv}. 
The connection with the (more often discussed) definition of these gauges in the Lagrangian framework is given in \App{app:lag}.

\subsubsection*{Spatially-flat gauge}
\label{sec:Spatially:Flat:def}
Let us start with the spatially-flat gauge, in which one sets $\delta\gamma_{ij}=0$. This implies that $\delta\gamma_1=\delta\gamma_2=0$.
In that gauge, phase-space reduction proceeds as follows. For $\delta\gamma_1$ and $\delta\gamma_2$ to remain zero, their equation of motion should vanish too (\ie $\delta\dot{\gamma}_1=\delta\dot{\gamma}_2=0)$, which from \Eq{eq:eom:gen} gives two constraint equations, namely
\bea
	& &\frac{2}{\sqrt{3}}kv^{2/3}\,\delta N_1+\frac{\sqrt{3}}{\Mp^2}v^{2/3}\theta\,\delta N+\frac{2}{\Mp^2}Nv^{1/3}\,\delta\pi_1 =0\, , \label{eq:const1flat} \\
	& &2\sqrt{\frac{2}{3}}kv^{2/3}\,\delta N_1-\frac{4}{\Mp^2}Nv^{1/3}\,\delta\pi_2=0\, . \label{eq:const2flat}
\eea
Moreover, when $\delta\gamma_A=0$, the linear constraints are given by
\bea
	\mathcal{D}^{(1)}&=&\pi_\phi\,\delta\phi-\frac{2}{\sqrt{3}}v^{2/3}\left(\delta\pi_1+\sqrt{2}\delta\pi_2\right)=0, \label{eq:diffeoflat} \\
	\mathcal{S}^{(1)}&=&\frac{\pi_\phi}{v}\,\delta\pi_\phi+vV_{,\phi}\,\delta\phi-\frac{\sqrt{3}}{\Mp^2}v^{2/3}\theta\,\delta\pi_1=0\, .\label{eq:scalarflat}
\eea
We thus have four constraint equations, which allow us to express $\delta N$, $\delta N_1$, $\delta\pi_1$ and $\delta\pi_2$ in terms of the other phase-space variables (namely $\delta\phi$ and $\delta\pi_\phi$), and one obtains
\bea
\label{eq:deltaNphi:flat}
\frac{\delta N}{N} & =& -\frac{\pi_\phi}{v\theta}\,\delta\phi\, ,\\
\label{eq:deltaN1phi:flat}
k\frac{\delta N_1}{N}&=&\left(\frac{3}{2\Mp^2}\frac{\pi_\phi}{v}-\frac{V_{,\phi}}{\theta}\right)\delta\phi-\frac{\pi_\phi}{v^2\theta}\,\delta\pi_\phi\, ,\\
\label{eq:pi1phi:flat} 
\delta\pi_1&=&\frac{\Mp^2}{\sqrt{3}v^{2/3}\theta}\left(\frac{\pi_\phi}{v}\,\delta\pi_\phi+vV_{,\phi}\,\delta\phi\right)\, , \\
 \label{eq:pi2phi}
\delta\pi_2&=&-\frac{\Mp^2}{\sqrt{6}}\frac{\pi_\phi}{v^{5/3}\theta}\,\delta\pi_\phi+\frac{\Mp^2}{\sqrt{6}}\left(\frac{3}{2\Mp^2}\frac{\pi_\phi}{v^{2/3}}-\frac{v^{1/3}V_{,\phi}}{\theta}\right)\delta\phi\, .
\eea
One thus has a single physical scalar degree of freedom, described by $\delta\phi$ and $\delta\pi_\phi$, the dynamics of which is given by the two last equations of \Eq{eq:eom:gen} where the above replacements are made. One can also check that, still with the above replacements, the equations of motion for $\delta\pi_1$ and $\delta\pi_2$, \ie the second and the fourth entries of \Eq{eq:eom:gen}, are automatically satisfied.

\subsubsection{Newtonian gauge}
\label{sec:Newtonian:def}
Let us now consider the Newtonian gauge, which corresponds to setting $\delta\gamma_2=\delta N_1=0$. For $\delta\gamma_2$ to remain zero, the third entry of \Eq{eq:eom:gen} has to vanish, which leads to $\delta\pi_2=0$. The gravitational anisotropic degree of freedom is therefore entirely frozen. Similarly, for $\delta\pi_2$ to remain zero, the fourth entry of \Eq{eq:eom:gen} has to vanish, which leads to 
\bea
\label{eq:deltaN:deltagamma1:Newtonian}
\frac{k^2}{\sqrt{6}v^{1/3}}\,\delta N +\frac{N}{6\sqrt{2}v}k^2\delta\gamma_1=0\, .
\eea
Moreover, the two linear constraints read
\bea
\label{eq:D1:Newtonian}
	\mathcal{D}^{(1)}&=&\pi_\phi\,\delta\phi+\frac{1}{2\sqrt{3}}v^{1/3}\theta\,\delta\gamma_1-\frac{2}{\sqrt{3}}v^{2/3}\delta\pi_1=0, \\
	\mathcal{S}^{(1)}&=&-\frac{\sqrt{3}}{\Mp^2}v^{2/3}\theta\,\delta\pi_1-\frac{v^{1/3}}{\sqrt{3}}\left(\frac{\pi_\phi^2}{v^2}-V+\Mp^2\frac{k^2}{v^{2/3}}\right)\delta\gamma_1+\frac{\pi_\phi}{v}\,\delta\pi_\phi+vV_{,\phi}\delta\phi=0. \nonumber \\
\eea
With the above three constraint equations, one can either fix $\delta N$, $\delta\gamma_1$ and $\delta\pi_1$, and work with $(\delta\phi,\delta\pi_\phi)$ as describing the remaining dynamical variable; or fix $\delta N$, $\delta\phi$ and $\delta\pi_\phi$, and work with $(\delta\gamma_1,\delta\pi_1)$ as describing the remaining dynamical variable; or any other combination.

Let us mention that an alternative definition of the Newtonian gauge is to start from the conditions $\delta\gamma_2=\delta\pi_2=0$, since the third entry of \Eq{eq:eom:gen} then implies that $\delta N_1=0$. 

\subsubsection{Generalised synchronous gauge}
\label{sec:Generalised:Synchronous:def}
The generalised synchronous gauges are such that neither the lapse function nor the shift vector are perturbed, $\delta N = \delta N_1=0$. They can be viewed as gauges in which the global time and space coordinate of the perturbed FLRW background exactly coincide with the time and space coordinate of the (strictly homogeneous and isotropic) FLRW space-time, irrespectively of the initial choice of the background lapse function. Choosing for instance the background lapse function to be cosmic time [so $N(\tau)=1$], this boils down to the standard synchronous gauge. 

Let us note that the conditions $\delta N = \delta N_1=0$ do not impose further constraints from the equations of motion~\eqref{eq:eom:gen}, contrary to what was obtained in the spatially-flat and Newtonian gauges. As a consequence, the generalised synchronous gauges are not entirely fixed and still contain two spurious gauge modes. 
\subsubsection{Uniform-expansion gauge}
\label{sec:Uniform:Expansion:Gauge:def}
As explained in \Refc{Pattison:2019hef}, the stochastic-$\delta N$ formalism~\cite{Vennin:2015hra, Fujita:2013cna} is formulated in the uniform-expansion gauge, in which the perturbation of the integrated expansion
\bea
\mathcal{N}_{\mathrm{int}} := \frac{1}{3} \int \nabla_\mu n^\mu\, n_\nu \dd x^{\nu} 
\eea
is set to zero. In this expression, $\nabla$ denotes the covariant derivative, and $n^{\mu}$ is the unit vector such that the form $n_\mu$ is orthogonal to the spatial hypersurfaces $\Sigma_\tau$ (see \App{app:expansion} for an explicit calculation of the expansion rate $ \nabla_\mu n^\mu$, in particular \Eq{eq:ExpRateGen}, and of the integrated expansion $\mathcal{N}_{\mathrm{int}}$).

The reason for setting $\delta \mathcal{N}_{\mathrm{int}}=0$ is that, in order to relate the large-scale curvature perturbation with the fluctuation in the number of \efolds $\mathcal{N}$, as implied by the $\delta N$ formalism, the Langevin equations of stochastic inflation have to be solved with the number of \efolds as the time variable, and this amounts to fixing $\mathcal{N}_{\mathrm{int}}$ across the different patches of the universe. 

In \App{app:expansion}, it is shown that at the background level, $\mathcal{N}_{\mathrm{int}} = \ln(v)/3$, see \Eq{eq:NintHom}, \ie the integrated expansion is nothing but the number of \efolds $\mathcal{N}$. At first order in the perturbation variables, one obtains $\delta \mathcal{N}_{\mathrm{int}} = \delta\gamma_1/(2\sqrt{3} v^{2/3}) + k \int \delta N_1\dd\tau/3$, see \Eq{eq:delta:N:int:CPT}. The uniform-expansion gauge thus corresponds to setting
\bea
\delta\gamma_1=\delta N_1=0\, .
\eea
Note that the vanishing of $\delta\dot{\gamma}_1$ in \Eq{eq:eom:gen} leads to an additional constraint equation, so together with the two first-order constraint equations, one can thus set five out of the eight variables. As a consequence, the uniform-expansion gauge is not entirely fixed and still contains one spurious gauge mode. This may seem a priori problematic~\cite{Figueroa:2021zah} but as we will show below in \Sec{sec:Uniform:Expansion:Gauge:SU:def}, in the separate-universe framework (where stochastic inflation is formulated), the gauge becomes unequivocally defined.
\section{Separate universe}
\label{sec:SepUniv}
Let us now describe the separate-universe approach~\cite{Salopek:1990jq, Sasaki:1995aw, Wands:2000dp, Lyth:2003im, Rigopoulos:2003ak, Lyth:2005fi} (also known as the quasi-isotropic approach~\cite{Lifshitz:1960, Starobinsky:1982mr, Comer:1994np, Khalatnikov:2002kn}), which consists in introducing local perturbations to the homogeneous and isotropic problem described in \Sec{ssec:bck}, as a proxy for the full perturbative problem studied in \Sec{sec:cosmopert}. Our goal is to establish this formalism (and the corresponding validity conditions) in the Hamiltonian framework, complementing analyses performed in the Lagrangian approach such as \Refc{Pattison:2019hef}. 
\subsection{Homogeneous and isotropic perturbations}
\label{sec:SepUniv:variables}
The starting point of the separate-universe approach is to perturb the homogeneous and isotropic background variables introduced in \Sec{ssec:bck}, namely $N\to N(\tau)+\overline{\delta N}$, $(v,\theta)\to[v(\tau)+\overline{\delta v},\theta(\tau)+\overline{\delta\theta}]$, and $(\phi,\pi_\phi)\to[\phi(\tau)+\overline{\delta\phi},\pi_\phi(\tau)+\overline{\delta\pi_\phi}]$. Hereafter, an overall bar denotes perturbations of the background variables, which a priori differ from the perturbation variables used in the full treatment of \Sec{sec:cosmopert}. As we will show, they however succeed in capturing their behaviour above the Hubble radius, \ie when $k\ll aH$.

In order to make this statement explicit, one must first determine to which background perturbations the variables introduced in \Sec{sec:cosmopert} correspond, \ie one must establish a ``dictionary'' between \CPT~and the separate universe approach. For obvious reasons, $\delta N$, $\delta\phi$ and $\delta\pi_\phi$ correspond to $\overline{\delta N}$, $\overline{\delta\phi}$ and $\overline{\delta\pi_\phi}$ respectively. Since the shift $N^i$ vanishes at the background level, there is no perturbed shift in the separate-universe approach, \ie $\overline{\delta N_1}=0$. For the gravitational sector, since $\gamma_{ij}(\tau)=v^{2/3}\widetilde{\gamma_{ij}}$ at the background level, one has $\gamma_{ij}\to (v+\overline{\delta v})^{2/3}\widetilde{\gamma_{ij}}$ in the separate universe, which leads to $\overline{\delta\gamma_{ij}}=[(v+\ovl{\delta v})^{2/3}-v^{2/3}] \widetilde{\gamma}_{ij}$. Making use of \Eq{eq:delta:gamma:A:delta:gamma:ij}, this gives rise to $\overline{\delta\gamma_1} = \sqrt{3}[(v+\ovl{\delta v})^{2/3}-v^{2/3}]  $ and $\overline{\delta\gamma_2}=0$.
Similarly, combining \Eqs{eq:pi_ij:pi_p} and~\eqref{eq:theta:def}, at the background level one has $\pi^{ij}(\tau)=v^{1/3}\theta\widetilde{\gamma}^{ij}/2$, which leads to $\overline{\delta\pi^{ij}}=[(v+\ovl{\delta v})^{1/3}(\theta+\ovl{\delta\theta})-v^{1/3}\theta]\widetilde{\gamma}^{ij}/2$. Making use of \Eq{eq:delta:gamma:A:delta:gamma:ij} again, this gives $\overline{\delta\pi_1}=\sqrt{3}[(v+\ovl{\delta v})^{1/3}(\theta+\ovl{\delta\theta})-v^{1/3}\theta]/2$ and $\overline{\delta\pi_2}=0$.
These formula are summarised in Table~\ref{table:correspondence}.
\begin{table}[h!]
\centering
\begin{tabular}{ |c|c| } 
 \hline
 \CPT~& separate-universe approach  \\
 \hline  \hline 
 $\delta N$ & $\overline{\delta N}$ \\ 
  \hline 
$\delta N_1$ & 0 \\ 
 \hline
 $\delta\gamma_1$ & $\overline{\delta\gamma_1}=\sqrt{3}[(v+\ovl{\delta v})^{2/3}-v^{2/3}] $\\
   \hline 
 $\delta\pi_1$ & $\overline{\delta\pi_1}=\frac{\sqrt{3}}{2}[(v+\ovl{\delta v})^{1/3}(\theta+\ovl{\delta\theta})-v^{1/3}\theta]$\\
 \hline
 $\delta\gamma_2$ & $0$\\
 \hline
 $\delta\pi_2$ & $0$\\
 \hline
 $\delta\phi$ & $\overline{\delta\phi}$\\
 \hline
 $\delta\pi_\phi$  & $\overline{\delta\pi_\phi}$\\
 \hline
\end{tabular}
\caption{Correspondence between variables in \CPT~and in the separate-universe approach.}
\label{table:correspondence}
\end{table}
One can see that the anisotropic degrees of freedom, $\delta\gamma_2$ and $\delta\pi_2$, as well as the shift, are simply absent in the separate-universe approach. One can also check that the transformation from $(\overline{\delta\gamma_1},\overline{\delta\pi_1})$ to $(\overline{\delta v},\overline{\delta\theta})$ is canonical, as it should.
\subsection{Dynamics of the background perturbations}
\label{sssec:sepunivgen}
The dynamics of the perturbations in the separate-universe approach can be obtained by plugging the replacement rules derived in \Sec{sec:SepUniv:variables} into the Hamiltonian~\eqref{eq:full:Hamiltonian}, whose contributions are given in \Eqs{eq:scgg}-\eqref{eq:dcsg}. This gives rise to 
\bea
	\mathcal{C}&=&\frac{-3}{4\Mp^2}v\theta^2\left(1+\frac{\ovl{\delta\gamma_1}}{\sqrt{3} v^{2/3}}\right)^{1/2}\left(1+\frac{2}{\sqrt{3}}\frac{\ovl{\delta\pi_1}}{v^{1/3}\theta}\right)^2\nonumber \\
	&&+\frac{\pi_\phi^2}{2v}\left(1+\frac{\ovl{\delta\gamma_1}}{\sqrt{3}v^{2/3}}\right)^{-3/2}\left(1+\frac{\ovl{\delta\pi_\phi}}{\pi_\phi}\right)^2+v\left(1+\frac{\ovl{\delta\gamma_1}}{\sqrt{3}v^{2/3}}\right)^{3/2}V\left(\phi+\ovl{\delta\phi}\right),
		\label{eq:Hamiltonian:SU:full}
\eea
where the smeared constraint is $C=\int \dd^3 \vec{x} N\mathcal{C}$, and where $\theta^2$ can be expressed using the background constraint equation~\eqref{eq:ConstHom}. Here we have parametrised the gravitational perturbations with $\ovl{\delta\gamma_1}$ and $\ovl{\delta\pi_1}$ instead of $\ovl{\delta v}$ and $\ovl{\delta\theta}$, to allow for a more direct comparison with \CPT. The two sets of variables are however simply related with the formulas given in Table~\ref{table:correspondence}, and below we will also provide the result in terms of $\ovl{\delta v}$ and $\ovl{\delta\theta}$, since they have the advantage of providing a simple interpretation as perturbations of the volume and of the expansion rate.
Note that we have also dropped the term proportional to $\partial_i\phi \partial_j \phi$ since $\delta\bar{\phi}$ is a homogeneous degree of freedom. Similarly, since only homogeneous and isotropic perturbations are included in the induced metric, its Ricci scalar vanishes, \ie $\mathcal{R}(\tau)=\ovl{\delta\mathcal{R}}=0$, which explains why the term proportional to $\mathcal{R}$ in \Eq{eq:scgg} is absent too.  

The next step is to expand \Eq{eq:Hamiltonian:SU:full} to the quadratic order in perturbations, see the discussion above \Eq{eq:Expanded:Hamiltonian}. It gives rise to the following Hamiltonian
\bea
	\ovl{C}\left[N+\overline{\delta N}\right]=N\,\mathcal{S}^{(0)}+\left(\overline{\delta N}\,\ovl{\mathcal{S}^{(1)}}+N\,\ovl{\mathcal{S}^{(2)}}\right) ,
\eea
where $\mathcal{S}^{(0)}$ is given by \Eqs{eq:BckgConsGrav} and~\eqref{eq:BckgConsPhi}, and the perturbed scalar constraint at linear and quadratic order is
\bea
	\ovl{\mathcal{S}^{(1)}}&=&-\frac{\sqrt{3}}{\Mp^2}v^{2/3}\theta\,\ovl{\delta\pi_1}-\frac{v^{1/3}}{\sqrt{3}}\left(\frac{\pi_\phi^2}{v^2}-V\right)\,\ovl{\delta\gamma_1}+\frac{\pi_\phi}{v}\,\ovl{\delta\pi_\phi}+vV_{,\phi}\,\ovl{\delta\phi}\, , \label{eq:s1su} \\
	\ovl{\mathcal{S}^{(2)}}&=&-\frac{v^{1/3}}{\Mp^2}\left(\ovl{\delta\pi_1}\right)^2+\frac{1}{3v^{1/3}}\left(\frac{\pi_\phi^2}{v^2}+\frac{V}{2}\right)\left(\ovl{\delta\gamma_1}\right)^2+\frac{1}{2v}\left(\ovl{\delta\pi_\phi}\right)^2+\frac{v}{2}V_{,\phi,\phi}\left(\ovl{\delta\phi}\right)^2 \nonumber \\
	&&-\frac{\theta}{2\Mp^2}\left(\ovl{\delta\pi_1}\right)\left(\ovl{\delta\gamma_1}\right)-\frac{\sqrt{3}}{2}v^{1/3}\left(\frac{\pi_\phi}{v^2}\,\ovl{\delta\pi_\phi}-V_{,\phi}\,\ovl{\delta\phi}\right)\,\ovl{\delta\gamma_1}\, .\label{eq:s2su}  
\eea
The separate-universe variables have to lie on the constraint $\ovl{\mathcal{S}^{(1)}}=0$, while $\ovl{\mathcal{S}^{(2)}}$ contributes to their dynamics, which Hamilton equations are given by
\bea
	&&\left\{\begin{array}{l}
		\ds\dot{\overline{\delta\gamma_1}}=-\frac{\sqrt{3}}{\Mp^2}v^{2/3}\theta\,\overline{\delta N}-\frac{N}{\Mp^2}\left({2v^{1/3}}\,\ovl{\delta\pi_1}+\frac{\theta}{2}\,\ovl{\delta\gamma_1}\right), \\
		\ds\dot{\overline{\delta\pi_1}}=\frac{v^{1/3}}{\sqrt{3}}\left(\frac{\pi^2_\phi}{v^2}-V\right)\ovl{\delta N}+N\left[-\frac{2}{3v^{1/3}}\left(\frac{\pi^2_\phi}{v^2}+\frac{V}{2}\right)\,\ovl{\delta\gamma_1}+\frac{\theta}{2\Mp^2}\,\ovl{\delta\pi_1}\right]
		\nonumber \\~~~~~~~~\ds
		+N\frac{\sqrt{3}}{2}v^{1/3}\left(\frac{\pi_\phi}{v^2}\,\ovl{\delta\pi_\phi}-V_{,\phi}\,\ovl{\delta\phi}\right),
	\end{array}\right. \label{eq:SepUnivFull}\\
	&& \\
	&&\left\{\begin{array}{l}
		\ds\dot{\overline{\delta\phi}}=\frac{\pi_\phi}{v}\,\overline{\delta N}+N\left(\frac{1}{v}\,\overline{\delta\pi_\phi}-\frac{\sqrt{3}}{2}\frac{\pi_\phi}{v^{5/3}}\,\ovl{\delta\gamma_1}\right), \\
		\ds\dot{\overline{\delta\pi_\phi}}=-vV_{,\phi}\,\overline{\delta N}-N\left(vV_{,\phi,\phi}\,\ovl{\delta\phi}+\frac{\sqrt{3}}{2}v^{1/3}V_{,\phi}\,\ovl{\delta\gamma_1}\right).
	\end{array}\right. \nonumber
\eea

By comparing those equations of motion with their \CPT~counterpart, \Eqs{eq:EOM}, one notices that the contribution involving the diffeomorphism constraint is absent in the SU. This is because $\mathcal{D}_i^{(1)}=i\, k_i\, \mathcal{D}^{(1)}$ is proportional to $k$, see \Eq{eq:D1i:D1}, so it indeed disappears at large scales. However, the constraint equation itself leads to a relationship between the perturbation variables that does not involve $k$, and which therefore contains non-trivial information even at large scales. The fact that it is lost in the separate-universe approach may therefore seem problematic a priori, and the consequences of this loss will be further analysed below. At this stage, let us simply notice that a SU version of the diffeomorphism constraint can still be defined using the correspondence table \ref{table:correspondence}:
\bea
\label{eq:diff1:cpt}
\ovl{\mathcal{D}}^{(1)} := \pi_\phi \ovl{\delta \phi}  + \frac{1}{2\sqrt{3}} v^{1/3} \theta \ovl{\delta\gamma}_1 - \frac{2}{\sqrt{3}} v^{2/3} \ovl{\delta \pi}_1\,.
\eea
Using \Eq{eq:SepUnivFull}, one can readily show that $\dot{\ovl{\mathcal{D}}}^{(1)}=0$ as long as the linear scalar constraint is satisfied $\ovl{\mathcal{S}}^{(1)}=0$. This implies that $\ovl{\mathcal{D}}^{(1)} $ is a conserved quantity in SU. 

It is worth stressing that the initial value of $\ovl{\mathcal{D}}^{(1)} $ is usually set by CPT, which is employed to describe cosmological perturbations before they cross out the Hubble radius. In CPT, $\mathcal{D}^{(1)}=0$, but this does not guarantee that $\ovl{\mathcal{D}}^{(1)} $ vanishes initially (hence at later time) since $\mathcal{D}^{(1)}$ and $\ovl{\mathcal{D}}^{(1)} $ generically differ.\footnote{We thank Diego Cruces for interesting discussions leading to this remark.} By comparing \Eqs{eq:diff1gen} and~\eqref{eq:diff1:cpt}, one notices that they coincide when $ 2 v^{1/3} \delta\pi_2+\theta \delta\gamma_2=0$. Therefore, by working in gauges where the anisotropic sector satisfies this constraint in CPT, one reinstates the diffeomorphism constraint in SU, $\ovl{\mathcal{D}}^{(1)}=0$. As we will see below, this condition is however not required for the SU approach to be reliable.

As mentioned above, it is also interesting to cast the result in terms of the variables $\ovl{\delta v}$ and $\ovl{\delta\theta}$ for the gravitational sector, and one finds that the scalar constraint is given by
\bea
\label{eq:S1:SU}
\ovl{\mathcal{S}^{(1)}} = -\frac{3v\theta}{2\Mp^2}\overline{\delta\theta}+vV_{,\phi}\overline{\delta\phi}+\frac{\pi^2_\phi}{v}\left(\frac{\overline{\delta\pi_\phi}}{\pi_\phi}-\frac{\overline{\delta v}}{v}\right),
	\eea 
that the diffeomorphism constraint reduces to
\bea
\ovl{\mathcal{D}}^{(1)}  = \pi_\phi \ovl{\delta\phi} - v \ovl{\delta\theta}\,,
\eea
and that the equations of motion read
\bea
	&&\left\{\begin{array}{l}
		\ds\dot{\overline{\delta v}}=-\frac{3}{2\Mp^2}v\theta\left(\overline{\delta N}+N\frac{\overline{\delta v}}{v}+N\frac{\overline{\delta\theta}}{\theta}\right), \\
		\ds\dot{\overline{\delta\theta}}=\frac{\pi^2_\phi}{v^2}\left(\overline{\delta N}-2N\frac{\overline{\delta v}}{v}+2N\frac{\overline{\delta\pi_\phi}}{\pi_\phi}\right),
	\end{array}\right. \label{eq:SepUnivVTheta}\\
	&&\left\{\begin{array}{l}
		\ds\dot{\overline{\delta\phi}}=\frac{\pi_\phi}{v}\,\left(\overline{\delta N}-N\frac{\overline{\delta v}}{v}+N\frac{\overline{\delta\pi_\phi}}{\pi_\phi}\right), \\
		\ds\dot{\overline{\delta\pi_\phi}}=-vV_{,\phi}\,\left(\overline{\delta N}+N\frac{\overline{\delta v}}{v}\right)-NvV_{,\phi,\phi}\,\overline{\delta\phi}\, .
	\end{array}\right. 
	\label{eq:SepUnivVTheta:2}
\eea
Note that in order to simplify the equation of motion for the perturbed expansion rate, we made use of the scalar constraint at the background level and at first order, $\mathcal{S}^{(0)}=\ovl{\mathcal{S}^{(1)}}=0$. \\

An important remark is that, while the above formulas have been obtained by plugging the correspondence relations given in Table \ref{table:correspondence} into the full Hamiltonian~\eqref{eq:full:Hamiltonian}-\eqref{eq:dcsg}, an alternative derivation would be to start from the Hamiltonian of the homogeneous and isotropic problem, \Eqs{eq:BckgConsGrav} and \eqref{eq:BckgConsPhi}, or even from the equations of motion of the homogeneous and isotropic problem, \ie \Eqs{eq:ConstHom}-\eqref{eq:Dot:theta} and~\eqref{eq:DotPhiPi}-\eqref{eq:DotPi}, and plug in the correspondence relations at these levels. In \App{app:pert}, we show that these two alternative procedures yield exactly the same equations. In other words, it is equivalent to (i) first perturb the system and then restrict the analysis to homogeneous and isotropic perturbations, and (ii) first impose homogeneity and isotropy and then perturb the reduced system.

Let us also note that, once the phase space has been reduced to the separate-universe degrees of freedom, the Hamiltonian~\eqref{eq:Hamiltonian:SU:full} is exact, \ie it does not contain any perturbative expansion. As a consequence, even though we have derived the relevant dynamical equations at leading order, one could treat the separate-universe non perturbatively, by imposing the vanishing of \Eq{eq:Hamiltonian:SU:full} (this is the scalar constraint equation) and using  \Eq{eq:Hamiltonian:SU:full} to derive the (non-linear) equations of motion.\footnote{We note that the equivalence with the alternative derivation consisting in including the separate-universe deviations in the homogeneous and isotropic Hamiltonian,  \Eqs{eq:BckgConsGrav} and \eqref{eq:BckgConsPhi}, or even directly in the homogeneous and isotropic equations of motion, \Eqs{eq:ConstHom}-\eqref{eq:Dot:theta} and~\eqref{eq:DotPhiPi}-\eqref{eq:DotPi}, also holds at the non-linear level, hence at all orders in the separate-universe perturbations (see \App{app:pert}).} Below we check the agreement between the separate-universe approach and standard \CPT~at leading order, but one should bear in mind that the separate-universe approach is non perturbative. 

\subsection{Fixing the gauge}
\label{sec:SU:gauge}
We end this section by mentioning that gauge fixing can also be performed in the separate-universe framework, which contains five variables, namely $\ovl{\delta N}$, $\ovl{\delta\phi}$, $\ovl{\delta\pi_\phi}$, $\ovl{\delta\gamma_1}$ and $\ovl{\delta\pi_1}$. Since the theory has a single Lagrange multiplier, namely the lapse function, a single scalar combination of the perturbation variables can be set to zero (compared to two in \CPT, see footnote~\ref{footnote:ChangeOfCoordinates}). The vanishing of its equation of motion then leads to a constraint equation, which, added to the scalar constraint equation, leaves two phase-space variables free, \ie one scalar physical degree of freedom (\ie the same number of physical degrees of freedom as in \CPT). As explained around \Eq{eq:MS:def} in the context of the full \CPT, this physical degree of freedom can be parametrised by the Mukhanov-Sasaki variable, which in the separate-universe framework reads
\bea
\label{eq:MS:SU:def}
\ovl{Q}_{{}_\mathrm{MS}} := \sqrt{\frac{v}{N}}\ovl{\delta\phi} + \frac{\Mp^2\pi_\phi}{\sqrt{3N}\theta v^{7/6}}\ovl{\delta\gamma_1}\, .
\eea
Making use of the above constraint and dynamical equations, it obeys the second-order equation of motion
\bea
\label{eq:Q:eom:SU}
\ddot{\ovl{Q}}_{{}_\mathrm{MS}} -\frac{\ddot{z}}{z}\ovl{Q}_{{}_\mathrm{MS}} =\left(4 \frac{\pi_\phi}{\theta v} V - 2  V_{,\phi}\right) \ovl{\mathcal{D}}^{(1)}=\sqrt{\frac{\epsilon_1}{2}}\epsilon_2 \Mp H^2 \ovl{\mathcal{D}}^{(1)}\, ,
\eea
where the second expression casts the right-hand side in terms of the first and second Hubble-flow parameters $\epsilon_2 \equiv \dd\ln\epsilon_1/\dd\ln(v^{1/3})$. This needs to be compared to its CPT counterpart, namely \Eq{eq:MS:CPT}. Two differences can be noticed. First, the term proportional to $k^2 Q_{{}_\mathrm{MS}}$ is absent in the SU, since gradient terms are indeed negligible at large scales. Second, a right-hand side involving the SU diffeomorphism constraint is present in \Eq{eq:Q:eom:SU}. As noted above, a specific constraint can be imposed in the anisotropic sector to make it vanish. Otherwise, $ \ovl{\mathcal{D}}^{(1)}$ is a constant, hence the right-hand side is either almost constant (as in slow-roll inflation), or decays (as in ultra slow-roll inflation, where it decays as $1/v$). In either case, it is much smaller than the left-hand side, which necessarily grows ($Q_{{}_\mathrm{MS}} \propto v^{1/3}$ hence  $\ddot{Q}_{{}_\mathrm{MS}} \propto v$, both in slow roll and ultra slow roll). As a consequence, the term arising from the diffeomorphism constraint can only affect sub-dominant modes on super-Hubble scales, which must be discarded in a gradient expansion anyway. We conclude that it does not jeopardise the SU approach.

Alternatively, let us see how the gauges introduced in \Sec{ssec:gaugefix} proceed in the separate-universe picture.
\subsubsection{Spatially-flat gauge}
In the spatially-flat gauge introduced in \Sec{sec:Spatially:Flat:def}, $\delta\gamma_{ij}=0$, which simply translates into $\ovl{\delta\gamma_1}=0$ in the separate-universe approach. Requiring that $\dot{\ovl{\delta\gamma_1}}=0$ in \Eq{eq:SepUnivFull}, together with the vanishing of the linear scalar constraint given in \Eq{eq:s1su}, then leads to
\bea
\label{eq:deltaNbar:SFgauge}
\overline{\delta N}& =& -\frac{2\Mp^2}{3}\frac{N}{\theta^2}\left(V_{,\phi}\,\overline{\delta\phi}+\frac{\pi_\phi}{v^2}\,\overline{\delta\pi_\phi}\right), \\
\ovl{\delta\pi_1}&=&\frac{\Mp^2 }{\sqrt{3}v^{2/3}\theta}\left(\frac{\pi_\phi}{v}\ovl{\delta\pi_\phi}+v V' \ovl{\delta\phi}\right).
\label{eq:deltapi1bar:SFgauge}
\eea
All variables can therefore be expressed in terms of $\ovl{\delta\phi}$ and $\ovl{\delta\pi_\phi}$ only, whose dynamics is given by \Eqs{eq:SepUnivVTheta:2} where the above replacements are made.
\subsubsection{Newtonian gauge}
The Newtonian gauge was defined in \Sec{sec:Newtonian:def} with the condition $\delta\gamma_2=\delta\pi_2=0$, or equivalently $\delta\gamma_2=\delta N_1=0$. Since the corresponding variables are already set to zero in the separate-universe approach, see Table~\ref{table:correspondence}, these conditions yield no prescription in the separate universe. 
\subsubsection{Generalised synchronous gauge}
The generalised synchronous gauge was defined in \Sec{sec:Generalised:Synchronous:def} with the condition $\delta N=0$, which here translates into $\ovl{\delta N}=0$. As already noticed in \Sec{sec:Generalised:Synchronous:def}, no further constraint is imposed from the equations of motion, so that gauge is not entirely fixed.
\subsubsection{Uniform-expansion gauge}
\label{sec:Uniform:Expansion:Gauge:SU:def}
In the uniform-expansion gauge introduced in \Sec{sec:Uniform:Expansion:Gauge:def}, $\delta N_1=\delta\gamma_1=0$, which simply translates into $\ovl{\delta\gamma_1}=0$ in the separate-universe approach. One thus obtains \Eqs{eq:deltaNbar:SFgauge} and~\eqref{eq:deltapi1bar:SFgauge} as in the separate-universe spatially-flat gauge, so the uniform-expansion gauge is unequivocally defined in the separate-universe framework.

\section{Separate universe versus cosmological-perturbation theory}
\label{sec:suvspert}
Having studied scalar fluctuations in \CPT, see \Sec{sec:cosmopert}, and in the separate-universe approach, see \Sec{sec:SepUniv}, we are now in a position where we can compare the two and derive the conditions under which the latter provides a reliable approximation of the former. This will be first done by leaving the gauge unfixed, where we will recover the conditions obtained in \Refc{Pattison:2019hef} from an analysis performed in the Lagrangian framework. We will then consider the gauges introduced in \Sec{ssec:gaugefix}, where we will show that the agreement between the gauge matching procedures is not always guaranteed, and that it sometimes requires specific matching prescriptions that we will establish.

\subsection{Arbitrary gauge}
\label{ssec:comp}
For the linear scalar constraint, one has to compare \Eq{eq:s1su} with \Eq{eq:scal1gen:simp} where the replacements outlined in Table~\ref{table:correspondence} are performed. One can see that the two constraints are the same, provided that
\bea
	\frac{k^2}{v^{2/3}}\ll\frac{1}{\Mp^2}\left|\frac{\pi_\phi^2}{v^2}-V\right|\, .\label{eq:lss1}
\eea
The case of the  diffeomorphism constraint was already discussed around \Eq{eq:diff1:cpt}.

For the quadratic scalar constraint, one has to compare \Eqs{eq:scalconst2} and~\eqref{eq:s2su}. For the terms proportional to $|\delta\phi|^2$ to match, one must impose
\bea
	\frac{k^2}{v^{2/3}}&\ll&\left|V_{,\phi,\phi}\right| \label{eq:lss3}\, ,
\eea
which implies that the physical wavenumber is much smaller than the mass of the scalar field.
The terms involving gravitational perturbations require more attention. They can be written in matricial form as  $(\delta \gamma_1, \delta\gamma_2) M (\delta \gamma_1^\star, \delta\gamma_2^\star) ^\mathrm{T} $, where
\bea
M=\begin{pmatrix}
\frac{1}{3v^{1/3}}\left(\frac{\pi_\phi^2}{v^2}+\frac{V}{2} - \frac{\Mp^2 k^2}{4v^{2/3}} \right) & \frac{\sqrt{2}\Mp^2}{24v}k^2 \\
 \frac{\sqrt{2}\Mp^2}{24v}k^2 & \frac{1}{3v^{1/3}}\left(\frac{\pi_\phi^2}{v^2}+\frac{V}{2} - \frac{\Mp^2 k^2}{8v^{2/3}} \right)
\end{pmatrix}
\label{eq:M:matrix:def}
\eea
is a symmetric matrix that can be read off from \Eq{eq:scalconst2}, and which eigenvalues are given by
\bea
\lambda_1=\frac{2}{3v^{1/3}}\left(\frac{\pi_\phi^2}{v^2}+\frac{V}{2}\right)
\quad\quad\text{and}\quad\quad
\lambda_2=\lambda_1-\frac{1}{4}\frac{\Mp^2 k^2}{v}\, .
\eea
For the term proportional to $k$ to play a negligible role, one thus has to impose
\bea
	\frac{k^2}{v^{2/3}}&\ll&\frac{1}{\Mp^2}\left|\frac{\pi_\phi^2}{v^2}+\frac{V}{2}\right|. \label{eq:lss2}
\eea
When the conditions~\eqref{eq:lss3} and~\eqref{eq:lss2} hold, \Eqs{eq:scalconst2} reduces to
\bea
\mathcal{S}^{(2)}(k\to 0)&\simeq & - \frac{v^{1/3}}{\Mp^2}\left|\delta\pi_1\right|^2 +\frac{1}{3v^{1/3}}\left(\frac{\pi^2_\phi}{v^2}+\frac{V}{2}\right)\left|\delta\gamma_1\right|^2 +\frac{\left|\delta\pi_\phi\right|^2}{2v} + \frac{v}{2} V_{,\phi,\phi} \left|\delta\phi\right|^2\nonumber \\ 
	&&-\frac{\theta}{2\Mp^2}\mathrm{Re}\left[\left({\delta\pi^\star_1}\right)\left({\delta\gamma_1}\right)\right]  -\frac{\sqrt{3}}{2}v^{1/3} \left\{\frac{\pi_\phi}{v^2}\mathrm{Re}\left[\left({\delta\pi^\star_\phi}\right)\left({\delta\gamma_1}\right)\right] -V_{,\phi} \mathrm{Re}\left[\left({\delta\phi^\star}\right)\left({\delta\gamma_1}\right)\right] \right\} \nonumber \\
	&& + \frac{2 v^{1/3}}{\Mp^2} \left|\delta\pi_2\right|^2 + \frac{1}{3v^{1/3}}\left(\frac{\pi^2_\phi}{v^2}+\frac{V}{2} \right) \left|\delta\gamma_2\right|^2 +\frac{\theta}{\Mp^2}\mathrm{Re}\left[\left({\delta\pi^\star_2}\right)\left({\delta\gamma_2}\right)\right] , 
\eea
which has to be compared to \Eq{eq:s2su}. 
The separate-universe quadratic constraint,  $\ovl{\mathcal{S}^{(2)}}$, can be formally matched to the two first lines of the above limit (note that cosmological perturbations are real-valued in real space). As expected, it is however unable to capture the last line of the above expression, which contains the anisotropic gravitational perturbations. Nonetheless, it is important to stress that in the above limit, these anisotropic degrees of freedom decouple from the isotropic ones. This is why, at large scales, the dynamics of the isotropic cosmological perturbations is independent of the anisotropic sector, and is thus correctly described by the separate-universe approach.
These considerations thus allow us to establish the following statement:\\ 

\textit{On large scales, the separate-universe framework, in which the homogeneous and isotropic problem is perturbed (either at the level of its Hamiltonian or at the level of its dynamical equations), is equivalent to the full \CPT~where the anisotropic degrees of freedom are set to zero.}\\

The next question is to determine whether or not it is legitimate to set the anisotropic degrees of freedom to zero, \ie under which condition the anisotropic degrees of freedom are negligible compared with the isotropic degrees of freedom. The answer to that question is necessarily gauge dependent, since the relative amplitude of both sets of degrees of freedom depends on the gauge. This is why, in the remaining part of this section, we will further investigate the gauges introduced in \Secs{ssec:gaugefix} and~\ref{sec:SU:gauge}. But before moving on to that discussion, two remarks are in order.

First, at the gauge-invariant level, one can compare the separate-universe approach and \CPT~by inspecting the equations of motion for the Mukhanov-Sasaki variable, \ie \Eqs{eq:Q:eom} and~\eqref{eq:Q:eom:SU}. Under the condition~\eqref{eq:lss3}, the latter reduces to the former, which confirms the validity of the separate-universe approach.

Second, the three conditions obtained on the amplitude of the wavenumber, \ie \Eqs{eq:lss1}, \eqref{eq:lss3} and~\eqref{eq:lss2}, can be summarised as follows. Upon writing $k=\sigma a H$, and using the Friedmann and Raychaudhuri equations~\eqref{eq:Friedmann:usual} and~\eqref{eq:Raychaudhury}, they give rise to
\bea
\label{eq:sigma:cond}
\sigma\ll \sqrt{\vert \eta \vert},  \sqrt{\vert 1+ 3w \vert}, \sqrt{\vert 1+ \frac{3}{5}w \vert }\, .
\eea
Here, $\eta\equiv V''/H^2$ is the so-called ``eta parameter'', which measures the squared mass of the field in Hubble units. In the context of inflation, it is given by $\eta\simeq 6\epsilon_1-3\epsilon_2/2$, where $\epsilon_1:=-\dd(\ln H)/(\dd\mathcal{N})$ and $\epsilon_2:=\dd(\ln \epsilon_1)/(\dd\mathcal{N})$ are the two first slow-roll parameters. It is therefore a small parameter. The quantity $w$ denotes the equation-of-state parameter, which in inflation differs from $-1$ by slow-roll corrections. Hence the second and third constraints are of order one. The most stringent constraint therefore comes from the eta parameter\footnote{It is worth noting that this might be different in a non-inflationary context, for instance when the universe transits from an accelerated expansion to a decelerated one (or vice-versa) for which $ \sqrt{\vert 1+ 3w \vert}$ vanishes. We also note that the last constraint is always of order one or larger unless one is considering matter contents violating the null energy condition, \ie $w<-1$.} and imposes to consider super-Hubble wavelengths. We finally stress that this set of conditions is gauge-dependent in the sense that some of them may not be mandatory in some specific gauges. For instance, the constraints~\eqref{eq:lss1} and~\eqref{eq:lss2} are not necessary when working in the uniform-expansion gauge or in the spatially-flat gauge, in which $\delta\gamma_1$ is imposed to be zero.
\subsection{Fixing the gauge}
\label{sec:US:vs:CPT:gauge}
Let us now compare the separate-universe approach and \CPT~in the few gauges discussed in \Secs{ssec:gaugefix} and~\ref{sec:SU:gauge}.
\subsubsection{Spatially-flat gauge}
The spatially-flat gauge is unequivocally defined both in \CPT~and in the separate-universe approach. However, the gauge-fixing procedure proceeds differently in these two frameworks. Indeed, even though the same expression is obtained for the perturbed momentum of the induced metric, see \Eqs{eq:pi1phi:flat} and~\eqref{eq:deltapi1bar:SFgauge}, it leads to different expressions for the perturbed lapse and shift, see \Eqs{eq:deltaNphi:flat} and~\eqref{eq:deltaNbar:SFgauge}. This clearly violates the correspondences of Table~\ref{table:correspondence}. Another manifestation of this mismatch comes from noticing that applying the correspondence of Table~\ref{table:correspondence} to \Eq{eq:deltaN1phi:flat} leads to a relationship between $\ovl{\delta\phi}$ and $\ovl{\delta\pi_\phi}$ that is clearly not satisfied in the separate-universe picture. The reason for these discrepancies can be traced back to the fact that $k\delta N_1$ is not $k$-suppressed~\cite{Cruces:2021iwq}, see \Eq{eq:deltaN1phi:flat} again.
We thus conclude that in the spatially-flat gauge, the separate universe approach does not lead to the appropriate gauge fixing.
\subsubsection{Newtonian gauge}
As explained in \Sec{sec:SU:gauge}, since the Newtonian gauge consists in freezing the anisotropic degrees of freedom, it does not lead to any relevant constraint in the separate-universe framework. 
This problem can be solved by considering an alternative definition of the Newtonian gauge by means of \Eqs{eq:deltaN:deltagamma1:Newtonian} and~\eqref{eq:D1:Newtonian}, \ie by imposing
\bea
\label{eq:Newtonian:alternative:1}
\frac{{\delta N}}{N}&=&-\frac{{\delta\gamma_1}}{2\sqrt{3} v^{2/3}}\, , \\
{\delta\pi_1} &=& \frac{\sqrt{3}\pi_\phi}{2 v^{2/3}}{\delta\phi} + \frac{\theta}{4v^{1/3}} {\delta\gamma_1}\, .
\label{eq:Newtonian:alternative:2}
\eea
The fact that these two conditions lead to the same definition of the Newtonian gauge as the one introduced in \Sec{sec:Newtonian:def} (namely $\delta\gamma_2=\delta N_1=0$, or equivalently $\delta\gamma_2=\delta\pi_2=0$), can be seen as follows. Combining \Eq{eq:Newtonian:alternative:2} with the vanishing of \Eq{eq:diff1gen} first leads to $\theta\delta\gamma_2+2v^{1/3}\delta\pi_2=0$. By differentiating this relationship with respect to time, using the equations of motion~\eqref{eq:eom:gen}, one obtains $N\delta\gamma_2(2\pi_\phi^2/v^2+4V+\Mp^2 k^2/v^{2/3})/6+N\theta v^{1/3}  \delta\pi_2/\Mp^2=0$, where we have used \Eq{eq:Newtonian:alternative:1} to simplify the result, together with the Friedmann equation~\eqref{eq:ConstHom}. The above two formulas then lead to $\delta\gamma_2=\delta\pi_2=0$, which indeed corresponds to the (original) definition of the Newtonian gauge.

The advantage of defining the Newtonian gauge with \Eqs{eq:Newtonian:alternative:1} and~\eqref{eq:Newtonian:alternative:2} is that these two relations (more precisely the barred version of them) give non-trivial constraints in the separate-universe framework. One may be concerned that the vanishing of the time derivative of \Eq{eq:Newtonian:alternative:2} in the separate-universe leads to an additional constraint equation, that would make the gauge over constrained. This is however not the case since \Eq{eq:Newtonian:alternative:2} comes from the vanishing of the diffeomorphism constraint, and as discussed in \Sec{sssec:sepunivgen}, in the separate universe one always has $\dot{\ovl{\mathcal{D}^{(1)}}}=0$. Furthermore, by construction, the gauge-fixing conditions~\eqref{eq:deltaN:deltagamma1:Newtonian} and~\eqref{eq:D1:Newtonian} are properly mapped through the correspondences of Table~\ref{table:correspondence}. This makes the Newtonian gauge well behaved from the separate-universe perspective, provided that the definition~\eqref{eq:Newtonian:alternative:1}-\eqref{eq:Newtonian:alternative:2} is employed.
\subsubsection{Generalised synchronous gauge}
As explained in \Secs{ssec:gaugefix} and~\ref{sec:SU:gauge}, the generalised synchronous gauges are under-constrained both in \CPT~and in the separate-universe approach. Let us note that, in the latter case, one can use the same trick as above in the Newtonian gauge, and add the barred version of \Eq{eq:Newtonian:alternative:2}, \ie ${\ovl{\mathcal{D}^{(1)}}}=0$, in the definition of the generalised synchronous gauge. Together with $\ovl{\delta N}=0$, this fully specifies that gauge in the separate-universe framework and makes it well behaved. This may also offer a way to cure the synchronous gauge in \CPT. This is because, as pointed out above, the condition ${\ovl{\mathcal{D}^{(1)}}}=0$ is equivalent to imposing $ \theta \delta\gamma_2+2v^{1/3} \delta\pi_2=0$ in the anisotropic sector of \CPT, which may fix the remaining gauge degrees of freedom. We plan to investigate this possibility in a future work.
\subsubsection{Uniform-expansion gauge}
As explained in \Secs{ssec:gaugefix} and~\ref{sec:SU:gauge}, the uniform-expansion gauge is not fully defined in \CPT, but it is unambiguous in the separate-universe approach. 
There are a priori several ways to complement the definition of that gauge in \CPT, for instance by further constraining the anisotropic sector (such that it does not lead to additional conditions in the separate-universe framework). However, the comparison with the separate-universe version of the uniform-expansion gauge does not depend on that choice since, as pointed out above, $\delta\gamma_2$ and $\delta\pi_2$ decouple from  the isotropic degrees of freedom in the large-scale limit.  This makes the uniform-expansion gauge well behaved from the separate-universe perspective (whatever its completion in \CPT).

\section{Conclusions}
\addtocontents{toc}{\protect\setcounter{tocdepth}{1}} 
\label{sec:conclusion}
In this work, we have presented a Hamiltonian, phase-space description of Cosmological-Perturbation Theory (CPT) and of the separate-universe approach, when the matter content of the universe is made of a scalar field and gravity is described with general relativity. The separate-universe approach consists in perturbing the reduced Hamiltonian of the homogeneous and isotropic problem, or equivalently, in perturbing the dynamical equations obtained for that same problem.\footnote{This equivalence is valid even at the non-perturbative level, as proven in \App{app:pert}.}

Our conclusion, stated at the end of \Sec{ssec:comp}, is that this matches \CPT~at leading order in perturbations when restricted to isotropic degrees of freedom (\ie when setting the anisotropic perturbations to zero, $\delta N_1=\delta\gamma_2=\delta\pi_2=0$), provided that one considers sufficiently large scales, \ie scales satisfying \Eq{eq:sigma:cond}. This result is non trivial since it implies that (i) phase-space reduction to the isotropic sector and (ii) derivation of the dynamical equations, are two commuting procedures on large scales. 
Since the dynamics of isotropic and anisotropic degrees of freedom decouple at large scales, we have shown that the separate-universe formalism provides an accurate description of the large-scale gauge-invariant combinations such as the Mukhanov-Sasaki variable. 

Note that we have not made any specific assumption about the background solution, hence the validity of the separate-universe approach has been established for all kind of cosmological evolution (slow-roll and non-slow-roll inflating --- in agreement with the conclusion of \Refc{Pattison:2019hef} but in contrast to what was found in \Refc{Cruces:2018cvq}, expanding, even contracting, \textit{etc}.). 

When calculations need to be performed in a given gauge, one should bear in mind that not all gauges are well suited for the separate-universe approach. More precisely, we have found that in the spatially-flat gauge, the gauge-fixing procedure fails in the separate-universe approach because of the important role the perturbed shift plays in the \CPT~version of that gauge. The Newtonian gauge is a priori ill-defined in the  separate-universe approach, but we have found an alternative (though perfectly equivalent at the level of \CPT) definition of that gauge that makes it unambiguous in the separate-universe approach, where the gauge-fixing procedure correctly reproduces \CPT. The synchronous gauges are ambiguous in both approaches, but they can be made well defined in the separate-universe approach by using a similar trick (which consists in further imposing that the diffeomorphism constraint vanishes as a gauge condition). Finally, the uniform-expansion gauge, which is employed in the stochastic-$\delta N$ formalism, is well defined in the separate-universe approach, where the gauge-fixing procedure correctly reproduces \CPT. We note that, among the different gauges that we considered, the separate-universe healthy gauges have in common that they impose the vanishing of the perturbed shift.

Let us now mention a few research directions this work opens up. First, although we have shown that the separate universe matches \CPT~ at leading order in perturbations only, our formulation allowed us to derive fully non-perturbative equations of motion in the separate universe, hence paving the way for investigating the matching with \CPT~at the next-to-leading order.
Second, our treatment of the gauge-invariant problem was restricted to deriving the equation of motion for the Mukhanov-Sasaki variable, but it remains to establish a systematic procedure that would provide all gauge-invariant parameterisations of the Hamiltonian phase space, both in \CPT~and in the separate-universe framework. Similarly, while we have exhibited examples of both problematic and healthy gauges in the separate-universe approach, building a formalism to study gauge transformations in the Hamiltonian picture should allow us to classify gauges in a more systematic way, and to derive generic criteria for them to (i) be unambiguous and (ii) feature a gauge-fixing procedure in the separate-universe approach that matches the one performed in \CPT. We will further investigate these aspects in forthcoming works.
 Third, as mentioned in \Sec{sec:intro}, a Hamiltonian description of the separate-universe dynamics is necessary for the stochastic-inflation formalism (at least in the absence of a phase-space attractor). Let us stress that in this context, there is no equivalent Lagrangian formulation, since the phase-space direction of the stochastic noise plays a crucial role, and it cannot be encoded in the Lagrangian approach. For instance, it is involved in determining whether stochastic effects break classical attractors~\cite{Grain:2017dqa}, in solving the vielbeins' frame ambiguity~\cite{Pinol:2020cdp}, or in describing the backreaction of quantum fluctuations in a phase of ultra-slow roll~\cite{Firouzjahi:2020jrj, Pattison:2021oen, Figueroa:2021zah}. This is why the present Hamiltonian formulation is a pre-requisite to using the stochastic formalism in the absence of a phase-space attractor, such as when slow roll is violated during inflation or in slowly contracting cosmologies.

\acknowledgments
We would like to thank Diego Cruces and David Wands for interesting discussions.

\appendix
\addtocontents{toc}{\protect\setcounter{tocdepth}{1}} 
\section{Connecting perturbations in the Lagrangian and Hamiltonian frameworks}
\label{app:lag}
In this appendix, we derive the relations between cosmological perturbations defined in the Hamiltonian framework and perturbations defined in the more usual Lagrangian approach. In practice, this implies to relate perturbations of the configuration variables in the Hamiltonian framework introduced in \Sec{ssec:dofpert} to the perturbations of the scalar field and of the four-dimensional metric in the Lagrangian approach. The case of the scalar field is straightforward since the perturbed configuration variable is nothing but $\delta\phi$ in the Lagrangian framework. For the gravitational sector, keeping only scalar degrees of freedom, the four-dimensional metric can be expanded as follows (see \eg \Refc{Malik:2008im})
\bea
\label{eq:perturbed:metric}
	\dd s^2&=&-\underbrace{N^2(\tau)\left(1+2A\right)}_{(N+\delta N)^2}\dd \tau^2+\underbrace{2{v^{2/3}}\partial_iB}_{2\gamma_{ij}\delta N^j}\dd x^i\dd\tau+\underbrace{v^{2/3}\left[\left(1+2C\right)\widetilde{\gamma}_{ij}+2\partial_i\partial_jE\right]}_{\gamma_{ij}+\delta\gamma_{ij}}\dd x^i\dd x^j, \nonumber \\
\eea
where $A,\,B,\,C$ and $E$ are four scalar functions, depending on both space and time, and upon which the perturbative expansion is performed (note that we use $p=v^{2/3}$ for the background metric). 

The first scalar function, $A$, is simply related to the lapse perturbation via $A=\delta N/N$, hence in Fourier space we have 
\bea
\delta N(\vec{k})= N(\tau)A(\vec{k})\, .
\eea
The second scalar function, $B$, generates perturbations in the shift vector, \ie $\delta N^i(\tau,\vec{x})={v^{2/3}}\gamma^{im}(\tau)\partial_m B(\tau,\vec{x})= \widetilde{\gamma}^{im}\partial_m B(\tau,\vec{x})$. In Fourier space, using \Eq{eq:deltaNi:deltaN1}, this leads to 
\bea
\delta N_1(\tau,\vec{k})=k B(\tau,\vec{k})\, .
\eea
Finally, the scalar functions $C$ and $E$ describe the isotropic and anisotropic perturbations of the metric. In Fourier space, \Eq{eq:perturbed:metric} implies that
\bea
	\delta\gamma_{ij}(\tau,\vec{k})=2v^{2/3}\left[C(\tau,\vec{k})\widetilde{\gamma}_{ij}-k_ik_j\,E(\tau,\vec{k})\right].
\eea
The configuration variables $\delta\gamma_1$ and $\delta\gamma_2$ are related to the induced metric through \Eq{eq:delta:gamma:A:delta:gamma:ij}, which gives rise to 
\bea
	\delta\gamma_1(\tau,\vec{k}) &=& \frac{2}{\sqrt{3}}v^{2/3}\left[3C(\tau,\vec{k})-k^2E(\tau,\vec{k})\right], \\
	\delta\gamma_2(\tau,\vec{k}) &=& -2\sqrt{\frac{2}{3}}v^{2/3}k^2E(\tau,\vec{k}).
\eea
We note that $\delta\gamma_2$ is a function of $E$ only. Hence in the Newtonian gauge (see \Sec{sec:Newtonian:def}), corresponding to the choice $B=E=0$, $\delta\gamma_2$ and $\delta N_1$ are set to zero, and the condition~\eqref{eq:deltaN:deltagamma1:Newtonian} leads to $A=-C$.
In the spatially-flat gauge (see \Sec{sec:Spatially:Flat:def}) where $C=E=0$, both $\delta\gamma_1$ and $\delta\gamma_2$ are set to zero. 
In the generalised synchronous gauge (see \Sec{sec:Generalised:Synchronous:def}), $A=B=0$, which leads to $\delta N=\delta N_1=0$. In the uniform-expansion gauge (see \Sec{sec:Uniform:Expansion:Gauge:def}), $\delta N_1=\delta\gamma_1=0$ implies that $B=3C-k^2E=0$ in the Lagrangian framework, which is in agreement with Eq.~(3.15) of \Refc{Pattison:2019hef}.
\section{Linear and quadratic constraints}
\label{app:gloss}
In this appendix, we expand the constraints up to quadratic order in scalar perturbations. 
At the background level (\ie in the homogeneous and isotropic setup studied in \Sec{ssec:bck}), we remind that the induced metric and its conjugated momentum are given by
\bea
\label{eq:gamma:ij:DOWN:app}
	\gamma_{ij}(\tau)&=&v^{2/3}\widetilde{\gamma}_{ij}\, , \\
	\pi^{ij}(\tau)&=&\frac{1}{2}v^{1/3}\theta\widetilde{\gamma}^{ij}=\frac{1}{2}v\theta\gamma^{ij}\, ,
\label{eq:pi:ij:UP:app}
\eea
where hereafter $\widetilde{\gamma}_{ij}=\mathrm{diag}[1,1,1]$ (\ie we consider the case of a spatially flat FLRW metric). Recalling that their indices are raised and lowered by the induced metric itself, one has
\bea
\label{eq:gamma:ij:UP:app}
	\gamma^{ij}(\tau)&=&v^{-2/3}\widetilde{\gamma}^{ij}\, , \\
	\pi_{ij}(\tau)&=&\frac{1}{2}v^{5/3}\theta\widetilde{\gamma}_{ij}=\frac{1}{2}v\theta\gamma_{ij}\, .
\eea
This also leads to $\pi:=\gamma_{ij}\pi^{ij}=3v\theta/2$. 
We remind that gravitational perturbations can be expanded according to
\bea
\label{eq:delta:gamma:ij:DOWN:app}
	\delta\gamma_{ij}(\tau,\vec{x})&=&\ds\int\frac{\dd^3\vec{k}}{(2\pi)^{3/2}}e^{i\vec{k}\cdot\vec{x}}\sum_{A=1}^2\delta\gamma_A(\tau,\vec{k})M^A_{ij}(\vec{k})\, , \\
	\delta\pi^{ij}(\tau,\vec{x})&=&\ds\int\frac{\dd^3\vec{k}}{(2\pi)^{3/2}}e^{i\vec{k}\cdot\vec{x}}\sum_{A=1}^2\delta\pi_A(\tau,\vec{k})M_A^{ij}(\vec{k})\, ,
\label{eq:delta:pi:ij:UP:app}
\eea
where $M_{ij}^1$ and $M_{ij}^2$ are the two matrices introduced in \Eq{eq:Mij:def}. Indices for $M^A_{ij}$ and $k_i$ are raised and lowered by the flat three-dimensional metric, $\widetilde{\gamma}_{ij}$. The $M_A$'s form an orthonormal basis, and they satisfy the following relations
\bea
\label{eq:MA:usefulrelations}
	M^A_{ij}M_{A'}^{ij}=\delta_{A,A'}, &~~~\widetilde\gamma^{ij}M^A_{ij}=\sqrt{3}\delta_{A,1}, &~~~\mathrm{and}~~~k^iM^A_{ij}=\sqrt{\frac{A}{3}}\,k_j\, ,
\eea
which will be useful in what follows.

From \Eq{eq:scgg}, let us split the gravitational part of the scalar constraint, $\mathcal{S}^{(\mathrm{G})}$, into a kinetic part and a potential part, respectively given by
\bea
\label{eq:calT:def}
	\mathcal{T}(\gamma_{ij},\pi^{mn})&:=&\frac{2}{\Mp^2\sqrt{\gamma}}\left(\pi^{ij}\pi_{ij}-\frac{1}{2}\pi^2\right)\, , \\
	\mathcal{W}(\gamma_{ij})&:=&-\frac{\Mp^2\sqrt{\gamma}}{2}\,\mathcal{R}(\gamma_{ij})\, ,
\label{eq:calW:def}
\eea
where we recall that $\gamma=\det(\gamma_{ij})=v^2$.
We note that $\mathcal{W}$ depends on the induced metric only, while the kinetic contribution $\mathcal{T}$ depends on both the momentum $\pi^{ij}$ and the induced metric $\gamma_{ij}$ (not only through $\sqrt{\gamma}$ but also via $\pi_{ij}=\gamma_{im}\gamma_{jn}\pi^{mn}$ and $\pi=\gamma_{ij}\pi^{ij}$).
A similar decomposition can be done for the scalar field contribution to the scalar constraint, $\mathcal{S}^{(\mathrm{\phi})}$, namely 
\bea
\label{eq:def:T}
	T(\pi_\phi)&=&\frac{\pi^2_\phi}{2\sqrt{\gamma}}\, , \\
	W(\phi)&=&\frac{\sqrt{\gamma}}{2}\gamma^{ij}\partial_i\phi\,\partial_j\phi+\sqrt{\gamma}V(\phi)\, ,
\label{eq:def:W}
\eea
see \Eq{eq:scsg}. From the considerations presented in \Sec{sec:Dynamics:Scalar:Perturbations}, the scalar constraint needs to be expanded up to quadratic order in perturbations, while the diffeomorphism constraint,
\bea
\label{eq:Diffeo:Constraints:App}
	\mathcal{D}_i & = & \pi_\phi\partial_i\phi-2\partial_m\left(\gamma_{ij}\pi^{jm}\right)+\pi^{mn}\partial_i\gamma_{mn}\, ,
\eea
see \Eqs{eq:dcgg} and~\eqref{eq:dcsg}, only needs to be expanded up to linear order in perturbations.
\subsection{Constraints at the background level} 
At the background level, the different constraints and their associated contributions for the gravitational sector are
\bea
	\mathcal{T}^{(0)}&=&\frac{-3}{4\Mp^2}v\theta^2, \\
	\mathcal{W}^{(0)}&=&0.
\eea
For the scalar-field sector, they read
\bea
	T^{(0)}&=&\frac{1}{2v}\pi^2_\phi, \\
	W^{(0)}&=&vV(\phi).
\eea
The diffeomorphism constraint is identically vanishing, \ie $\mathcal{D}_i=0$.

\subsection{Constraints at first order} 
At linear order in perturbation theory, one has
\bea
	\delta\gamma^{ij}&=&- v^{-4/3}\ds\int\frac{\dd^3k}{(2\pi)^{3/2}}e^{i\vec{k}\cdot\vec{x}}\sum_{A=1}^2\delta\gamma_A M_A^{ij}, \\
	\delta\pi_{ij}&=&v^{4/3}\ds\int\frac{\dd^3k}{(2\pi)^{3/2}}e^{i\vec{k}\cdot\vec{x}}\sum_{A=1}^2\left(\delta\pi_A+v^{-1/3}\theta\delta\gamma_A\right)M^A_{ij},
\eea
see \Eqs{eq:delta:gamma:ij:UP} and~\eqref{eq:delta:pi:ij:DOWN}. 
\subsubsection{Scalar constraint}
We start with the kinetic part of the scalar gravitational constraint. Linearising it at first order gives
\bea
\label{eq:T1:interm}
	\mathcal{T}^{(1)}=-\frac{\delta\gamma}{2\gamma}\mathcal{T}^{(0)}+\frac{2}{\Mp^2 v}\left[\delta\pi^{ij}\,\pi_{ij}+\pi^{ij}\,\delta\pi_{ij}-\pi\left(\delta\gamma_{ij}\,\pi^{ij}+\gamma_{ij}\,\delta\pi^{ij}\right)\right].
\eea
This requires to compute the following contractions for the induced metric,
\bea
	\gamma^{ij}\delta\gamma_{ij}&=&\frac{\sqrt{3}}{v^{2/3}}\,\delta\gamma_1, \\
	\gamma_{ij}\delta\gamma^{ij}&=&-\gamma^{ij}\delta\gamma_{ij}=-\frac{\sqrt{3}}{v^{2/3}}\,\delta\gamma_1, \\
	\delta\gamma&=&\gamma\,\gamma^{ij}\,\delta\gamma_{ij}=\sqrt{3}v^{4/3}\,\delta\gamma_1,
	\label{eq:delta:gamma}
\eea
where the last equation requires to use the identity $\ln(\det \gamma_{ij}) = \mathrm{Tr}(\ln \gamma_{ij})$. The contractions involving the conjugate momentum are given by
\bea
	\pi_{ij}\delta\pi^{ij}&=&\frac{\sqrt{3}}{2}v^{5/3}\theta\,\delta\pi_1, \\
	\pi^{ij}\delta\pi_{ij}&=&\frac{\sqrt{3}}{2}v^{1/3}\theta\,\left(v^{4/3}\delta\pi_1+v\theta\delta\gamma_1\right),
\eea
and cross-contractions read
\bea
\label{eq:piij:deltagammaij}
	\pi^{ij}\delta\gamma_{ij}&=&\frac{\sqrt{3}}{2}v^{1/3}\theta\,\delta\gamma_1, \\
	\gamma_{ij}\delta\pi^{ij}&=&\sqrt{3}v^{2/3}\,\delta\pi_1.
\label{eq:gammaij:deltapiij}
\eea
Plugging the above expressions into \Eq{eq:T1:interm} leads to
\bea
\label{eq:calT1:app:final}
	\mathcal{T}^{(1)}=\frac{-\sqrt{3}}{\Mp^2}v^{1/3}\theta\left(\frac{\theta}{8}\delta\gamma_1+v^{1/3}\delta\pi_1\right).
\eea
Note that only  the isotropic part of the perturbations, \ie $\delta\gamma_1$ and $\delta\pi_1$, contributes to the kinetic part, $\mathcal{T}^{(1)}$.

The linearised potential term is $\mathcal{W}^{(1)}\propto \left(\delta\gamma/2\gamma\right)\mathcal{R}+\delta\gamma^{ij}\,\mathcal{R}_{ij}+\gamma^{ij}\delta\mathcal{R}_{ij}$, where the Ricci tensor is defined in terms of Christoffel symbols $\gamma^k_{ij}$ as:
\bea
\label{eq:Ricci:tensor:Christoffel}
\mathcal{R}_{ij} = \partial_k \gamma_{ij}^k - \partial_i \gamma_{k j}^k + \gamma_{k\ell}^k \gamma_{ij}^\ell - \gamma_{i\ell}^k \gamma_{k j}^\ell\, .
\eea
In a spatially flat FLRW metric, the three-dimensional Ricci tensor and Ricci scalar are zero, \ie $\mathcal{R}_{ij}=0=\mathcal{R}$. This is why the perturbed potential term at first order reduces to $\mathcal{W}^{(1)}=-(\Mp^2/2)\sqrt{\gamma}\gamma^{ij}\,\delta\mathcal{R}_{ij}$. Because the Christoffel symbols vanish on the background, the variation of the Ricci tensor is related to the variation of the Christoffel symbols via $\delta\mathcal{R}_{ij}=\partial_\ell\delta\gamma^\ell_{ij}-\partial_j\delta\gamma^\ell_{\ell i}$. 
Finally, using the full expression for the Christoffel symbol
\bea
\label{eq:Christoffel:def}
\gamma_{i j}^k = \frac{1}{2} \gamma^{k\ell} \left(\partial_i \gamma_{j\ell}+\partial_j \gamma_{i\ell} - \partial_\ell \gamma_{ij}\right)\,,
\eea
one can compute its first order perturbation,
\bea
\delta \gamma_{i j}^k = \frac{1}{2 v^{2/3}} \tilde{\gamma}^{k\ell} \left(\partial_i \delta\gamma_{j\ell}+\partial_j \delta\gamma_{i\ell} - \partial_\ell \delta\gamma_{ij}\right)\, .
\eea
Combining the above results, one obtains
\bea
	\mathcal{W}^{(1)}=-\frac{\Mp^2v}{2}\,\left(\gamma^{i\ell}\gamma^{jm}-\gamma^{ij}\gamma^{\ell m}\right)\partial_\ell\partial_m\delta\gamma_{ij}.
\eea
In Fourier space, plugging \Eqs{eq:gamma:ij:UP:app} and~\eqref{eq:delta:gamma:ij:DOWN:app} into this formula leads to
\bea
\label{eq:calW1:app:final}
	\mathcal{W}^{(1)}=-\frac{\Mp^2}{\sqrt{3}}\left(\frac{k^2}{v^{1/3}}\right)\left(\delta\gamma_1-\frac{1}{\sqrt{2}}\delta\gamma_2\right)\, ,
\eea
where \Eq{eq:MA:usefulrelations} has been used.

For the scalar-field contribution, the first-order linearised scalar constraint has contributions
\bea
	T^{(1)}&=&-\frac{\delta\gamma}{2\gamma}T^{(0)}+\frac{1}{v}\pi_\phi\,\delta\pi_\phi, \\
	W^{(1)}&=&\frac{\delta\gamma}{2\gamma}W^{(0)}+vV_{,\phi}\,\delta\phi,
\eea
where $\delta\gamma$ is given by \Eq{eq:delta:gamma}, leading to
\bea
\label{eq:T1:app:final}
	T^{(1)}&=&-\frac{\sqrt{3}}{4}\frac{\pi_\phi^2}{v^{5/3}}\,\delta\gamma_1+\frac{\pi_\phi}{v}\,\delta\pi_\phi, \\
	W^{(1)}&=&\frac{\sqrt{3}}{2}v^{1/3}V\,\delta\gamma_1+vV_{,\phi}\,\delta\phi.
\label{eq:W1:app:final}
\eea	
	
\subsubsection{Diffeomorphism constraint}
Perturbing \Eq{eq:Diffeo:Constraints:App} at first order gives
\bea
	\mathcal{D}_i^{(1)}(\tau,\vec{x})=\pi_\phi\partial_i\delta\phi+\pi^{mn}\partial_i\delta\gamma_{mn}-2\pi^{jm}\partial_m\delta\gamma_{ij}-2\gamma_{ij}\partial_m\delta\pi^{jm},
\eea
which in Fourier space reads
\bea
\label{eq:D1:interm}
\mathcal{D}_i^{(1)}(\tau,\vec{k})=ik_i\left(\pi_\phi\delta\phi+\pi^{mn}\delta\gamma_{mn}\right)-2ik_m\left(\pi^{jm}\delta\gamma_{ij}+\gamma_{ij}\delta\pi^{jm}\right).
\eea
The contraction $\pi^{mn}\delta\gamma_{mn}$ was computed in \Eq{eq:piij:deltagammaij}, while the second term in \Eq{eq:D1:interm} can be computed from \Eqs{eq:gamma:ij:DOWN:app}, \eqref{eq:pi:ij:UP:app}, \eqref{eq:delta:gamma:ij:DOWN:app}, \eqref{eq:delta:pi:ij:UP:app} and~\eqref{eq:MA:usefulrelations}, which leads to 
\bea
	\mathcal{D}^{(1)}_i(\vec{k})=ik_i\left[\pi_\phi\delta\phi+\frac{1}{\sqrt{3}}v^{1/3}\theta\left(\frac{1}{2}\delta\gamma_1-\sqrt{2}\delta\gamma_2\right)-\frac{2}{\sqrt{3}}v^{2/3}\left(\delta\pi_1+\sqrt{2}\delta\pi_2\right)\right].
\label{eq:D1:app:final}
\eea
Note that it is independent of the perturbations of the scalar-field momentum, $\delta\pi_\phi$.

\subsection{Constraints at quadratic order}

Let us now derive the expression of the scalar constraint at second order. A first remark is that the expressions~\eqref{eq:delta:gamma:ij:UP} and~\eqref{eq:delta:pi:ij:DOWN} for $\delta\gamma^{ij}$ and $\delta\pi_{ij}$ are valid only at linear order. In what follows, we will not need their expressions at second order, since $\delta\gamma^{ij}$ and $\delta\pi_{ij}$ always appear multiplied by perturbative quantities. For the sake of completeness, however, let us mention that those expressions can be derived by expanding $\gamma^{ij}\gamma_{j\ell}=\delta^i_\ell = [\gamma^{ij}(\tau)+\delta\gamma^{ij}][\gamma_{j\ell}(\tau)+\delta\gamma_{j\ell}(\tau)]$, leading to 
\bea
\delta\gamma^{ij} 
= - \gamma^{im}(\tau)\gamma^{\ell j}(\tau)\delta\gamma_{m\ell}-\gamma^{\ell j}(\tau)\delta\gamma^{im}\delta\gamma_{m\ell}\, .
\eea
Note that this expression is exact (\ie at all orders).
At first order, it reduces to $\delta_1\gamma^{ij}= - \gamma^{im}(\tau)\gamma^{\ell j}(\tau)\delta\gamma_{m\ell}$ and one recovers \Eq{eq:delta:gamma:ij:UP}. At second order, one finds
\bea
\delta_1\gamma^{ij} + \delta_2 \gamma^{ij}= - \gamma^{im}(\tau)\gamma^{\ell j}(\tau)\delta\gamma_{m\ell}+\gamma^{\ell j}(\tau) \gamma^{ik}(\tau)\gamma^{mn}(\tau)
\delta\gamma_{kn}\delta\gamma_{m\ell}\, ,
\eea
where the notations $\delta_1$ and $\delta_2$ refer to the first- and second-order perturbations respectively (elsewhere in this article the notation $\delta_1$ is not used since there is no possible confusion about the order at which a given expression is valid, but we employ it in the few following equations since first- and second-order quantities are considered simultaneously).
In Fourier space, this leads to
\bea
 \kern-2em
\delta_1\gamma^{ij} + \delta_2 \gamma^{ij}  &= &-\frac{1}{v^{4/3}}\left(\delta\gamma_1 M_1^{ij}+\delta\gamma_2 M_2^{ij}\right)
\nonumber \\ & & \kern-2em
+\frac{1}{\sqrt{3}v^2}\left[\left\vert \delta\gamma_1\right\vert^2 M_1^{ij}+\left\vert \delta\gamma_2\right\vert^2\left(M_1^{ij}+\frac{M_2^{ij}}{\sqrt{2}}\right)+\left(\delta\gamma_1\delta\gamma_2^\star + \delta\gamma_1^\star \delta\gamma_2\right)M_2^{ij}\right] .
\label{eq:delta:gamma:ij:UP:2ndOrder}
\eea

For the conjugated momentum, similarly, one has to expand the relation $\pi_{ij} = \gamma_{mi}\gamma_{nj}\pi^{mn}$, leading to 
\bea
\delta\pi_{ij}&=& \gamma_{mi}(\tau)\gamma_{nj}(\tau)\delta\pi^{mn}+\gamma_{mi}(\tau)\pi^{mn}(\tau)\delta\gamma_{nj}+\gamma_{nj}(\tau)\pi^{mn}(\tau)\delta\gamma_{mi}
\nonumber \\ & &
+\gamma_{mi}(\tau)\delta\gamma_{nj}\delta\pi^{mn}+\gamma_{nj}(\tau)\delta\gamma_{mi}\delta\pi^{mn}+\pi^{mn}(\tau)\delta\gamma_{mi}\delta\gamma_{nj}
\nonumber \\ & &
+\delta\gamma_{mi}\delta\gamma_{nj}\delta\pi^{mn}\, .
\eea
In this expression, which is again exact (hence valid at all orders), the first line corresponds to first-order terms and leads to \Eq{eq:delta:pi:ij:DOWN}, the second line corresponds to second-order terms, and the third line to the third-order term. Computing the second-order term in Fourier space as before, one obtains, at second order
\bea
\delta_1\pi_{ij}+ \delta_2 \pi_{ij} &=&\left(v^{4/3}\delta\pi_1+v\theta\delta\gamma_1\right) M^1_{ij}+\left(v^{4/3}\delta\pi_2+v\theta\delta\gamma_2\right) M^2_{ij}
\nonumber \\ & &
+\frac{2}{\sqrt{3}}v^{2/3} \left(\delta\gamma_1\delta\pi_1^\star+ \delta\gamma_2\delta\pi_2^\star+\frac{\theta}{4v^{1/3}}\left\vert\delta\gamma_1\right\vert^2+\frac{\theta}{4v^{1/3}}\left\vert\delta\gamma_2\right\vert^2\right) M^1_{ij}
\nonumber \\ & &
+\frac{2}{\sqrt{3}}v^{2/3} \left[\delta\gamma_1\delta\pi_2^\star + \delta\gamma_2\delta\pi_1^\star+\frac{\delta\gamma_2\delta\pi_2^\star}{\sqrt{2}}  +\frac{\theta}{4v^{1/3}}\left( \delta\gamma_1\delta\gamma_2^\star+\delta\gamma_2\delta\gamma_1^\star +\frac{\left\vert \delta\gamma_2\right\vert^2}{\sqrt{2}}  \right)\right] M^2_{ij}\, .
\nonumber \\ & &
\label{eq:delta:pi:ij:DOWN:2ndOrder}
\eea

\subsubsection{Gravitational contribution} \label{app:GravContrib}
Let us first consider the kinetic part of the gravitational sector. Expanding \Eq{eq:calT:def} to quadratic order in perturbations gives
\bea
	\mathcal{T}^{(2)}&=&\frac{2}{\Mp^2v}\left\{\left[\frac{3}{8}\left(\frac{\delta\gamma}{\gamma}\right)^2-\frac{1}{2}\frac{\delta_2\gamma}{\gamma}\right]\mathcal{K}^{(0)}-\frac{1}{2}\frac{\delta\gamma}{\gamma}\mathcal{K}^{(1)}+\mathcal{K}^{(2)}\right\},
	\label{eq:calT2:interm}
\eea
where $\delta_2\gamma$ stands for the quadratic-order perturbations of $\gamma$, and the $\mathcal{K}^{(i)}$'s are given by 
\bea
	\mathcal{K}^{(0)}&=&\left(\gamma_{im}\gamma_{jn}-\frac{1}{2}\gamma_{ij}\gamma_{mn}\right)\pi^{ij}\,\pi^{mn}, \\
	\mathcal{K}^{(1)}&=&2\left[\left(\gamma_{im}\gamma_{jn}-\frac{1}{2}\gamma_{ij}\gamma_{mn}\right)\pi^{ij}\,\delta\pi^{mn}+\pi^{mn}\pi^{ij}\left(\gamma_{im}\,\delta\gamma_{jn}-\frac{1}{2}\gamma_{ij}\,\delta\gamma_{mn}\right)\right],
\eea
and
\bea
	\mathcal{K}^{(2)}&=&\left(\gamma_{im}\gamma_{jn}-\frac{1}{2}\gamma_{ij}\gamma_{mn}\right)\delta\pi^{ij}\,\delta\pi^{mn}+4\gamma_{im}\pi^{ij}\,\left(\delta\gamma_{jn}\,\delta\pi^{mn}\right)-\left(\delta\gamma_{ij}\,\pi^{ij}\right)\left(\gamma_{mn}\,\delta\pi^{mn}\right) \nonumber \\
	&&-\gamma_{ij}\pi^{ij}\,\left(\delta\gamma_{mn}\,\delta\pi^{mn}\right)+\left(\delta\gamma_{im}\,\delta\gamma_{jn}-\frac{1}{2}\,\delta\gamma_{ij}\,\delta\gamma_{mn}\right)\pi^{ij}\pi^{mn}.
\eea
The above is obtained using a Taylor expansion of $\sqrt{\gamma}$. We remind that $\delta\gamma/\gamma=\gamma^{ij}\delta\gamma_{ij}=\sqrt{3}\,\delta\gamma_1/v^{2/3}$. The quadratic perturbation of the determinant can be obtained using $\det(\gamma_{ij}+\delta\gamma_{ij})=\exp\{\mathrm{Tr}[\ln(\gamma_{ij}+\delta\gamma_{ij})]\}$.\footnote{This can be done as follows. We first write \Eqs{eq:gamma:ij:DOWN:app} and~\eqref{eq:delta:gamma:ij:DOWN:app} as
\bea
	\boldsymbol{\gamma}+\delta\boldsymbol{\gamma}=v^{2/3}\left(\boldsymbol{I}+\frac{\delta\gamma_1}{v^{2/3}}\boldsymbol{M}_1+\frac{\delta\gamma_2}{v^{2/3}}\boldsymbol{M}_2\right),
\eea
where the bold notation denotes (three-by-three) matrices with lowered indices, and $\boldsymbol{I}$ is the identity matrix. The properties~\eqref{eq:MA:usefulrelations} lead to $\mathrm{Tr}(\boldsymbol{M}_1)=\sqrt{3}$, $\mathrm{Tr}(\boldsymbol{M}_2)=0$, and $\mathrm{Tr}(\boldsymbol{M}_A\boldsymbol{M}_{A'})=\delta_{A,A'}$. We then conveniently rewrite
\bea
	\det(\boldsymbol{\gamma}+\delta\boldsymbol{\gamma})=v^2\exp\left\{\mathrm{Tr}\left[\ln\left(\boldsymbol{I}+\frac{\delta\gamma_1}{v^{2/3}}\boldsymbol{M}_1+\frac{\delta\gamma_2}{v^{2/3}}\boldsymbol{M}_2\right)\right]\right\}.
\eea
Since we are interested in the quadratic expansion of the determinant, it is sufficient to compute the logarithm matrix at quadratic order using its expression as an expansion. It gives 
\bea
	\ln\left(\boldsymbol{I}+\frac{\delta\gamma_1}{v^{2/3}}\boldsymbol{M}_1+\frac{\delta\gamma_2}{v^{2/3}}\boldsymbol{M}_2\right)=\frac{1}{v^{2/3}}\left[\delta\gamma_1\boldsymbol{M}_1+\delta\gamma_2\boldsymbol{M}_2-\frac{1}{2v^{2/3}}\left(\delta\gamma_1\boldsymbol{M}_1+\delta\gamma_2\boldsymbol{M}_2\right)^2\right].
\eea
The rest is straightforward as it consists in tracing (which is a linear operation) and expanding the exponential. We obtain
\bea
	\det(\boldsymbol{\gamma}+\delta\boldsymbol{\gamma})=v^2\left(1+\frac{\sqrt{3}}{v^{2/3}}\delta\gamma_1+\frac{1}{v^{4/3}}\left|\delta\gamma_1\right|^2-\frac{1}{2v^{4/3}}\left|\delta\gamma_2\right|^2\right).
	\label{eq:DetExpansion}
\eea
Note that we also recover the expression of the first-order expansion of the determinant, \Eq{eq:delta:gamma}.} It gives $\delta_2\gamma=v^{2/3}(\left|\delta\gamma_1\right|^2-\frac{1}{2}\left|\delta\gamma_2\right|^2)$. We then derive the expression of each term appearing in \Eq{eq:calT2:interm} in Fourier space. After a lengthy though straightforward calculation, we obtain
\bea
	\left(\frac{\delta\gamma}{\gamma}\right)^2\mathcal{K}^{(0)}&=&-\frac{9}{8}v^{2/3}\theta^2\,\left|\delta\gamma_1\right|^2\, , \\
	\frac{\delta_2\gamma}{\gamma}\mathcal{K}^{(0)}&=&-\frac{3}{8}v^{2/3}\theta^2\left[\left|\delta\gamma_1\right|^2-\frac{1}{2}\left|\delta\gamma_2\right|^2\right]\, ,\\
	\frac{\delta\gamma}{\gamma}\mathcal{K}^{(1)}&=&-\frac{3}{2}v^{2/3}\theta\left[v^{1/3}\,\left(\delta\gamma_1\right)\left(\delta\pi_1^\star\right)+\frac{1}{2}\theta\,\left|\delta\gamma_1\right|^2\right]\, ,\\
	\mathcal{K}^{(2)}&=&\frac{v^{4/3}}{2}\left(-\left|\delta\pi_1\right|^2+2\left|\delta\pi_2\right|^2\right)+v\theta\left[-\left(\delta\gamma_1\right)\left(\delta\pi^\star_1\right)+\frac{1}{2}\left(\delta\gamma_2\right)\left(\delta\pi^\star_2\right)\right] \nonumber\\
	&&+\frac{v^{2/3}\theta^2}{8}\left[-\left|\delta\gamma_1\right|^2+2\left|\delta\gamma_2\right|^2\right]. 
\eea
Combining the above results, one thus has
\bea
	\mathcal{T}^{(2)}&=&\frac{v^{1/3}}{\Mp^2}\left(-\left|\delta\pi_1\right|^2+2\left|\delta\pi_2\right|^2\right)+\frac{\theta}{2\Mp^2}\left[-\left(\delta\gamma_1\right)\left(\delta\pi_1^\star\right)+2\left(\delta\gamma_2\right)\left(\delta\pi_2^\star\right)\right] \nonumber \\
	&&+\frac{1}{32\Mp^2}\frac{\theta^2}{v^{1/3}}\left(\left|\delta\gamma_1\right|^2+10\left|\delta\gamma_2\right|^2\right)\, ,
\label{eq:calT2:app:final}
\eea
where $\theta^2$ in the second line can be replaced with the background scalar constraint equation~\eqref{eq:ConstHom}.
We stress that there is no coupling between the two gravitational degrees of freedom in the kinetic term of the gravitational scalar constraint. 

Then, for the potential part of the gravitational sector, expanding \Eq{eq:calW:def} at second order leads to 
\bea
\label{eq:calW:2:interm}
	\mathcal{W}^{(2)}=\frac{-\Mp^2\sqrt{\gamma}}{4} \frac{\delta\gamma}{\gamma} \gamma^{ij}\,\delta\mathcal{R}_{ij}-\frac{\Mp^2\sqrt{\gamma}}{2}\left(\delta\gamma^{ij}\,\delta\mathcal{R}_{ij}+\gamma^{ij}\,\delta_2\mathcal{R}_{ij}\right),
\eea
where $\delta_2\mathcal{R}_{ij}$ denotes the quadratic expansion of the Ricci tensor and where we have used that the Ricci tensor vanishes at the background level. For the first term, $\delta\gamma$ is given in \Eq{eq:delta:gamma} and $\delta\mathcal{R}_{ij}$ was already computed below \Eq{eq:Ricci:tensor:Christoffel}, leading to
\bea
	\frac{-\Mp^2\sqrt{\gamma}}{4} \frac{\delta\gamma}{\gamma} \gamma^{ij}\,\delta\mathcal{R}_{ij}=-\frac{\Mp^2\sqrt{\gamma}}{4}\left(\gamma^{ij}\,\delta\gamma_{ij}\right)\left(\gamma^{im}\gamma^{jn}-\gamma^{ij}\gamma^{mn}\right)\partial_m\partial_n\delta\gamma_{ij}.
\eea
In Fourier space, using the decomposition~\eqref{eq:delta:gamma:ij:DOWN:app}, this reduces to
\bea
	\frac{-\Mp^2\sqrt{\gamma}}{4}\frac{\delta\gamma}{\gamma}\gamma^{ij}\,\delta\mathcal{R}_{ij}=-\frac{\Mp^2}{2v}k^2\left[\left|\delta\gamma_1\right|^2-\frac{1}{\sqrt{2}}\left(\delta\gamma_1\right)\left(\delta\gamma_2^\star\right)\right].
\eea
We then consider each contribution involved in the second term of \Eq{eq:calW:2:interm}. The first one is given by
\bea
	\delta\gamma^{ij}\,\delta\mathcal{R}_{ij}=\delta\gamma^{ij}\left[\partial_i\partial_n\left(\gamma^{nm}\,\delta\gamma_{mj}\right)-\frac{1}{2}\gamma^{mn}\partial_m\partial_n\left(\delta\gamma_{ij}\right)-\frac{1}{2}\partial_i\partial_j\left(\gamma^{mn}\,\delta\gamma_{mn}\right)\right]\, ,
\eea
hence its contribution to $\mathcal{W}^{(2)}$ in Fourier space is
\bea
	\frac{-\Mp^2\sqrt{\gamma}}{2}\delta\gamma^{ij}\,\delta\mathcal{R}_{ij}=\frac{\Mp^2}{3v}k^2\left[\left|\delta\gamma_1\right|^2-\frac{1}{4}\left|\delta\gamma_2\right|^2-\frac{1}{\sqrt{2}}\left(\delta\gamma^\star_1\right)\left(\delta\gamma_2\right)+\frac{1}{2\sqrt{2}}\left(\delta\gamma_1\right)\left(\delta\gamma_2^\star\right)\right]. \nonumber \\
\eea
From \Eq{eq:Ricci:tensor:Christoffel}, the quadratic expansion of the Ricci tensor can be expressed in terms of the linear and the quadratic expansions of the Christoffel symbols. This gives
\bea
	\delta_2\mathcal{R}_{ij}=\partial_\ell\delta_2\gamma^\ell_{ij}-\partial_j\delta_2\gamma^\ell_{\ell i}+\delta\gamma^\ell_{\ell m}\,\delta\gamma^m_{ij}-\delta\gamma^\ell_{mi}\,\delta\gamma^m_{\ell j}.
\eea
The first two terms involving the quadratic expansion of the Christoffel symbols are total derivatives and, as such, they do not contribute to the equations of motion (since $\delta_2\mathcal{R}_{ij}$ multiplies background functions in the action). The contribution from the last two terms is rather cumbersome though it partially simplifies by computing $\gamma^{ij}\delta_2\mathcal{R}_{ij}$ directly, and using that $M_{2}$ is traceless and symmetric, and that $M_{1}$ is a pure trace (hence  symmetric). We finally arrive at 
\bea
\frac{-\Mp^2\sqrt{\gamma}}{2}\,\gamma^{ij}\delta_2\mathcal{R}_{ij}=\frac{\Mp^2}{12v}k^2\left[\left|\delta\gamma_1\right|^2+\frac{1}{2}\left|\delta\gamma_2\right|^2+{\sqrt{2}}\left(\delta\gamma_1\right)\left(\delta\gamma_2^\star\right)-2{\sqrt{2}}\left(\delta\gamma^\star_1\right)\left(\delta\gamma_2\right)\right], \nonumber \\
\eea
up to total-derivative terms. Combining the above results, one obtains the following expression for the potential part of the gravitational sector
\bea
\mathcal{W}^{(2)}=\frac{\Mp^2}{24v}k^{2}\left[-2\left|\delta\gamma_1\right|^2-\left|\delta\gamma_2\right|^2+{10}{\sqrt{2}}\left(\delta\gamma_1\right)\left(\delta\gamma_2^\star\right)-8{\sqrt{2}}\left(\delta\gamma^\star_1\right)\left(\delta\gamma_2\right)\right].
\label{eq:calW2:app:final}
\eea
Let us note that unlike the kinetic term, the potential term couples the two types of gravitational perturbations.

\subsubsection{Scalar-field contribution}
The contributions from the scalar field follow from expanding \Eqs{eq:def:T} and~\eqref{eq:def:W} at second order, leading to
\bea
	T^{(2)}&=&\frac{1}{2\sqrt{\gamma}}\left(\delta\pi_\phi\right)^2-\frac{\pi_\phi}{2\sqrt{\gamma}}\frac{\delta\gamma}{\gamma}\delta\pi_\phi+\left[\frac{3}{8}\left(\frac{\delta\gamma}{\gamma}\right)^2-\frac{1}{2}\frac{\delta_2\gamma}{\gamma}\right]\frac{\pi^2_\phi}{2\sqrt{\gamma}}\, , \\
	W^{(2)}&=&\frac{\sqrt{\gamma}}{2}\gamma^{ij}\left(\partial_i\delta\phi\right)\left(\partial_j\delta\phi\right)+\frac{\sqrt{\gamma}}{2}\,V_{,\phi,\phi}\,\left(\delta\phi\right)^2+\frac{\sqrt{\gamma}V_{,\phi}}{2}\frac{\delta\gamma}{\gamma}\delta\phi \nonumber  \\
	&&+\left[\frac{1}{2}\frac{\delta_2\gamma}{\gamma}-\frac{1}{8}\left(\frac{\delta\gamma}{\gamma}\right)^2\right]\sqrt{\gamma}V\, . 
\eea
We note that the perturbed scalar field is coupled to the perturbed induced metric only through its determinant. In Fourier space, these two contributions read
\bea
\label{eq:T2:app:final}
	T^{(2)}&=&\frac{1}{2v}\left|\delta\pi_\phi\right|^2-\frac{\sqrt{3}}{2}\frac{\pi_\phi}{v^{5/3}}\left(\delta\pi_\phi\,\delta\gamma_1^\star\right)+\frac{1}{4v^{1/3}}\left(\frac{\pi_\phi}{v}\right)^2\left(\frac{5}{4}\left|\delta\gamma_1\right|^2+\frac{1}{2}\left|\delta\gamma_2\right|^2\right), \\
	W^{(2)}&=&\frac{v}{2}\left(\frac{k^2}{v^{2/3}}+V_{,\phi,\phi}\right)\left|\delta\phi\right|^2+\frac{\sqrt{3}}{2}v^{1/3}V_{,\phi}\,\left(\delta\phi\,\delta\gamma_1^\star\right)+\frac{1}{4}\frac{V}{v^{1/3}}\left(\frac{1}{2}\left|\delta\gamma_1\right|^2-\left|\delta\gamma_2\right|^2\right). \nonumber \\
	\label{eq:W2:app:final}
\eea
One can see that the scalar-field perturbations are solely coupled to the isotropic component of the metric perturbations, $\delta\gamma_1$, as a result of being only coupled to the perturbed determinant at linear order.
\section{Perturbed background equations}
\label{app:pert}
In this appendix, we derive the dynamical equations for the separate-universe approach by directly perturbing the background equations of motion, \ie by plugging the replacements $N\to N(\tau)+\ovl{\delta N}$, $v\to v(\tau)+\ovl{\delta v}$, $\theta\to \theta(\tau)+\ovl{\delta\theta}$, $\phi\to\phi(\tau)+\ovl{\delta\phi}$, and $\pi_\phi\to\pi_\phi(\tau)+\ovl{\delta\pi_\phi}$ into $C^{(0)}[N]=N\mathcal{S}^{(0)}$, where $\mathcal{S}^{(0)}$ is given by \Eqs{eq:BckgConsGrav} and~\eqref{eq:BckgConsPhi}. It gives rise to 
\bea
	C^{(0)}[N]\to C\left[N+\overline{\delta N}\right]=N\,\mathcal{S}^{(0)}+\left(\overline{\delta N}\,{\mathcal{S}^{(1)}_\mathrm{bckg}}+N\,{\mathcal{S}^{(2)}_\mathrm{bckg}}\right) ,
\eea
where the perturbed scalar constraint at linear and quadratic order is
\bea
	\mathcal{S}^{(1)}_\mathrm{bckg}&=&-\frac{3}{4\Mp^2}\left(\theta^2\overline{\delta v}+2v\theta\overline{\delta\theta}\right)-\frac{\pi^2_\phi}{2v^2}\overline{\delta v}+ V(\phi)\overline{\delta v}+\frac{\pi_\phi}{v}\overline{\delta\pi_\phi}+ v V_{,\phi}\overline{\delta\phi} \nonumber \\
	&=&-\frac{\pi^2_\phi}{v^2}\,\overline{\delta v}-\frac{3}{2\Mp^2}v\theta\,\overline{\delta\theta}+\frac{\pi_\phi}{v}\overline{\delta\pi_\phi}+ v V_{,\phi}\overline{\delta\phi}\, , \\
	\mathcal{S}^{(2)}_\mathrm{bckg}&=&-\frac{3}{4\Mp^2}\left[2\theta\overline{\delta v}\,\overline{\delta\theta}+v\left(\overline{\delta\theta}\right)^2\right]+\frac{\pi^2_\phi}{2v^3}\left(\overline{\delta v}\right)^2-\frac{\pi_\phi}{v^2}\overline{\delta v}\,\overline{\delta\pi_\phi}+V_{,\phi}\overline{\delta v}\,\overline{\delta\phi} \nonumber \\
	&&+\frac{1}{2v}\left(\overline{\delta\pi_\phi}\right)^2+\frac{1}{2}v\,V_{,\phi,\phi}\left(\overline{\delta\phi}\right)^2\, .
\eea
Note that the second line for $\mathcal{S}^{(1)}_\mathrm{bckg}$ was obtained using the scalar constraint at the background level, \ie $\mathcal{S}^{(0)}=0$. 
One can easily check that it matches \Eq{eq:S1:SU}, hence it gives the same constraint equation. Moreover, the Hamilton equations derived from the above Hamiltonian also match \Eqs{eq:SepUnivVTheta} and~\eqref{eq:SepUnivVTheta:2}, if the linear scalar constraint equation, $\mathcal{S}^{(1)}_\mathrm{bckg}=0$, is used to simplify the equation of motion for $\ovl{\delta\theta}$. It is also straightforward to verify that these equations of motion can be obtained by directly perturbing the constraint and dynamical equations of the homogeneous and isotropic problem, namely \Eqs{eq:ConstHom}-\eqref{eq:DotPi}.

This argument can be generalized to the non-perturbative case. Let us consider the non-perturbative-isotropic sector of the constraint, that is the SU constraint Eq. \eqref{eq:Hamiltonian:SU:full}, that we rewrite in terms of $\ovl{\delta v}$ and $\ovl{\delta \theta}$ for simplicity:
\bea
\mathcal{C}=\frac{\left(\pi_\phi + \ovl{\delta \pi_\phi} \right)^2}{2 \left(v + \ovl{\delta v} \right)} + \left(v + \ovl{\delta v} \right) V(\phi + \ovl{\delta \phi}) - \frac{3}{4 \Mp^2} (v + \ovl{\delta v}) \left(\theta + \ovl{\delta \theta}\right)^2
\eea
One can readily see that this expression can alternatively be obtained by including the separate-universe deviations into the FLRW contraints~\eqref{eq:BckgConsGrav} and~\eqref{eq:BckgConsPhi}. The equations of motion are therefore the same:
\bea
\dot{\left(v + \ovl{\delta v}\right)} &=& -\left(N+\ovl{\delta N} \right) \frac{3}{2 \Mp^2} (v + \ovl{\delta v}) \left(\theta + \ovl{\delta \theta}\right)  \,,\\
\dot{\left(\theta + \ovl{\delta \theta}\right)} &=& \left(N+\ovl{\delta N} \right) \left[ \frac{\left(\pi_\phi + \ovl{\delta \pi_\phi} \right)^2}{2 \left(v + \ovl{\delta v} \right)^2} -  V(\phi + \ovl{\delta \phi}) + \frac{3}{4 \Mp^2} \left(\theta + \ovl{\delta \theta}\right)^2 \right] \,,\\
\dot{\left(\phi + \ovl{\delta \phi}\right)} &=& \left(N+\ovl{\delta N} \right) \frac{\left(\pi_\phi + \ovl{\delta \pi_\phi} \right)}{ \left(v + \ovl{\delta v} \right)}  \,,\\
\dot{\left(\pi_\phi + \ovl{\delta \pi_\phi}\right)} &=& - \left(N+\ovl{\delta N} \right) \left(v + \ovl{\delta v} \right) V_{,\phi}(\phi + \ovl{\delta \phi})  \,,
\eea
which indeed coincide with the FLRW equations of motion~\eqref{eq:Dotv}, \eqref{eq:Dot:theta}, \eqref{eq:DotPhiPi} and~\eqref{eq:DotPi} once the deviations $\ovl{\delta v}, \ovl{\delta \theta}, \ovl{\delta \phi}, \ovl{\delta \pi_\phi}$ are included.
In other words, the reduction to the isotropic sector can be equivalently performed at the level of the full Hamiltonian or in the FLRW theory, and this equivalence holds at all orders.

\section{Expansion rate}
\label{app:expansion}

\subsection{Definition}
\label{app:expansion:def}

The expansion rate of spatial hypersurfaces is defined as  
\bea
\Theta:=\nabla_\mu n^\mu\, ,
\eea
where $n^\mu:=(-1/N,N^i/N)$ is the unit vector orthogonal to the hypersurfaces $\Sigma_\tau$, and $\nabla_\mu$ is the four-dimensional covariant derivative. 
Let us see how it is related to the phase-space variables used in the Hamiltonian formalism.

We first recall that the ADM metric~\eqref{eq:metric:ADM} is given by
\bea
\label{eq:ADM:metric:low:component}
g_{00}=-N^2+N_i N^i
\, ,\quad\quad
g_{0i}=N_i
\, ,\quad\quad
g_{ij} = \gamma_{ij}\, ,
\eea
the inverse of which reads
\bea
\label{eq:ADM:metric:high:component}
g^{00}=-\frac{1}{N^2}
\, ,\quad\quad
g^{0i}=\frac{N^i}{N^2}
\, ,\quad\quad
g^{ij} = \gamma^{ij} - \frac{N^i N^j}{N^2}\, ,
\eea
where the indices of the shift vector $N^i$ are raised and lowered by the induced metric $\gamma_{ij}$. This gives rise to $n_\mu=(N,\vec{0})$, so $n$ is indeed orthogonal to $\Sigma_\tau$, and one can check that $n_\mu n^\mu=-1$. This also allows one to introduce the integrated amount of expansion,
\bea
\label{eq:Nint:def}
\mathcal{N}_{\mathrm{int}} = \int\frac{\Theta}{3} n_\mu\dd x^\mu =- \frac{1}{3}\int\Theta N\dd\tau\, .
\eea
Expanding the covariant derivative in terms of the Christoffel symbols, one has
\bea
\label{eq:Theta:interm1}
\Theta = \partial_\mu n^\mu + \Gamma_{\mu \nu}^\mu n^\nu\, ,
\eea
where $\Gamma_{\mu\nu}^\rho$ is given by a similar expression as in \Eq{eq:Christoffel:def} but where the full metric is used instead, namely
\bea
\label{eq:Christoffel:full:def}
\Gamma_{\mu\nu}^\rho = \frac{1}{2} g^{\rho\sigma} \left(\partial_\mu g_{\nu\sigma}+\partial_\nu g_{\mu\sigma} - \partial_\sigma g_{\mu\nu}\right)\, .
\eea
If all indices are chosen to be spatial, this implies that
\bea
\Gamma_{i j}^k &=& \frac{1}{2} g^{ k \sigma} \left(\partial_i g_{j\sigma}+\partial_j g_{i\sigma} - \partial_\sigma g_{ij}\right) \nonumber \\
&=&
 \frac{1}{2} g^{ k 0} \left(\partial_i g_{j0}+\partial_j g_{i0} - \partial_0 g_{ij}\right)+
  \frac{1}{2} g^{ k \ell} \left(\partial_i g_{j\ell}+\partial_j g_{i\ell} - \partial_\ell g_{ij}\right)
  \nonumber \\
&=&
  \frac{N^k}{2N^2}\left(\partial_i N_j + \partial_j N_i-\dot{\gamma}_{ij}\right)-
  \frac{N^{ k }N^{\ell}}{2N^2}  \left(\partial_i \gamma_{j\ell}+\partial_j \gamma_{i\ell} - \partial_\ell \gamma_{ij}\right)+
  \frac{1}{2} \gamma^{ k \ell} \left(\partial_i \gamma_{j\ell}+\partial_j \gamma_{i\ell} - \partial_\ell \gamma_{ij}\right)
    \nonumber \\
\eea
where we have used \Eqs{eq:ADM:metric:low:component}-\eqref{eq:ADM:metric:high:component} in the last line.
The last term matches \Eq{eq:Christoffel:def}, hence 
\bea
\Gamma_{ij}^k = \frac{N^k}{2N^2}\left(\partial_i N_j + \partial_j N_i-\dot{\gamma}_{ij}\right)
-
  \frac{N^{ k }N^{\ell}}{2N^2}  \left(\partial_i \gamma_{j\ell}+\partial_j \gamma_{i\ell} - \partial_\ell \gamma_{ij}\right)+\gamma_{ij}^k\, .
\eea
This formula can be used when expanding \Eq{eq:Theta:interm1} along temporal and spatial indices, and one obtains
\bea
\Theta &= &\frac{\dot{N}}{N^2}-\frac{N^i\partial_i N}{N^2}+\frac{N^{i} N^j}{N^3}\partial_i N_j -\frac{\dot{\gamma_{ij}}}{2N^3}N^i N^j
-\frac{\Gamma_{00}^0+\Gamma_{i0}^i}{N}
\nonumber \\ & &
+\frac{N^i}{N}\Gamma_{0i}^0
-\frac{N^i N^j N^\ell}{2N^3}\partial_j \gamma_{i\ell}
 + \frac{1}{N}\left(\partial_i N^i + \gamma_{ij}^i N^j\right)\, .
\eea
In the last term of this expression, one recognises the covariant derivative with respect to the induced metric $\gamma_{ij}$, which we denote $D$, \ie the last term is given by $(D_i N^i)/N$. The other terms require to compute some of the Christoffel symbols. Plugging \Eqs{eq:ADM:metric:low:component} and~\eqref{eq:ADM:metric:high:component} into \Eq{eq:Christoffel:full:def}, one finds
\bea
\Gamma_{00}^0 &=& \frac{\dot{N}}{N}-\frac{N^i}{2N^2}\partial_i\left(-N^2+N_j N^j\right)+\frac{\dot{\gamma}_{ij}}{2N^2} N^i N^j\, ,\\
\Gamma_{0i}^i &=&\frac{N^i}{2N^2}\partial_i\left(-N^2+N_j N^j\right)+\frac{\dot{\gamma}_{ij}}{2}\left(\gamma^{ij}-\frac{N^i N^j}{N^2}\right) ,\\
N^i\Gamma_{0i}^0&=& -\frac{N^i}{2N^2}\partial_i\left(-N^2+N_j N^j\right) + \frac{\dot{\gamma}_{ij}}{2N^2} N^i N^j\, .
\eea
Combining the above results, one obtains\footnote{This expression matches the trace of the extrinsic curvature, see \eg \Refs{thieman_book,Malik:2008im}.}
\bea
\label{eq:Theta:interm:2}
\Theta = \frac{1}{2N}\left(2 D_i N^i-\dot{\gamma}_{ij} \gamma^{ij}\right)\, .
\eea
This formula can be further simplified as follows. The equation of motion for $\gamma_{ij}$ can be obtained from the Hamiltonian~\eqref{eq:full:Hamiltonian}-\eqref{eq:dcsg}, and one finds
\bea\label{eq:gammaijdot}
\dot{\gamma}_{ij} = \frac{\partial C}{\partial \pi^{ij}}=\frac{2N}{\Mp^2\sqrt{\gamma}}\left(2\pi_{ij}-\pi\gamma_{ij}\right)+2\gamma_{\ell i}\partial_j N^\ell + N^\ell \partial_\ell \gamma_{ij}\, ,
\eea
where we have first used integration by parts to deal with the term involving the gradient of $\pi^{ij}$ in $C$. When contracted with the induced metric, this gives rise to
\bea
\label{eq:gamma:gammadot}
\gamma^{ij}\dot{\gamma}_{ij} = - \frac{2N}{\Mp^2\sqrt{\gamma}}\gamma_{ij}\pi^{ij}+2\partial_i N^i + N^i \gamma^{\ell m}\partial_i\gamma_{\ell m}\, .
\eea
Moreover, the term $D_i N^i$ in \Eq{eq:Theta:interm:2} can be expanded along the Christoffel symbols~\eqref{eq:Christoffel:def}, which gives rise to
\bea
D_i N^i &=& \partial_i N^i + \gamma_{ij}^i N^j
\nonumber \\ &=&
\partial_i N^i + \frac{1}{2}\gamma^{i\ell}\left(\partial_j \gamma_{i\ell}+\partial_{i}\gamma_{j\ell}-\partial_\ell \gamma_{ij}\right)N^j\, .
\eea
The last two terms correspond to contracting an object that is symmetric in $i$ and $\ell$, namely $\gamma_{i\ell}$, with an antisymmetric combination, namely $\partial_{i}\gamma_{j\ell}-\partial_\ell \gamma_{ij}$. Therefore, they give a vanishing contribution, so one has
\bea
\label{eq:Di:Ni}
D_i N^i = \partial_i N^i + \frac{N^j}{2}\gamma^{i\ell}\partial_j \gamma_{i\ell} \, .
\eea
Plugging \Eqs{eq:gamma:gammadot} and~\eqref{eq:Di:Ni} into \Eq{eq:Theta:interm:2} finally leads to
\bea
\label{eq:ExpRateGen}
\Theta = \frac{\gamma_{ij} \pi^{ij}}{\Mp^2\sqrt{\gamma}}\, .
\eea
\subsection{Expansion rate at the background level}
\label{app:Theta:background}
In homogeneous and isotropic cosmologies, using the formulas established in \Sec{ssec:bck}, \Eq{eq:ExpRateGen} reduces to
\bea
	\Theta=-\frac{\dot{v}}{Nv}=\frac{3}{2\Mp^2}\,\theta\, , \label{eq:ThetaHom}
\eea
so it is directly proportional to the momentum $\theta$. The integrated amount of expansion, see \Eq{eq:Nint:def}, is given by
\bea
\label{eq:NintHom}
\mathcal{N}_{\mathrm{int}} = \frac{1}{3}\ln(v)\, .
\eea
Recalling that $v=a^3$ where $a$ is the FLRW scale factor, $\mathcal{N}_{\mathrm{inf}}$ is nothing but the number of $e$-folds (hence the notation).
\subsection{Expansion rate at first order}
\label{App:Theta:FirstOrder}
\subsubsection{Cosmological perturbation theory}
By plugging \Eqs{eq:delta:gamma}, \eqref{eq:piij:deltagammaij} and~\eqref{eq:gammaij:deltapiij} into the first-order perturbation of \Eq{eq:ExpRateGen}, one obtains
\bea
\delta\Theta=\frac{\sqrt{3}}{v^\frac{2}{3} \Mp^2}\left(v^\frac{1}{3}\delta\pi_1-\frac{\theta}{4}\delta\gamma_1\right) \, .
\label{eq:expratepert}
\eea
For the integrated amount of expansion, upon perturbing \Eq{eq:Nint:def}, one has
\bea
\delta \mathcal{N}_{\mathrm{inf}} &=& -\frac{1}{3}\int \left(\delta\Theta N + \Theta\delta N\right)\dd\tau
\nonumber \\ &=&
-\frac{1}{3}\int\left[\frac{\sqrt{3}N}{v^{2/3} \Mp^2}\left(v^{1/3}\delta\pi_1-\frac{\theta}{4}\delta\gamma_1\right)+\frac{3\theta}{2\Mp^2}\delta N\right]\dd\tau\, ,
\eea
where we have made use of \Eqs{eq:ThetaHom} and~\eqref{eq:expratepert}. This expression can be further simplified as follows. First, let us make use of the equation of motion for $\delta\gamma_1$, namely the first entry of \Eq{eq:EOM}, to express $\delta N$ in terms of $\delta\dot{\gamma_1}$ and the other phase-space variables. This gives rise to
\bea
\delta \mathcal{N}_{\mathrm{int}} &=& \int\left[\frac{k}{3}\delta N_1+\frac{1}{2\sqrt{3}v^{2/3}}\left(\delta\dot{\gamma_1}+\frac{N\theta}{\Mp^2}\delta\gamma_1\right)\right]\dd\tau\, .
\label{eq:deltaN:int:interm}
\eea
Second, making use of \Eq{eq:Dotv}, one can show that
\bea
\frac{\partial}{\partial\tau} \left(\frac{\delta\gamma_1}{v^{2/3}}\right) = \frac{\delta\dot{\gamma}_1}{v^{2/3}}-\frac{2}{3}\delta\gamma_1 \frac{\dot{v}}{v^{5/3}} = \frac{1}{v^{2/3}}\left(\delta\dot{\gamma}_1+\frac{N\theta}{\Mp^2}\delta\gamma_1\right) .
\eea
One readily recognises the last term in \Eq{eq:deltaN:int:interm}, which can therefore be integrated and one obtains
\bea
\label{eq:delta:N:int:CPT}
\delta\mathcal{N}_{\mathrm{int}} = \frac{1}{2\sqrt{3}v^{2/3}}\delta\gamma_1 + \frac{k}{3}\int\delta N_1\dd\tau\, .
\eea

\subsubsection{Separate universe}
The same considerations as those presented above can be applied to the separate-universe framework, where one starts from the same ADM metric where the replacements outlined in Table~\ref{table:correspondence} are performed. At the background level, one recovers \Eqs{eq:ThetaHom} and~\eqref{eq:NintHom}. At first order in perturbations, one obtains the barred version of \Eq{eq:expratepert} for the expansion rate, namely
\bea
\ovl{\delta\Theta}=\frac{\sqrt{3}}{v^\frac{2}{3} \Mp^2}\left(v^\frac{1}{3}\ovl{\delta\pi_1}-\frac{\theta}{4}\ovl{\delta\gamma_1}\right) \, .
\eea
This is expected since \Eq{eq:expratepert} shows that the expansion rate only involves isotropic degrees of freedom. For the integrated amount of expansion, one finds
\bea
\delta\ovl{\mathcal{N}}_{\mathrm{int}} =  \frac{1}{2\sqrt{3}v^{2/3}}\ovl{\delta\gamma_1}\, ,
\eea
which indeed corresponds to the barred version of \Eq{eq:delta:N:int:CPT}.

\subsection{Expansion rate at quadratic order}
\subsubsection{Cosmological perturbation theory}
From Eq. \eqref{eq:ExpRateGen}, one can also compute the second-order perturbation of the expansion rate:
\bea
\delta_2 \Theta &=& \frac{\pi}{\Mp^2} \delta_2\left(\frac{1}{\sqrt{\gamma}}\right) - \frac{1}{2\Mp^2} \frac{\delta \gamma}{\gamma^{3/2}} \delta\pi + \frac{1}{\Mp^2 \sqrt{\gamma}} \delta \gamma_{ij} \delta \pi^{ij} \\
&=& \frac{\pi}{\Mp^2 \sqrt{\gamma}} \left[ \frac{3}{8}\left(\frac{\delta\gamma}{\gamma}\right)^2 -\frac{1}{2} \frac{\delta_2 \gamma}{\gamma} \right] -\frac{1}{2\Mp^2} \frac{\delta \gamma}{\gamma^{3/2}} \left( \delta \gamma_{ij} \pi^{ij} + \gamma_{ij} \delta \pi^{ij} \right) \nonumber\\
&\,&+ \frac{1}{\Mp^2 \sqrt{\gamma}} \delta\gamma_{ij} \delta\pi^{ij} \,. 
\eea
Plugging in the results of Eqs. \eqref{eq:deltaNi:deltaN1}, \eqref{eq:delta:gamma:ij:DOWN:app}, \eqref{eq:delta:pi:ij:UP:app}, \eqref{eq:piij:deltagammaij}, \eqref{eq:gammaij:deltapiij} and \eqref{eq:DetExpansion}, one gets:
\bea
\label{eq:delta2:Theta:CPT}
\delta_2 \Theta &=& \frac{1}{\Mp^2 v} \left(\frac{3\theta}{16 v^{1/3}} + 1 \right) \left\vert \delta\gamma_1 \right\vert^2 + \frac{1}{\Mp^2 v} \left(\frac{3\theta}{8 v^{1/3}} + 1 \right) \left\vert \delta\gamma_2 \right\vert^2 - \frac{3}{2\Mp^2 v} \delta\gamma_1 \delta\pi_1^\star \,,
\eea
where we also used the expression for $\pi=3v\theta/2$ and the orthonormality of the basis $\left(M^1_{ij},M^2_{ij}\right)$.
\subsubsection{Separate universe}
The calculation can be reproduced in the separate-universe approach, starting from the replacements outlined in Table~\ref{table:correspondence}. One obtains
\bea
\ovl{\delta_2 \Theta} = \frac{1}{\Mp^2 v} \left(\frac{3\theta}{16 v^{1/3}} + 1 \right) \left\vert \ovl{\delta\gamma_1} \right\vert^2 - \frac{3}{2\Mp^2 v} \ovl{\delta\gamma_1}\, \ovl{\delta\pi_1}^\star \, ,
\eea
which indeed reduces to \Eq{eq:delta2:Theta:CPT} under those same replacements. 

\bibliographystyle{JHEP}
\bibliography{SepUniv}

\providecommand{\href}[2]{#2}\begingroup\raggedright\begin{thebibliography}{10}

\bibitem{Salopek:1990jq}
D.S.~Salopek and J.R.~Bond, \emph{{Nonlinear evolution of long wavelength
  metric fluctuations in inflationary models}},
  \href{https://doi.org/10.1103/PhysRevD.42.3936}{\emph{Phys. Rev.} {\bfseries
  D42} (1990) 3936}.

\bibitem{Sasaki:1995aw}
M.~Sasaki and E.D.~Stewart, \emph{{A General analytic formula for the spectral
  index of the density perturbations produced during inflation}},
  \href{https://doi.org/10.1143/PTP.95.71}{\emph{Prog. Theor. Phys.} {\bfseries
  95} (1996) 71} [\href{https://arxiv.org/abs/astro-ph/9507001}{{\ttfamily
  astro-ph/9507001}}].

\bibitem{Wands:2000dp}
D.~Wands, K.A.~Malik, D.H.~Lyth and A.R.~Liddle, \emph{{A New approach to the
  evolution of cosmological perturbations on large scales}},
  \href{https://doi.org/10.1103/PhysRevD.62.043527}{\emph{Phys. Rev.}
  {\bfseries D62} (2000) 043527}
  [\href{https://arxiv.org/abs/astro-ph/0003278}{{\ttfamily
  astro-ph/0003278}}].

\bibitem{PhysRevD.68.103515}
D.H.~Lyth and D.~Wands, \emph{Conserved cosmological perturbations},
  \href{https://doi.org/10.1103/PhysRevD.68.103515}{\emph{Phys. Rev. D}
  {\bfseries 68} (2003) 103515}.

\bibitem{PhysRevD.68.123518}
G.I.~Rigopoulos and E.P.S.~Shellard, \emph{Separate universe approach and the
  evolution of nonlinear superhorizon cosmological perturbations},
  \href{https://doi.org/10.1103/PhysRevD.68.123518}{\emph{Phys. Rev. D}
  {\bfseries 68} (2003) 123518}.

\bibitem{Lyth:2005fi}
D.H.~Lyth and Y.~Rodriguez, \emph{{The Inflationary prediction for primordial
  non-Gaussianity}},
  \href{https://doi.org/10.1103/PhysRevLett.95.121302}{\emph{Phys. Rev. Lett.}
  {\bfseries 95} (2005) 121302}
  [\href{https://arxiv.org/abs/astro-ph/0504045}{{\ttfamily
  astro-ph/0504045}}].

\bibitem{Tanaka:2021dww}
T.~Tanaka and Y.~Urakawa, \emph{{Anisotropic separate universe and Weinberg's
  adiabatic mode}},
  \href{https://doi.org/10.1088/1475-7516/2021/07/051}{\emph{JCAP} {\bfseries
  07} (2021) 051} [\href{https://arxiv.org/abs/2101.05707}{{\ttfamily
  2101.05707}}].

\bibitem{Lifshitz:1960}
E.M.~Lifshitz and I.M.~Khalatnikov, \emph{{About singularities of cosmological
  solutions of the gravitational equations. I}}, {\emph{ZhETF} {\bfseries 39}
  (1960) 149}.

\bibitem{Starobinsky:1982mr}
A.A.~Starobinsky, \emph{{Isotropization of arbitrary cosmological expansion
  given an effective cosmological constant}}, {\emph{JETP Lett.} {\bfseries 37}
  (1983) 66}.

\bibitem{PhysRevD.49.2759}
G.L.~Comer, N.~Deruelle, D.~Langlois and J.~Parry, \emph{Growth or decay of
  cosmological inhomogeneities as a function of their equation of state},
  \href{https://doi.org/10.1103/PhysRevD.49.2759}{\emph{Phys. Rev. D}
  {\bfseries 49} (1994) 2759}.

\bibitem{Khalatnikov_2002}
I.M.~Khalatnikov, A.Y.~Kamenshchik and A.A.~Starobinsky, \emph{Comment about
  quasi-isotropic solution of einstein equations near the cosmological
  singularity},
  \href{https://doi.org/10.1088/0264-9381/19/14/322}{\emph{Classical and
  Quantum Gravity} {\bfseries 19} (2002) 3845–3849}.

\bibitem{Pattison:2018bct}
C.~Pattison, V.~Vennin, H.~Assadullahi and D.~Wands, \emph{{The attractive
  behaviour of ultra-slow-roll inflation}},
  \href{https://doi.org/10.1088/1475-7516/2018/08/048}{\emph{JCAP} {\bfseries
  08} (2018) 048} [\href{https://arxiv.org/abs/1806.09553}{{\ttfamily
  1806.09553}}].

\bibitem{Miranda:2019ara}
T.~Miranda, E.~Frion and D.~Wands, \emph{{Stochastic collapse}},
  \href{https://doi.org/10.1088/1475-7516/2020/01/026}{\emph{JCAP} {\bfseries
  01} (2020) 026} [\href{https://arxiv.org/abs/1910.10000}{{\ttfamily
  1910.10000}}].

\bibitem{Grain:2020wro}
J.~Grain and V.~Vennin, \emph{{Unavoidable shear from quantum fluctuations in
  contracting cosmologies}},
  \href{https://arxiv.org/abs/2005.04222}{{\ttfamily 2005.04222}}.

\bibitem{Wands:1998yp}
D.~Wands, \emph{{Duality invariance of cosmological perturbation spectra}},
  \href{https://doi.org/10.1103/PhysRevD.60.023507}{\emph{Phys. Rev. D}
  {\bfseries 60} (1999) 023507}
  [\href{https://arxiv.org/abs/gr-qc/9809062}{{\ttfamily gr-qc/9809062}}].

\bibitem{Finelli:2001sr}
F.~Finelli and R.~Brandenberger, \emph{{On the generation of a scale invariant
  spectrum of adiabatic fluctuations in cosmological models with a contracting
  phase}}, \href{https://doi.org/10.1103/PhysRevD.65.103522}{\emph{Phys. Rev.
  D} {\bfseries 65} (2002) 103522}
  [\href{https://arxiv.org/abs/hep-th/0112249}{{\ttfamily hep-th/0112249}}].

\bibitem{Khoury:2001wf}
J.~Khoury, B.A.~Ovrut, P.J.~Steinhardt and N.~Turok, \emph{{The Ekpyrotic
  universe: Colliding branes and the origin of the hot big bang}},
  \href{https://doi.org/10.1103/PhysRevD.64.123522}{\emph{Phys. Rev. D}
  {\bfseries 64} (2001) 123522}
  [\href{https://arxiv.org/abs/hep-th/0103239}{{\ttfamily hep-th/0103239}}].

\bibitem{Barrau:2013ula}
A.~Barrau, T.~Cailleteau, J.~Grain and J.~Mielczarek, \emph{{Observational
  issues in loop quantum cosmology}},
  \href{https://doi.org/10.1088/0264-9381/31/5/053001}{\emph{Class. Quant.
  Grav.} {\bfseries 31} (2014) 053001}
  [\href{https://arxiv.org/abs/1309.6896}{{\ttfamily 1309.6896}}].

\bibitem{Brandenberger:2016vhg}
R.~Brandenberger and P.~Peter, \emph{{Bouncing Cosmologies: Progress and
  Problems}}, \href{https://doi.org/10.1007/s10701-016-0057-0}{\emph{Found.
  Phys.} {\bfseries 47} (2017) 797}
  [\href{https://arxiv.org/abs/1603.05834}{{\ttfamily 1603.05834}}].

\bibitem{Agullo:2016tjh}
I.~Agullo and P.~Singh, \emph{{Loop Quantum Cosmology}},  in \emph{{Loop
  Quantum Gravity}: {The First 30 Years}}, A.~Ashtekar and J.~Pullin, eds.,
  pp.~183--240, WSP (2017),
  \href{https://doi.org/10.1142/9789813220003\_0007}{DOI}
  [\href{https://arxiv.org/abs/1612.01236}{{\ttfamily 1612.01236}}].

\bibitem{Langlois:1994ec}
D.~Langlois, \emph{{Hamiltonian formalism and gauge invariance for linear
  perturbations in inflation}},
  \href{https://doi.org/10.1088/0264-9381/11/2/011}{\emph{Class. Quant. Grav.}
  {\bfseries 11} (1994) 389}.

\bibitem{Domenech:2017ems}
G.~Dom\`enech and M.~Sasaki, \emph{{Hamiltonian approach to second order gauge
  invariant cosmological perturbations}},
  \href{https://doi.org/10.1103/PhysRevD.97.023521}{\emph{Phys. Rev. D}
  {\bfseries 97} (2018) 023521}
  [\href{https://arxiv.org/abs/1709.09804}{{\ttfamily 1709.09804}}].

\bibitem{Grain:2019vnq}
J.~Grain and V.~Vennin, \emph{{Squeezing formalism and canonical
  transformations in cosmology}},
  \href{https://doi.org/10.1088/1475-7516/2020/02/022}{\emph{JCAP} {\bfseries
  2002} (2020) 022} [\href{https://arxiv.org/abs/1910.01916}{{\ttfamily
  1910.01916}}].

\bibitem{Starobinsky:1982ee}
A.A.~Starobinsky, \emph{{Dynamics of Phase Transition in the New Inflationary
  Universe Scenario and Generation of Perturbations}},
  \href{https://doi.org/10.1016/0370-2693(82)90541-X}{\emph{Phys. Lett.}
  {\bfseries B117} (1982) 175}.

\bibitem{Starobinsky:1986fx}
A.A.~Starobinsky, \emph{{Stochastic de Sitter (inflationary) stage in the early
  universe}}, \href{https://doi.org/10.1007/3-540-16452-9_6}{\emph{Lect.Notes
  Phys.} {\bfseries 246} (1986) 107}.

\bibitem{Starobinsky:1994bd}
A.A.~Starobinsky and J.~Yokoyama, \emph{{Equilibrium state of a selfinteracting
  scalar field in the De Sitter background}},
  \href{https://doi.org/10.1103/PhysRevD.50.6357}{\emph{Phys. Rev.} {\bfseries
  D50} (1994) 6357} [\href{https://arxiv.org/abs/astro-ph/9407016}{{\ttfamily
  astro-ph/9407016}}].

\bibitem{Tsamis:2005hd}
N.~Tsamis and R.~Woodard, \emph{{Stochastic quantum gravitational inflation}},
  \href{https://doi.org/10.1016/j.nuclphysb.2005.06.031}{\emph{Nucl. Phys. B}
  {\bfseries 724} (2005) 295}
  [\href{https://arxiv.org/abs/gr-qc/0505115}{{\ttfamily gr-qc/0505115}}].

\bibitem{Finelli:2010sh}
F.~Finelli, G.~Marozzi, A.~Starobinsky, G.~Vacca and G.~Venturi,
  \emph{{Stochastic growth of quantum fluctuations during slow-roll
  inflation}},
  \href{https://doi.org/10.1103/PhysRevD.82.064020}{\emph{Phys.Rev.} {\bfseries
  D82} (2010) 064020} [\href{https://arxiv.org/abs/1003.1327}{{\ttfamily
  1003.1327}}].

\bibitem{Garbrecht:2013coa}
B.~Garbrecht, G.~Rigopoulos and Y.~Zhu, \emph{{Infrared correlations in de
  Sitter space: Field theoretic versus stochastic approach}},
  \href{https://doi.org/10.1103/PhysRevD.89.063506}{\emph{Phys. Rev. D}
  {\bfseries 89} (2014) 063506}
  [\href{https://arxiv.org/abs/1310.0367}{{\ttfamily 1310.0367}}].

\bibitem{Onemli:2015pma}
V.~Onemli, \emph{{Vacuum Fluctuations of a Scalar Field during Inflation:
  Quantum versus Stochastic Analysis}},
  \href{https://doi.org/10.1103/PhysRevD.91.103537}{\emph{Phys. Rev. D}
  {\bfseries 91} (2015) 103537}
  [\href{https://arxiv.org/abs/1501.05852}{{\ttfamily 1501.05852}}].

\bibitem{Burgess:2015ajz}
C.~Burgess, R.~Holman and G.~Tasinato, \emph{{Open EFTs, IR effects \&
  late-time resummations: systematic corrections in stochastic inflation}},
  \href{https://doi.org/10.1007/JHEP01(2016)153}{\emph{JHEP} {\bfseries 01}
  (2016) 153} [\href{https://arxiv.org/abs/1512.00169}{{\ttfamily
  1512.00169}}].

\bibitem{Kamenshchik:2021tjh}
A.Y.~Kamenshchik, A.A.~Starobinsky and T.~Vardanyan, \emph{{Massive scalar
  field in de Sitter spacetime: a two-loop calculation and a comparison with
  the stochastic approach}},
  \href{https://arxiv.org/abs/2109.05625}{{\ttfamily 2109.05625}}.

\bibitem{Grain:2017dqa}
J.~Grain and V.~Vennin, \emph{{Stochastic inflation in phase space: Is slow
  roll a stochastic attractor?}},
  \href{https://doi.org/10.1088/1475-7516/2017/05/045}{\emph{JCAP} {\bfseries
  1705} (2017) 045} [\href{https://arxiv.org/abs/1703.00447}{{\ttfamily
  1703.00447}}].

\bibitem{Vennin:2015hra}
V.~Vennin and A.A.~Starobinsky, \emph{{Correlation Functions in Stochastic
  Inflation}}, \href{https://doi.org/10.1140/epjc/s10052-015-3643-y}{\emph{Eur.
  Phys. J.} {\bfseries C75} (2015) 413}
  [\href{https://arxiv.org/abs/1506.04732}{{\ttfamily 1506.04732}}].

\bibitem{Fujita:2013cna}
T.~Fujita, M.~Kawasaki, Y.~Tada and T.~Takesako, \emph{{A new algorithm for
  calculating the curvature perturbations in stochastic inflation}},
  \href{https://doi.org/10.1088/1475-7516/2013/12/036}{\emph{JCAP} {\bfseries
  1312} (2013) 036} [\href{https://arxiv.org/abs/1308.4754}{{\ttfamily
  1308.4754}}].

\bibitem{Pattison:2017mbe}
C.~Pattison, V.~Vennin, H.~Assadullahi and D.~Wands, \emph{{Quantum diffusion
  during inflation and primordial black holes}},
  \href{https://doi.org/10.1088/1475-7516/2017/10/046}{\emph{JCAP} {\bfseries
  10} (2017) 046} [\href{https://arxiv.org/abs/1707.00537}{{\ttfamily
  1707.00537}}].

\bibitem{Biagetti:2018pjj}
M.~Biagetti, G.~Franciolini, A.~Kehagias and A.~Riotto, \emph{{Primordial Black
  Holes from Inflation and Quantum Diffusion}},
  \href{https://doi.org/10.1088/1475-7516/2018/07/032}{\emph{JCAP} {\bfseries
  07} (2018) 032} [\href{https://arxiv.org/abs/1804.07124}{{\ttfamily
  1804.07124}}].

\bibitem{Ezquiaga:2019ftu}
J.M.~Ezquiaga, J.~García-Bellido and V.~Vennin, \emph{{The exponential tail of
  inflationary fluctuations: consequences for primordial black holes}},
  \href{https://doi.org/10.1088/1475-7516/2020/03/029}{\emph{JCAP} {\bfseries
  03} (2020) 029} [\href{https://arxiv.org/abs/1912.05399}{{\ttfamily
  1912.05399}}].

\bibitem{Nakao:1988yi}
K.-i.~Nakao, Y.~Nambu and M.~Sasaki, \emph{{Stochastic Dynamics of New
  Inflation}},
  \href{https://doi.org/10.1143/PTP.80.1041}{\emph{Prog.Theor.Phys.} {\bfseries
  80} (1988) 1041}.

\bibitem{PhysRevD.46.2408}
S.~Habib, \emph{Stochastic inflation: Quantum phase-space approach},
  \href{https://doi.org/10.1103/PhysRevD.46.2408}{\emph{Phys. Rev. D}
  {\bfseries 46} (1992) 2408}.

\bibitem{Rigopoulos:2005xx}
G.I.~Rigopoulos, E.P.S.~Shellard and B.J.W.~van Tent, \emph{{Non-linear
  perturbations in multiple-field inflation}},
  \href{https://doi.org/10.1103/PhysRevD.73.083521}{\emph{Phys. Rev.}
  {\bfseries D73} (2006) 083521}
  [\href{https://arxiv.org/abs/astro-ph/0504508}{{\ttfamily
  astro-ph/0504508}}].

\bibitem{Tolley:2008na}
A.J.~Tolley and M.~Wyman, \emph{{Stochastic Inflation Revisited: Non-Slow Roll
  Statistics and DBI Inflation}},
  \href{https://doi.org/10.1088/1475-7516/2008/04/028}{\emph{JCAP} {\bfseries
  0804} (2008) 028} [\href{https://arxiv.org/abs/0801.1854}{{\ttfamily
  0801.1854}}].

\bibitem{Weenink:2011dd}
J.~Weenink and T.~Prokopec, \emph{{On decoherence of cosmological perturbations
  and stochastic inflation}},
  \href{https://arxiv.org/abs/1108.3994}{{\ttfamily 1108.3994}}.

\bibitem{Firouzjahi:2018vet}
H.~Firouzjahi, A.~Nassiri-Rad and M.~Noorbala, \emph{{Stochastic Ultra Slow
  Roll Inflation}},
  \href{https://doi.org/10.1088/1475-7516/2019/01/040}{\emph{JCAP} {\bfseries
  01} (2019) 040} [\href{https://arxiv.org/abs/1811.02175}{{\ttfamily
  1811.02175}}].

\bibitem{Ezquiaga:2018gbw}
J.M.~Ezquiaga and J.~García-Bellido, \emph{{Quantum diffusion beyond
  slow-roll: implications for primordial black-hole production}},
  \href{https://doi.org/10.1088/1475-7516/2018/08/018}{\emph{JCAP} {\bfseries
  08} (2018) 018} [\href{https://arxiv.org/abs/1805.06731}{{\ttfamily
  1805.06731}}].

\bibitem{Pattison:2019hef}
C.~Pattison, V.~Vennin, H.~Assadullahi and D.~Wands, \emph{{Stochastic
  inflation beyond slow roll}},
  \href{https://doi.org/10.1088/1475-7516/2019/07/031}{\emph{JCAP} {\bfseries
  1907} (2019) 031} [\href{https://arxiv.org/abs/1905.06300}{{\ttfamily
  1905.06300}}].

\bibitem{Pattison:2021oen}
C.~Pattison, V.~Vennin, D.~Wands and H.~Assadullahi, \emph{{Ultra-slow-roll
  inflation with quantum diffusion}},
  \href{https://doi.org/10.1088/1475-7516/2021/04/080}{\emph{JCAP} {\bfseries
  04} (2021) 080} [\href{https://arxiv.org/abs/2101.05741}{{\ttfamily
  2101.05741}}].

\bibitem{thieman_book}
T.~Thiemann, \emph{{Modern Canonical Quantum General Relativity}}, Cambridge
  University Press (2008).

\bibitem{Arnowitt:1962hi}
R.L.~Arnowitt, S.~Deser and C.W.~Misner, \emph{{The Dynamics of general
  relativity}}, \href{https://doi.org/10.1007/s10714-008-0661-1}{\emph{Gen.
  Rel. Grav.} {\bfseries 40} (2008) 1997}
  [\href{https://arxiv.org/abs/gr-qc/0405109}{{\ttfamily gr-qc/0405109}}].

\bibitem{Bojowald:2010qpa}
M.~Bojowald, \emph{{Canonical Gravity and ApplicationsCosmology, Black Holes,
  and Quantum Gravity}}, Cambridge University Press (12, 2010).

\bibitem{Liddle:1994dx}
A.R.~Liddle, P.~Parsons and J.D.~Barrow, \emph{{Formalizing the slow roll
  approximation in inflation}},
  \href{https://doi.org/10.1103/PhysRevD.50.7222}{\emph{Phys. Rev.} {\bfseries
  D50} (1994) 7222} [\href{https://arxiv.org/abs/astro-ph/9408015}{{\ttfamily
  astro-ph/9408015}}].

\bibitem{Mukhanov:1990me}
V.F.~Mukhanov, H.A.~Feldman and R.H.~Brandenberger, \emph{{Theory of
  cosmological perturbations. Part 1. Classical perturbations. Part 2. Quantum
  theory of perturbations. Part 3. Extensions}},
  \href{https://doi.org/10.1016/0370-1573(92)90044-Z}{\emph{Phys. Rept.}
  {\bfseries 215} (1992) 203}.

\bibitem{Malik:2008im}
K.A.~Malik and D.~Wands, \emph{{Cosmological perturbations}},
  \href{https://doi.org/10.1016/j.physrep.2009.03.001}{\emph{Phys. Rept.}
  {\bfseries 475} (2009) 1} [\href{https://arxiv.org/abs/0809.4944}{{\ttfamily
  0809.4944}}].

\bibitem{Lifshitz:1945du}
E.~Lifshitz, \emph{{Republication of: On the gravitational stability of the
  expanding universe}},
  \href{https://doi.org/10.1007/s10714-016-2165-8}{\emph{J. Phys. (USSR)}
  {\bfseries 10} (1946) 116}.

\bibitem{Colas:2021llj}
T.~Colas, J.~Grain and V.~Vennin, \emph{{Four-mode squeezed states: two-field
  quantum systems and the symplectic group $\mathrm{Sp}(4,\mathbb{R})$}},
  \href{https://arxiv.org/abs/2104.14942}{{\ttfamily 2104.14942}}.

\bibitem{Mukhanov:1981xt}
V.F.~Mukhanov and G.V.~Chibisov, \emph{{Quantum Fluctuations and a Nonsingular
  Universe}}, {\emph{JETP Lett.} {\bfseries 33} (1981) 532}.

\bibitem{Kodama:1984ziu}
H.~Kodama and M.~Sasaki, \emph{{Cosmological Perturbation Theory}},
  \href{https://doi.org/10.1143/PTPS.78.1}{\emph{Prog. Theor. Phys. Suppl.}
  {\bfseries 78} (1984) 1}.

\bibitem{Fidler:2015npa}
C.~Fidler, C.~Rampf, T.~Tram, R.~Crittenden, K.~Koyama and D.~Wands,
  \emph{{General relativistic corrections to $N$-body simulations and the
  Zel'dovich approximation}},
  \href{https://doi.org/10.1103/PhysRevD.92.123517}{\emph{Phys. Rev. D}
  {\bfseries 92} (2015) 123517}
  [\href{https://arxiv.org/abs/1505.04756}{{\ttfamily 1505.04756}}].

\bibitem{Figueroa:2021zah}
D.G.~Figueroa, S.~Raatikainen, S.~Rasanen and E.~Tomberg, \emph{{Implications
  of stochastic effects for primordial black hole production in ultra-slow-roll
  inflation}},  \href{https://arxiv.org/abs/2111.07437}{{\ttfamily
  2111.07437}}.

\bibitem{Lyth:2003im}
D.H.~Lyth and D.~Wands, \emph{{Conserved cosmological perturbations}},
  \href{https://doi.org/10.1103/PhysRevD.68.103515}{\emph{Phys.Rev.} {\bfseries
  D68} (2003) 103515} [\href{https://arxiv.org/abs/astro-ph/0306498}{{\ttfamily
  astro-ph/0306498}}].

\bibitem{Rigopoulos:2003ak}
G.I.~Rigopoulos and E.P.S.~Shellard, \emph{{The separate universe approach and
  the evolution of nonlinear superhorizon cosmological perturbations}},
  \href{https://doi.org/10.1103/PhysRevD.68.123518}{\emph{Phys. Rev.}
  {\bfseries D68} (2003) 123518}
  [\href{https://arxiv.org/abs/astro-ph/0306620}{{\ttfamily
  astro-ph/0306620}}].

\bibitem{Comer:1994np}
G.~Comer, N.~Deruelle, D.~Langlois and J.~Parry, \emph{{Growth or decay of
  cosmological inhomogeneities as a function of their equation of state}},
  \href{https://doi.org/10.1103/PhysRevD.49.2759}{\emph{Phys.Rev.} {\bfseries
  D49} (1994) 2759}.

\bibitem{Khalatnikov:2002kn}
I.~Khalatnikov, A.Y.~Kamenshchik and A.A.~Starobinsky, \emph{{Comment about
  quasiisotropic solution of Einstein equations near cosmological
  singularity}},
  \href{https://doi.org/10.1088/0264-9381/19/14/322}{\emph{Class.Quant.Grav.}
  {\bfseries 19} (2002) 3845}
  [\href{https://arxiv.org/abs/gr-qc/0204045}{{\ttfamily gr-qc/0204045}}].

\bibitem{Cruces:2021iwq}
D.~Cruces and C.~Germani, \emph{{Stochastic inflation at all order in slow-roll
  parameters: foundations}},
  \href{https://arxiv.org/abs/2107.12735}{{\ttfamily 2107.12735}}.

\bibitem{Cruces:2018cvq}
D.~Cruces, C.~Germani and T.~Prokopec, \emph{{Failure of the stochastic
  approach to inflation beyond slow-roll}},
  \href{https://doi.org/10.1088/1475-7516/2019/03/048}{\emph{JCAP} {\bfseries
  03} (2019) 048} [\href{https://arxiv.org/abs/1807.09057}{{\ttfamily
  1807.09057}}].

\bibitem{Pinol:2020cdp}
L.~Pinol, S.~Renaux-Petel and Y.~Tada, \emph{{A manifestly covariant theory of
  multifield stochastic inflation in phase space: solving the discretisation
  ambiguity in stochastic inflation}},
  \href{https://doi.org/10.1088/1475-7516/2021/04/048}{\emph{JCAP} {\bfseries
  04} (2021) 048} [\href{https://arxiv.org/abs/2008.07497}{{\ttfamily
  2008.07497}}].

\bibitem{Firouzjahi:2020jrj}
H.~Firouzjahi, A.~Nassiri-Rad and M.~Noorbala, \emph{{Stochastic nonattractor
  inflation}}, \href{https://doi.org/10.1103/PhysRevD.102.123504}{\emph{Phys.
  Rev. D} {\bfseries 102} (2020) 123504}
  [\href{https://arxiv.org/abs/2009.04680}{{\ttfamily 2009.04680}}].

\end{thebibliography}\endgroup
\end{document}